\documentclass[a4paper,fleqn,usenatbib]{mnras}

\usepackage{newtxtext,newtxmath}
\usepackage[T1]{fontenc}
\usepackage{ae,aecompl}

\usepackage{graphicx}	% Including figure files
\usepackage{amsmath}	% Advanced maths commands
\usepackage{amssymb}	% Extra maths symbols
\title[Galactic outflows in simulations of disc galaxies]{On the effect of galactic outflows in cosmological simulations of disc galaxies}

\author[M. Valentini et al.]
{Milena Valentini$^{1,2}$\thanks{E-mail: milena.valentini@sissa.it}, 
Giuseppe Murante$^{2}$\thanks{E-mail: murante@oats.inaf.it},
Stefano Borgani$^{2,3,4}$\thanks{E-mail: borgani@oats.inaf.it},
\newauthor
Pierluigi Monaco$^{2,3}$,
Alessandro Bressan$^{1}$,
and Alexander M. Beck$^5$
\\ ~ \\
\footnotesize 
$^{1}$ SISSA - International School for Advanced Studies, via Bonomea 265, I-34136 Trieste, Italy\\
$^{2}$ INAF - Osservatorio Astronomico di Trieste, via Tiepolo 11, I-34131 Trieste, Italy\\
$^{3}$ Astronomy Unit, Department of Physics, University of Trieste,
via Tiepolo 11, I-34131 Trieste, Italy\\
$^{4}$ INFN - National Institute for Nuclear Physics, Via Valerio 2, I-34127 Trieste, Italy\\
$^{5}$ University Observatory Munich, Scheinerstr. 1, D-81679 Munich, Germany
}

\date{Accepted 2017 May 30.  Received 2017 May 27; in original form 2016 November 3}

\pubyear{2017}

\begin{document}
\label{firstpage}
\pagerange{\pageref{firstpage}--\pageref{lastpage}}
\maketitle

\begin{abstract}

\noindent
We investigate the impact of galactic outflow modelling on the formation and evolution of a disc galaxy, 
by performing a suite of cosmological simulations with zoomed-in initial conditions (ICs) of a Milky Way-sized halo.  
We verify how sensitive the general properties of the simulated galaxy are to the way in which stellar 
feedback triggered outflows are implemented, keeping ICs, simulation code and star formation 
(SF) model all fixed.
We present simulations that are based on a version of the GADGET3 code where our 
sub-resolution model is coupled with an advanced implementation of smoothed particle hydrodynamics 
that ensures a more accurate fluid sampling and an improved description of gas mixing and 
hydrodynamical instabilities. 
We quantify the strong interplay between the adopted hydrodynamic scheme and the sub-resolution 
model describing SF and feedback.
We consider four different galactic outflow models, including the one introduced by \citet{DVS2012} and a 
scheme that is inspired by the \citet{SpringelHernquist2003} model.
We find that the sub-resolution prescriptions adopted to generate galactic outflows are the main
shaping factor of the stellar disc component at low redshift.
The key requirement that a feedback model must have to be successful in producing a 
disc-dominated galaxy is the ability to regulate the high-redshift SF (responsible for the formation 
of the bulge component), the cosmological infall of gas from the large-scale environment, and gas 
fall-back within the galactic radius at low redshift, in order to avoid a too high SF rate at $z=0$. 

\end{abstract}


%%%%%%%%%%%%%%%%%%%%%%%%%%%%%%%%%%%%%%%%%%%%%%%%%%

%%%%%%%%%%%%%%%%% BODY OF PAPER %%%%%%%%%%%%%%%%%%

\section{Introduction} 
\label{sec:introduction}

The formation of disc galaxies with a limited bulge and a dominant disc 
component is a challenging task in cosmological simulations of galaxy formation
\citep[e.g.][]{Mayer2008,Scannapieco2012}. The astrophysical processes involved in
the cosmological formation and evolution of these galaxies span a wide
dynamical range of scales, from the $\sim$parsec (pc) scales typical of star
formation (SF) to the $\sim$Mpc scales where gravitational instabilities
drive the evolution of the dark matter (DM) component and the
hierarchical assembly of galaxy-sized haloes, passing through the
$\sim$kpc scales where galactic winds distribute the feedback energy
from SF processes to the surrounding gas, thereby
regulating its accretion and subsequent SF. Sub-resolution
models are thus essential to model processes that take place at
scales not resolved by cosmological hydrodynamic simulations of galaxy
formation, but that affect the evolution on scales that are
explicitly resolved. In general, these sub-resolution models resort to
a phenomenological description of rather complex processes, through a
suitable choice of parameters, that need to be carefully tuned by
asking simulations to reproduce basic observational properties of the
galaxy population. 

Stellar feedback, that triggers galactic outflows, is a crucial
component in simulations that want to investigate galaxy formation. It
contributes to determine the overall evolution of galaxies
\citep[][M15 hereafter]{Stinson2013,Aumer2013,Marinacci2014,muppi2014}, 
to shape their morphology \citep{Crain2015,Hayward2015}, to set their sizes 
and properties \citep{IllustrisVogel2014, schaye2015eagle}. Additionally,
stellar feedback driven winds are expected to expel low angular momentum 
gas at high redshift from the innermost galaxy regions and to foster its 
following fall-back, once further angular momentum has been gained from 
halo gas \citep{Brook2012,Ubler2014,Teklu2015,Genel2015}. 
Galactic outflows contribute to the continuous interaction between galaxies 
and their surrounding medium, and allow for the interplay between stars 
and different components of the interstellar medium (ISM). Using 
non-cosmological simulations of isolated galaxies, \citet{Gatto2016} recently 
pointed out that strongly star-forming regions in galactic discs are indeed
able to launch supernova (SN)-driven outflows and to promote the
development of a hot volume-filling phase. They also noted a positive 
correlation between outflow efficiency, i.e. the mass loading factor, and 
the hot gas volume-filling fraction. Moreover, \citet{Su2016} have recently 
shown that stellar feedback is the primary component in determining the 
star formation rate (SFR) and the ISM structure, and that microphysical 
diffusion processes and magnetic fields only play a supporting role.
\defcitealias{muppi2014}{M15}

Despite the general consensus on the need of stellar feedback to form
spiral galaxies that are not too much centrally concentrated and have
a dominant disc component, it is still unclear which are the sub-resolution 
models able to capture an effective description of feedback energy
injection, and how to select the best ones among them. 
Different stellar feedback prescriptions are succeeding in promoting the 
formation of disc galaxies and a variety of approaches have been so far 
proposed to numerically describe galactic outflows \citepalias[see][and references therein]{muppi2014}.

For instance, \citet{Ubler2014} contrasted galaxy formation models
that adopt different stellar feedback prescriptions: a weak feedback
version \citep{Oser2010} that promotes spheroid formation, and a
strong feedback scheme \citep{Aumer2013} that favours disc
formation. They also compared the predicted mass assembly histories
for systems with different virial masses. \citet{Agertz2015}
investigated the effect of different SF and stellar feedback
efficiencies on the evolution of a set of properties of a late-type
galaxy: they highlighted that morphology, angular momentum, baryon
fraction and surface density profiles are sensitively affected by the
analysed parameters. A grasp on the comparison between distinct
galactic outflow models would reveal vital information on the effects
and consequences of diverse schemes, especially when these are
implemented within the same prescriptions for SF and cooling, in the
same code.

The Aquila comparison project \citep{Scannapieco2012} brought out the
crucial importance of the description of the ISM below the resolution
achievable in simulations and showed how different schemes for SF and
stellar feedback lead to the formation of galaxies having rather
different properties.  Besides the importance of the sub-resolution
prescriptions, \citet{Scannapieco2012} pointed out that a
non-negligible role in determining the simulation results is also
played by the choice of the underlying hydrodynamical scheme, either
Eulerian or Lagrangian smoothed particle hydrodynamics (SPH).

Great effort has been therefore devoted to improving numerical
techniques and enhancing hydrodynamical solvers. Advanced SPH schemes
have been proposed: modern versions of SPH adopt more accurate kernel
functions, include time-step limiters and introduce correction terms,
such as artificial conduction (AC) and viscosity (AV), in order to properly
deal with fluid mixing, to accurately describe the spread of entropy
and to follow hydrodynamical instabilities \citep{Price2008,Saitoh2009,Cullen2010,
dehnen2012,Durier2012,Pakmor2012,Price2012,Valdarnini2012,beck2015}. 
Other enhanced SPH flavours adopt a pressure-entropy formulation, in 
addition \citep{Saitoh2013,Hopkins2013,Schaller2015}.

In this paper we will use the version of SPH presented by
\citet{beck2015}, that has been implemented in the developer version
of the GADGET3 code \citep{springel2005}. This implementation of the
SPH has been applied to simulations of galaxy clusters by
\citet{Rasia2015}, who analysed the cool-core structure of clusters 
\citep[see also][]{sembolini2016a, sembolini2016b, Biffi2015}.
Interestingly, \citet{Schaller2015} analysed the influence of the
hydrodynamic solver on a subset of the Eagle simulations. They showed
that their improved hydrodynamic scheme does not affect galaxy masses,
sizes and SFRs in all haloes, but the most massive ones; moreover,
they highlighted that the higher the resolution is, the more
responsive to the accuracy of the hydrodynamic scheme simulation
outcomes are expected to be.

The analysis that we present in this paper has been carried out with
two main goals. First, the improved SPH implementation by
\cite{beck2015} has been introduced into the TreePM+SPH code GADGET3
that we use to perform simulations of galaxy formation.  Such an
implementation has shown to be able to alleviate a number of problems
that historically affect \textquotedblleft
traditional\textquotedblright $\:$ SPH flavours, e.g. the spurious loss of
angular momentum caused by the AV in shear flows,
and the presence of a numerical surface tension force that dumps
important hydrodynamical instabilities such as the Kelvin-Helmholtz
one. We now want to test the interplay between this new SPH
implementation in cosmological simulations of galaxy formation, and
our adopted sub-resolution model, named MUPPI (MUlti Phase Particle Integrator), for SF and
stellar feedback (\citealt[][M10 hereafter]{muppi2010}; \citetalias{muppi2014}).
\defcitealias{muppi2010}{M10}

Secondly, we explore a different implementation of feedback models
within the MUPPI scheme to highlight the variation in the final results 
that we obtain by modifying only the description of galactic outflows. 
We implement and test three different models for the production of
galactic outflows, besides the standard one already used by \citetalias{muppi2014}. 
One of these models, that was introduced by \citet{DVS2012}, is adapted 
to MUPPI. Another one is similar in spirit to the galactic outflow model originally 
introduced by \citet{SpringelHernquist2003}, again fitted for MUPPI. The 
third one is a variation over our standard galactic outflow prescription.

The main questions that we want to address in this paper can be
summarized as follows:
how much are results of a sub-grid scheme affected by the hydrodynamic
solver? How sensitive are the properties of a simulated galaxy when
different galactic outflow models are implemented within the same SF
model? What are the requirements for a galactic outflow
model to lead to the formation of a late-type galaxy with a limited bulge 
and a dominant disc component?

In Section \ref{sec:simCode} we describe the code on which the
simulations are based: the main modifications introduced by the new
hydrodynamical scheme are highlighted (Section \ref{sec:beck}) and
the main features of the MUPPI sub-grid model are reviewed (Section
\ref{sec:muppi}). In Sections \ref{sec:coupling} and \ref{sec:AC} we provide the
description of the coupling of the improved hydrodynamic scheme with
the MUPPI model. In Section \ref{sec:confronto} we introduce the
different galactic outflow models that we implemented, in order to
investigate and compare their key features when included in MUPPI. In 
Section \ref{sec:simms} the suite of simulations is described. In
Section \ref{sec:bestGal} we present and discuss results from cosmological
simulations that lead to the formation of disc galaxies, while we draw our 
main conclusions in Section \ref{sec:conclusions}.

\section{Numerical models}
\label{sec:simCode}

In this paper, we use a version of the TreePM+SPH GADGET3 code, a
non-public evolution of the GADGET2 code \citep{springel2005}. In this
version of the code, the standard flavour of SPH has been
improved by the implementation of hydrodynamics presented in
\citet{beck2015}. Furthermore, the code includes the description 
of a multi-phase ISM and the resulting SF model as described in
\citetalias{muppi2010} and \citetalias{muppi2014}. In this paper, we present for the first time 
simulations that are based on a version of GADGET3 where our SF 
model is interfaced with the improved implementation of SPH. 
We describe in this section the relevant aspects of
both SPH and SF implementations.

\subsection{The improved SPH scheme}
\label{sec:beck}
We briefly recall the main features of this implementation and refer
the reader to \citet{beck2015} for further details. We adopt an
entropy-density formulation of the SPH \citep{SpringelHernquist02}. We
carry out SPH interpolation by using a Wendland $C^4$ kernel function
\citep{dehnen2012} with 200 neighbours, instead of the standard cubic
spline (with $64$ neighbours). The functional form of the new
interpolating function reads:
\begin{equation}
w(q) \:=\: \frac{495}{32 \pi}(1-q)^6(1+6q+\frac{35}{3}q^2)\,\,.
\end{equation}
Here, $w(q) = h_{\rm i}^3 {W_{\rm i j} (x_{\rm i j}, h_{\rm i})}$, where 
$q= x_{\rm i j} / h_{\rm i}$, i.e. the ratio of the module of the distance between two
particles $x_{\rm i j}= | \mathbf{x}_{\rm i} - \mathbf{x}_{\rm j} |$ and the smoothing
length $h_{\rm i}$ assigned to the position of the $i$th particle. $W_{\rm i j}$ 
is the smoothing kernel employed in the evaluation of variables in the SPH 
formalism. The smoothing is adaptive, such that the product $h_{\rm i} \rho_{\rm i}$, 
with $\rho_{\rm i}$ the density of the $i$th particle, keeps roughly constant. 
Thanks to this improvement, pairing instability is avoided \citep{dehnen2012}
and a gain in accuracy in quantity estimates over the standard cubic
spline is achieved.

Besides the new kernel function, the most important improvements in
this new formulation of SPH mainly consist in an advanced treatment of
AV and AC.

The inclusion of the AV correction \citep{Cullen2010,beck2015}
allows us to suppress AV far from shocks and where unwanted. 
New second-order estimates for velocity gradients and curls are crucial 
to this purpose: improved computations of the Balsara switch \citep{balsara95} 
and of factors that reveal the presence of a shock control the action of AV, 
by removing it in purely shear flows, and by limiting its effect where the relative 
motion of gas particles appears convergent because of vorticity or rotation effects, 
as e.g. in Keplerian flows.

The AC term is implemented so as to promote the diffusion of entropy
among particles at contact discontinuities, thus removing the spurious
surface tension. As a result, it allows to follow hydrodynamical
instabilities more efficiently than in standard SPH. The contribution
of AC to the equation of the specific internal energy reads:
\begin{equation}
\biggl(\frac{d u_{\rm i}}{dt} \biggr)_{\rm cond} \:=\: - \frac{1}{2} \sum_{\rm j} 
\frac{m_{\rm j}}{\rho_{\rm ij}}(u_{\rm i} - u_{\rm j}) \: \alpha^{\rm c}_{\rm ij} v^{\rm sign, c}_{\rm ij} \bar{F}_{\rm ij}\,\,,
\label{AC1}
\end{equation}
where double indices refer to symmetrized variables; for instance
$\rho_{\rm ij}=(\rho_{\rm i}+\rho_{\rm j})/2$.  The term
$\bar{F}_{\rm ij}=[F_{\rm ij}(h_{\rm i})+F_{\rm ij}(h_{\rm j})]/2$ represents the symmetrized
scalar part of the kernel gradient, which is defined as
$F_{\rm ij}(h_{\rm i}) \mathbf{x_{\rm ij}}/x_{\rm ij}=\nabla_{\rm i} W_{\rm ij}(h_{\rm i})$.

In equation (\ref{AC1}), $\alpha^{\rm c}_{\rm ij}=(\alpha^{\rm c}_{\rm i} + \alpha^{\rm c}_{\rm j})/2$ is
the symmetrized AC coefficient, it depends on the gradient of thermal
energy, with:
\begin{equation}
\alpha^{\rm c}_{\rm i} \: =\: - \frac{h_{\rm i}}{3} \frac{| \nabla u |_{\rm i}}{|u_{\rm i}|}\,\,.
\label{AC2}
\end{equation}

The signal velocity $v^{\rm sign, c}_{\rm ij}$ used in the AC is given by \citep{Price2008}:
\begin{equation}
v^{\rm sign, c}_{\rm ij} \: =\: - \sqrt{\frac{|P_{\rm i} - P_{\rm j}|}{\rho_{\rm ij}}}\,\,,
\label{AC3}
\end{equation}
and a limiter on the AC is taken into account, since the total thermal pressure gradient 
is corrected for the contribution from gravitationally induced pressure gradients 
\citep[see][for further details]{beck2015}.

Moreover, in this enhanced implementation a time-step limiting particle wake-up 
scheme \citep[see][and references therein for a thorough description]{beck2015} has 
been adopted: it alleviates high discrepancies in the size of time-steps among nearby 
particles. Indeed, the code integrates quantities of different particles on an individual 
particle time-step basis: single time-steps are computed according to hydrodynamical 
properties of particles. Then, at a given time-step, some {\sl active} particles have their 
physical properties updated, while other particles with longer time-step are {\sl inactive} 
and will be considered only for later integrations. However, inactive particles in underdense 
regions may not be able to realize an abrupt variation of entropy or velocity of a nearby 
active particle, and to act accordingly. As a spurious net result, inactive particles are prone 
to gathering and clumping in regions where they remain almost fixed. The wake-up scheme 
acts in the following way: by comparing the signal velocity\footnote{The local signal velocity 
    $v^{\rm sign}_{\rm i}$ is the maximum value that the pairwise signal velocity $v^{\rm sign}_{\rm ij}$ 
    assumes within the smoothing length $h_{\rm i}$ among different particle pairs. The pairwise signal 
    velocity reads: $\, v^{\rm sign}_{\rm ij} =\: c_{\rm i}^{\rm s} + c_{\rm j}^{\rm s} + \beta \mu_{\rm ij}\,$,
    where $c^{\rm s}$ is the sound speed of each particle, $\mu_{\rm ij}=\mathbf{v}_{\rm ij} \cdot 
    \mathbf{x}_{\rm ij}/x_{\rm ij}$, $\mathbf{v}_{\rm ij}$ being the difference of velocity between 
    particles $i$ and $j$, and $\beta=3$ has been chosen.} 
$v_{\rm ij}^{\rm sign}$ between particles $i$ and its neighbour $j$ to the maximum 
signal velocity of $j$ and any possible contributing particle within its smoothing length 
(i.e. $v_{\rm j}^{\rm sign}$), an inactive neighbouring particle can be promoted to be 
active if $v_{\rm ij}^{\rm sign} > f_{\rm w} v_{\rm j}^{\rm sign}$, 
where $f_{\rm w}=3$ has been adopted from hydrodynamical tests \citep{beck2015}. 
In this way, time-steps of some particles that were inactive are adapted according to 
the local signal velocity $v_{\rm j}^{\rm sign}$ and just woken-up particles are then 
taken into account in the current time-step.

\subsection{Star formation and feedback sub-grid model: MUPPI}
\label{sec:muppi}

The sub-grid MUPPI model \citepalias[see][]{muppi2010,muppi2014} represents a 
multiphase ISM and accounts for SF and stellar feedback, both in thermal and
kinetic forms. A multiphase particle is the constitutive element of the
model: it is made up of a hot and a cold gas components in pressure
equilibrium, plus a virtual stellar component. A set of ordinary
differential equations describes mass and energy flows between the
different components.

A gas particle is eligible to become multi-phase whenever its density
rises above a density threshold and its temperature falls below a 
temperature threshold ($T_{\rm thresh}=10^5$ K). 
We choose $n_{\rm thres}=0.01$ 
cm$^{-3}$ as particle number density threshold, the adopted mean molecular weight 
being $\mu \sim 0.6$ \citepalias[see][]{muppi2010}; such a threshold corresponds to 
$\rho_{\rm thres} \simeq 1.5 \cdot 10^5$ M$_{\odot}$ kpc$^{-3}$ and to a density 
$n_{\rm H} \sim 0.0045$ cm$^{-3}$ in units of the number of hydrogen particles 
per cm$^3$ (log$_{10}$ (n$_{\rm H}$[cm$^{-3}$]) $\sim -2.3$), the assumed fraction of neutral 
hydrogen being $0.76$ (see also Table \ref{tab2} for the relevant parameters of 
the sub-resolution model).  
SPH density and temperature are assigned to the hot
phase when the multiphase stage begins. Cooling is then allowed,
according to the density and the metallicity of the gas particle.  Hot
gas can condense and cool because of radiative losses into a cold
phase, a fraction of which in turn evaporates because of destruction
of molecular clouds. A portion $f_{\rm mol}$ of the cold gas mass
$M_{\rm c}$ is expected to be in the molecular phase: of that, a fraction
$f_{\ast}$ will be converted into stars. Therefore, the SFR associated to 
that particles is:
\begin{equation}
\centering
\dot{M}_{\rm sf} \:=\: f_{\ast} \: \frac{f_{\rm mol} \, M_{\rm c}}{t_{\rm dyn}} \,.
\label{eq:sfr}
\end{equation}
\noindent
Here, $t_{\rm dyn}$ is the dynamical time of the cold phase. The SFR is
directly proportional to the molecular fraction $f_{\rm mol}$, that is
computed according to the phenomenological prescription by
\citet{blitz2006}:
\begin{equation}
\centering
f_{\rm mol} \:=\: \frac{1}{1+P_0/P} \,\,, 
\label{eq:f_mol}
\end{equation}
\noindent
where $P_0$ is the pressure of the ISM at which
$f_{\rm mol}=0.5$. According to this prescription, the hydrodynamic
pressure of the gas particle is
used as the ISM pressure $P$ entering in the phenomenological 
relation above. Furthermore, a fraction of the
star-forming gas is restored into the hot phase. 

SF is implemented according to the stochastic model introduced by 
\citet{SpringelHernquist2003}. A star particle of mass $M_{\ast}$ is created 
if the probability 
\begin{equation}
\centering
p \:=\: \frac{M}{M_{\ast}} \Biggl[ 1 - {\text {exp} } \Biggl( - \frac{\Delta M_{\ast}}{M}  \Biggr) \Biggr] \,\,, 
\label{eq:SF}
\end{equation}
\noindent
exceeds a randomly generated number in the interval [0,1]. In equation (\ref{eq:SF}), 
$M$ is the total initial mass of a gas particle, and $\Delta M_{\ast}$ is
the mass of the multiphase star-forming particle that has been
converted into stars in a time-step. Each star particle is spawned with mass 
$M_{\ast} = M/N_{\ast}$, where $N_{\ast}$ is the number of stellar generations, 
i.e. the number of star particles generated by each gas particle \citepalias[we adopt 
$N_{\ast}=4$; see also ][\citealt{tornatore2007}]{muppi2010}.
Sources of energy
that counterbalance the cooling process are: energy directly injected
into the ambient ISM by SN explosions (a fraction $f_{\rm fb, local}$
of $E_{\rm SN}$, the energy provided by each SN), energy provided 
by thermal feedback from dying massive
stars belonging to neighbouring star-forming particles ($f_{\rm fb,
  out} \cdot E_{\rm SN}$) and the hydrodynamical source term which accounts for
shocks and heating or cooling due to compression or expansion of the
gas particle.

A gas particle exits its multiphase stage whenever its density drops
below $0.2 \rho_{\rm thresh}$ or after a time interval $t_{\rm clock}$ that
is chosen to be proportional to the dynamical time $t_{\rm dyn}$. 
When a gas particle exits a multiphase stage, it has a probability
$P_{\rm kin}$ of being \textquotedblleft
kicked\textquotedblright $\:$ and to become a wind particle for a
time interval $t_{\rm wind}$. Both $P_{\rm kin}$ and $t_{\rm wind}$
are free parameters of the outflow model. This scheme relies on the
physical idea that stellar winds are powered by type-II SN explosions,
once the molecular cloud out of which stars formed has been
destroyed. Wind particles are decoupled from the surrounding medium
for the aformentioned temporal interval $t_{\rm wind}$. During this
time, they receive kinetic energy from neighbouring star-forming gas
particles, as described below. Nonetheless, the wind stage can be
concluded before $t_{\rm wind}$ if the particle density drops below a
chosen threshold, $0.3 \rho_{\rm thresh}$.

Stellar feedback is taken into account both in thermal \citepalias{muppi2010} and 
kinetic \citepalias{muppi2014} forms. Concerning thermal feedback, each 
star-forming particle delivers to neighbours the following amount of thermal energy
in a given time-step: 
\begin{equation}
\centering
\Delta E_{\rm heat}= f_{\rm fb, out} \: E_{\rm SN} \frac{\Delta M_{\ast}}{M_{\ast, \rm SN}}\,.
\label{eq:thFB}
\end{equation}
Here, $M_{\ast, \rm SN}$ is the mass in stars that is required on
average to have a single type-II SN event 
and $\Delta M_{\ast}$ is
the mass of the multiphase star-forming particle that has been
converted into stars. In the original implementation, the star-forming
particle shares its thermal feedback energy among neighbours within a
cone whose half-opening angle is $\theta$. The origin of the cone lies 
on the particle itself and its axis is aligned according to minus the particle's 
density gradient. Each energy donor weights its contribution to eligible particles 
using the SPH kernel, where the distance from the axis of the cone is
considered instead of the radial distance between particle pairs.

\begin{figure}
\begin{center}
\includegraphics[trim=5.4cm 2.1cm 5.7cm 3.cm, clip, width=0.461\textwidth]{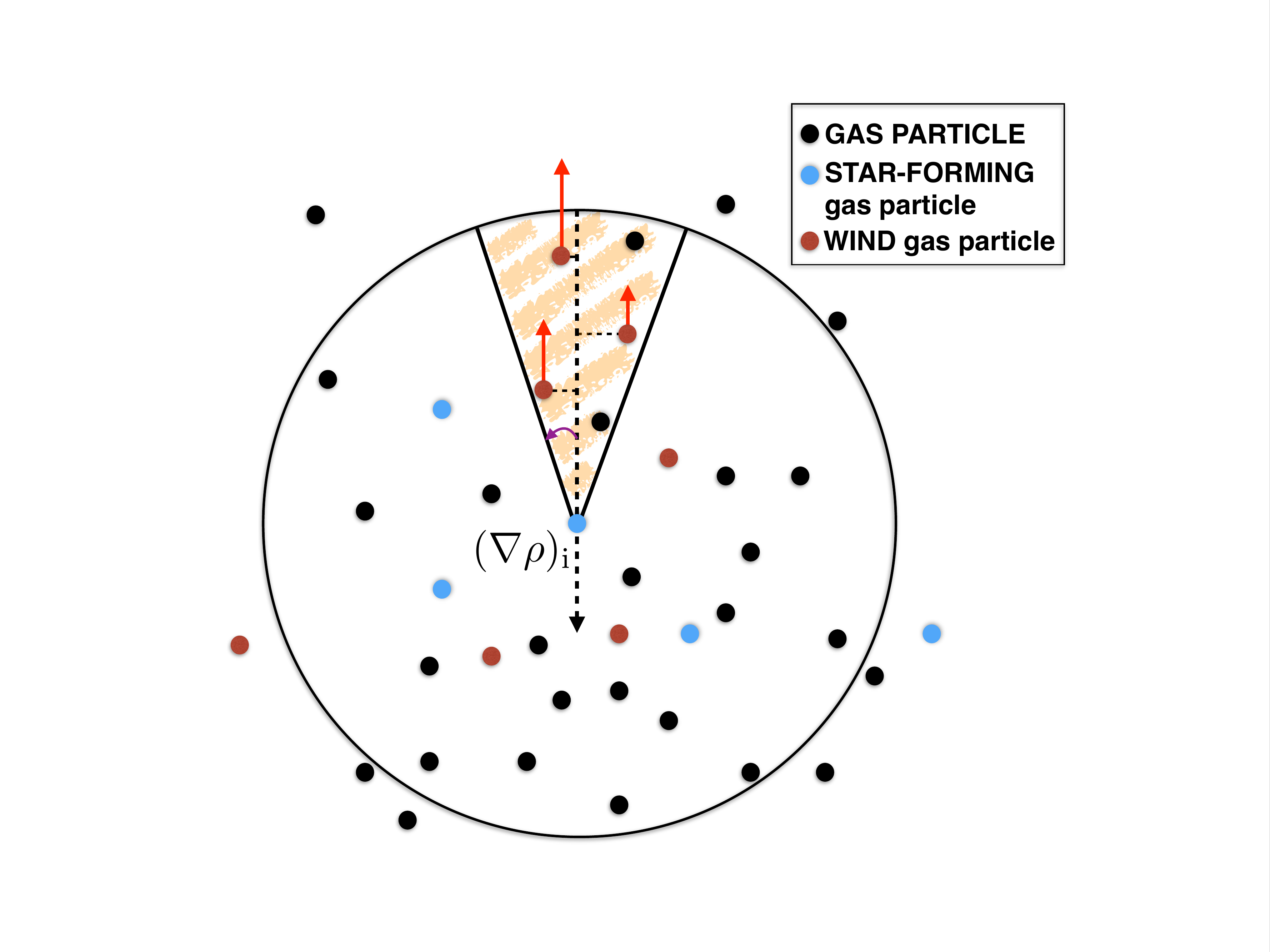}
\end{center}
\caption{Cartoon depicting the original galactic outflow model 
implemented in MUPPI: each star-forming particle provides kinetic feedback 
energy to neighbour wind particles that are located within a cone of a given 
opening angle (the purple arrow highlights $\theta$, the half-opening angle). 
The axis of the cone is aligned as minus the density gradient 
of the energy donor particle. Energy contributions are weighted according to the 
distance between wind particles and the axis of the cone.
}
\label{fig:completeFBorig}
\end{figure}

Kinetic feedback is implemented so that each star-forming particle can
provide $f_{\rm fb, kin} \cdot E_{\rm SN}$ as feedback energy. This
amount of energy is distributed to wind particles lying inside both the cone and
the smoothing length of the star-forming particle, and the delivering mechanism is the
same as the thermal scheme. Thus, outflowing energy is modelled so as to leave the
star-forming particle through the least-resistance direction \citep{Monaco2004}. 
Note that only gas particles that were selected to become wind particles are allowed to 
receive kinetic energy (see Fig. \ref{fig:completeFBorig}).

If there are no particles in the cone, the total amount of thermal
energy is given to the particle nearest to the axis
\citepalias{muppi2010,muppi2014}. This does not happen with the kinetic
energy; in this case, if no eligible wind particle can receive it, the
energy is not assigned. 

All our simulations include the metallicity-dependent cooling as
illustrated by \cite{wiersma2009}, and self-consistently describe
chemical evolution according to the model originally introduced by
\citet{tornatore2007}, to which we refer for a thorough
description. The effect of a uniform time-dependent ionizing cosmic
background is also included following \cite{HaardtMadau2001}. Star
particles are treated as simple stellar populations with a
\citet{kroupa93} initial mass function (IMF) in the mass range $[0.1,
100]$ M$_{\odot}$. Mass-dependent stellar lifetimes are computed 
according to \citet{PadovaniMatteucci1993}. We account for the 
contribution of different sources to the production of metals, namely 
SNIa, SNII, and asymptotic giant branch (AGB) stars. We trace the contribution to
enrichment of $15$ elements (H, He, C, N, O, Ne, Na, Mg, Al, Si, S,
Ar, Ca, Fe and Ni), each atomic species providing its own contribution
to the cooling rate.

\subsection{Coupling MUPPI with the improved SPH}
\label{sec:coupling}
An important point in combining the MUPPI SF model and the SPH scheme in
which an AC term is included concerns the effect of AC on the thermal
structure of the ISM. In fact, we should remind that AC is artificial and 
introduced as a switch in the energy equation to overcome intrinsic 
limitations of standard SPH. It does not describe a physical process
involving thermal conduction. Therefore, we want to suppress, or turn-off, 
AC in all the unwanted situations (as described below). One of these 
situations concerns the description of the thermal structure of the ISM, 
as provided by a sub-resolution model.

In this context, we remind that our sub-resolution MUPPI model does not
rely upon an effective equation of state to describe the ISM, thus we
do not impose the pressure of multi-phase particles to be a function of
the density by adopting a polytropic equation for $P(\rho)$, as
e.g. in \citet{SpringelHernquist2003}, and \citet{SchayeDV2008}.

Within our MUPPI model, each multiphase particle samples a portion of
the ISM separately. During the multiphase stage of a particle, 
external properties as density and pressure can change. The hot phase 
of the particle can also receive energy from neighbouring star-forming particles, 
and from SF occurring inside the particle itself. As a consequence, particles' 
average temperature can significantly change during the multiphase stage. 
Since the beginning of the multi-phase stage is not synchronized among
neighbouring particles, they can be in different evolutionary stages
and have different thermodynamical properties.

\begin{figure}
\newcommand{\captionfonts}{\small}
\hspace{-2.1ex}
\includegraphics[trim=0.4cm 1.0cm 0.2cm 0.05cm, clip, width=0.51\textwidth]{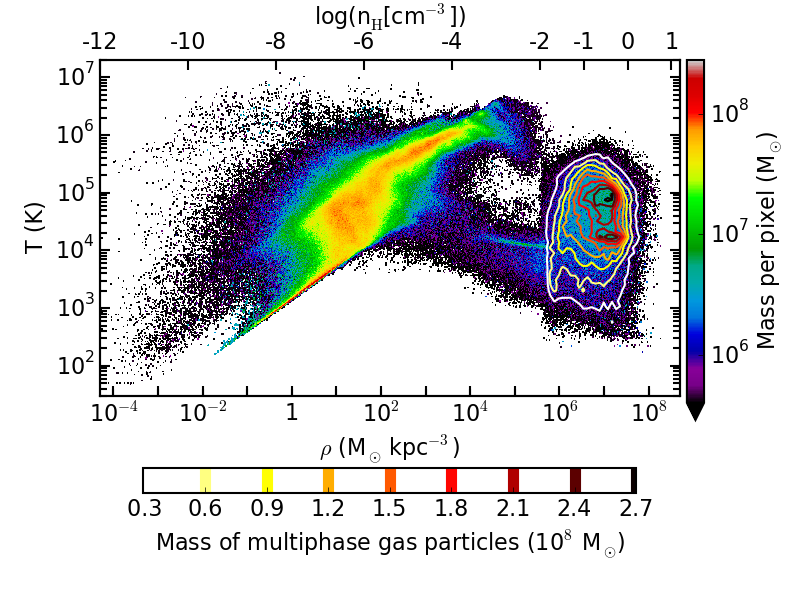} 
\caption{Phase diagram of all the gas particles (within the entire Lagrangian
  region) of the AqC5-newH simulation (see Table \ref{simmHC}). 
  Contours describe the mass distribution of multiphase particles. 
  The phase diagram is shown at redshift $z=0$. 
  Colours encode the gas mass per density-temperature bin (colour scale on the right). 
  Colours of contours are encoded as shown in the bottom colour scale. 
  Bin size is independent for the two distributions, in order to better display the region where they overlap.}
\label{AqC5_newHydro_MP_PD} 
\end{figure}

This is shown in Fig. \ref{AqC5_newHydro_MP_PD}, where we analyse the mass 
distribution in the density-temperature phase diagram of all gas particles (within the entire 
Lagrangian region) for a simulation of ours (AqC5-newH; see Section
\ref{sec:simms} for a description of the simulations used in this
work) at redshift $z=0$: multiphase particles (whose mass distribution is
shown by contours) mainly scatter across a cloud that spans three
orders of magnitude in density and almost four in temperature. The
temperature plotted in Fig. \ref{AqC5_newHydro_MP_PD} is, for
multiphase particles, the mass-weighted average of the (fixed)
temperature of the cold phase and that of the hot phase. SPH
temperature is considered for single-phase particles and the plotted
density is the SPH estimate for all particles. Colours in Fig. \ref{AqC5_newHydro_MP_PD} 
encode the gas mass per density-temperature bin.

Clearly, such a spread would not appear if multiphase particles obeyed 
an effective equation of state. It is a consequence of the fact that
MUPPI does not use a solution of the equations describing the ISM 
obtained under an equilibrium hypothesis, as e.g. in
\citet{SpringelHernquist2003}. Instead we follow the dynamical
evolution of the ISM, whose average energy depends on its past history
and on the age of the sampled ISM portion. This is the reason why two
neighbouring multiphase particles, having the same density, can have
quite different internal energies. Note that, in the most active
regions of the galaxy, the energy balance of the gas is dominated by
the ISM (sub-grid) physics, mainly via thermal feedback. The use of AC
among multiphase particles would smear their properties independently
of the evolution of the ISM they sample. Thus, AC must not be used
when the gas particles are multi-phase.

When gas particles exit their multiphase stage in MUPPI, they inherit
a mean temperature that has memory of the past multiphase stage.
AC between former multiphase particles and surrounding neighbours 
would smear local properties of the galaxy over wide regions, without 
preserving the peculiar thermal structure of the galaxy system that the 
sub-resolution model aims to account for. This is another situation where 
AC is clearly unwanted.

Finally, wind particles must also be excluded from AC, since we treat them 
as hydrodynamically decoupled from the rest of the gas. We note that, 
when they recouple, there can be a significant difference in internal
energy with the medium where they end up.

\subsection{A new switch for AC}
\label{sec:AC}

Therefore, we implemented a prescription that allows AC to be active
only when well-defined conditions on the physical properties of the
considered particles are met:
\begin{itemize}
\item[1 -] AC is switched off for both multiphase particles and wind particles 
as soon as they enter the wind stage. 
\item[2 -] Multiphase particles that exit their multiphase stage are not 
allowed to artificially conduct.
\item[3 -] All the non-multiphase particles with AC off (therefore,
  former multiphase particles and wind particles, too) are allowed to
  artificially conduct again whenever their temperature differs by
  20 per cent at most from the mass-weighted temperature of neighbouring
  particles.
\end{itemize}

Formally, our AC limiter reads as:
\begin{equation}
\centering
0.8 \cdot T_{\rm mw, Ngb}  \leq T_{\rm AC, off\rightarrow on} \leq  1.2 \cdot T_{\rm mw, Ngb} \,\,,
\label{eq:ACswitch}
\end{equation}
where $T_{\rm AC, off\rightarrow on}$ is the temperature of the particle that is
not conducting in the present time-step,
while $T_{\rm mw, Ngb}$ is the mass-weighted temperature of the 
particle's neighbours within the smoothing length.

The physical motivation behind this switch is the following.
Multiphase particles have their AC switched off, the hydrodynamics
being not able to consistently describe physical processes at the
unresolved scales that these particles capture. As soon as a
multiphase particle exits its multiphase stage, it will retain
physical properties close to the ones it had in advance for some
time. For this reason, past multiphase particles keep the AC switched
off. When the particle, not sampling anymore the ISM, reaches thermal
equilibrium with its surrounding environment, it is allowed to
artificially conduct again. We checked the dependence of our results
on the exact value of the aformentioned percentage (20 per cent): we
found that the precise temperature range within which a particle is
allowed to conduct again (i.e. $[0.8, 1.2]$ in our default case) has
not a crucial impact on the evolution of the gas, as long as it is
narrow enough.

\section{Implementation of galactic outflow models}
\label{sec:confronto}

The \citetalias{muppi2014} original kinetic stellar feedback implemented in MUPPI
successfully produced the massive outflows required to avoid excessive
SF and resulted in a lower central mass concentration of the
simulated galaxies \citepalias[][\citealt{barai2015}]{muppi2014}. 
This stellar feedback algorithm produces realistic galactic winds, in terms of 
wind velocities and mass loading factor \citep{barai2015}.

We consider here three further numerical implementations of galactic
outflows: the first one (FB1 hereafter; Section $\ref{sec:DvS}$) is
based on the scheme originally proposed by \citet{DVS2012}, the second
one (FB2 hereafter; Section $\ref{sec:FB2}$) is a modified version of
the \citetalias{muppi2014} original kinetic stellar feedback in MUPPI, where particles
that are eligible to produce galactic outflows are selected according
to a different prescription. The third one (FB3 hereafter; Section
$\ref{sec:FB3}$) is a new and distinct model, where the mass loading
factor is directly imposed.  This third scheme is similar in spirit to
that proposed by \citet{SpringelHernquist2003}, adapted to our ISM
model.

Our main aim in introducing these three additional schemes is to
verify how sensitive the general properties of the simulated
galaxy are to the way in which outflows are modelled, while keeping
initial conditions (ICs), simulation code, and SF model all
fixed. Moreover, we want to investigate how two popular outflow
models, namely those by \citet{DVS2012} and 
\citet{SpringelHernquist2003}, perform when implemented within MUPPI.

\subsection{FB1: Dalla Vecchia \& Schaye model}
\label{sec:DvS}

The model proposed by \citet{DVS2012} assumes that thermal energy
released by SN explosions is injected in the surrounding medium using
a selection criterion: to guarantee the effectiveness of the feedback,
the gas particles that experience feedback have to be heated up to a
threshold temperature. Such a temperature increase, $\Delta T$,
ensures that the cooling time of a heated particle is longer than its
sound-crossing time, so that heated gas is uplifted before it radiates 
all its energy. It is worth noting that stellar feedback is actually 
implemented as a thermal feedback in this model, since gas particles 
are heated as a result of the energy provided by type-II SNe, but the 
effective outcome is a galactic wind, because thermal energy is 
converted into momentum and hence outflows originate.

Particles have a probability to be heated by nearby star-forming
particles that depends directly on the available amount
$\varepsilon_{\rm SN II}$ of thermal energy per unit stellar mass
released by star-forming particles of mass $m_{\ast}$ as they explode
as SNe. Once the temperature increase $\Delta T$ of gas particles
receiving feedback energy has been set, the probability $p_{\rm i}$ 
that a gas particle is heated is determined by the fraction $f_{\rm fb, kin}$ 
of the total amount of type-II SNe energy that is injected on average,
i.e.:
\begin{equation}
\centering
p_{\rm i} \:=\: \frac{f_{\rm fb, kin} \, \varepsilon_{\rm SN II} \, m_{\ast}} 
{\Delta \varepsilon_{\rm i} \sum_{\rm j=1}^{N_{\rm ngb}} m_{\rm j}}\,\,, 
\label{eq:dvs}
\end{equation}
where $\Delta \varepsilon_{\rm i}$ is the thermal energy per unit mass that
corresponds to the temperature jump $\Delta T_{\rm i}$ (see below). 
Here, $\varepsilon_{\rm SNII}$ only refers to the fraction of SN energy 
that was given as {\it kinetic} energy in the original model. The thermal energy 
exchange of MUPPI remains unchanged: this is needed, since our model 
requires a thermal heating of the hot phase of multiphase particles. 
Eligible gas particles are all the neighbours ($N_{\rm ngb}$) of star-forming particles 
within their smoothing length, defined (similarly to the SPH smoothing length) as
the radius of a sphere containing $N_{\rm ngb}$ gas particles \footnote{Multiphase 
particles are also eligible particles. Thermal energy is deposited in each particle over all 
the gas mass, i.e. not only in the hot phase. Moreover, if a gas particle that has been 
selected to receive energy was in a multiphase stage, then it is forced to exit it.}.

The temperature increase $\Delta T_{\rm i}$, or equivalently the 
temperature at which the gas particle is heated when it experiences 
feedback, is a free parameter of the model. \citet{DVS2012} choose 
$\Delta T=10^{7.5}$ K as their fiducial temperature increase and found 
no significative dependence among the values explored, provided that 
they are above a given threshold that depends on mass resolution and 
that ensures Bremsstrahlung to dominate the radiative cooling. 
They also found that cooling losses make their feedback 
scheme inefficient for gas density above a critical threshold, this 
floor mainly depending on the temperature threshold described above, 
on the number of neighbouring particles and on the average (initial) 
mass of the gas particles. Moreover, when this feedback scheme is 
adopted in cosmological simulations \citep{schaye2015eagle, Crain2015}, 
a $3 \cdot 10^7$ Myr delay between SF and the release of
feedback energy is adopted. Since star particles are expected to move
away from the dense regions where they formed, such a time delay helps
in delivering energy to gas having lower densities and thus lower
radiative losses.

In our implementation of this model, we choose $\Delta T=10^{7.5}$ K 
as the minimum allowed temperature increase. Then, we account for the presence 
of gas particles whose density is higher than the maximum density for which this feedback 
model is expected to be effective (see discussion above), by computing the temperature 
increase that is needed to guarantee effectiveness. We indeed prefer to adopt a temperature
threshold ($\Delta T_{\rm i}$) that varies\footnote{
We show and discuss results of a simulation where we fix the temperature jump to the 
reference value of $T=10^{7.5}$ K in Appendix \ref{appendixB}.
} as a function of the gas density 
($\rho_{\rm i}$) instead of introducing a delay as a further free parameter. 
Therefore, we calculate the temperature that each eligible particle has to 
reach in order to make the feedback effective \citep[from equation $17$ of
][]{DVS2012}, by taking into account the number of neighbour particles
that our kernel function adopts, the average (initial) mass of gas
particles and a ratio $t_{\rm c}/t_{\rm s}=25$ between the cooling and the sound
crossing time. After carrying out extensive tests, we found that this value is required to have 
a feedback that is effective in producing a realistic disc galaxy. The difference with respect to 
that adopted by \citet{DVS2012}, $t_{\rm c}/t_{\rm s}=10$, 
should not surprise given that the same outflow model is here applied on a completely different 
sub-resolution model for the ISM and SF. 
Hence, in our implementation, the temperature jump of gas 
particles receiving energy is set to max[$\Delta T=10^{7.5}$ K, $\Delta T_{\rm i}$]. 

We implemented the feedback efficiency adopted in the reference simulation of the
Eagle set \citep{schaye2015eagle}, where the proposed feedback efficiency increases 
with the gas density $n$, computed as soon as a new star particle is spawned, and decreases 
with the metallicity $Z$, according to:
\begin{equation}
f_{\rm fb, kin}(n, Z) = f_{\rm fb, kin}^{\rm \,\,\,min} + \frac{f_{\rm fb, kin}^{\rm \,\,\,max} -  f_{\rm fb, kin}^{\rm \,\,\,min}}
{1+ \bigl( \frac{Z}{0.1 \cdot Z_{\odot}}\bigr) ^{n_{\rm Z}} \bigl( \frac{n}{ n_{H, 0}}\bigr) ^{-n_{\rm n}}  }  \,\,\,,
\label{FkEagle}
\end{equation}
\noindent
where $Z_{\odot}=0.0127$. Here, $n_{\rm H, 0}=0.67$ cm$^{-3}$, $n_{\rm Z}=0.87$, and
$n_{\rm n}=0.87$ are free parameters of the model. The above analytical expression 
foretells a $f_{\rm fb, kin}$ that ranges between a high-redshift $f_{\rm fb, kin}^{\rm \,\,\,max}$ and 
a low-redshift $f_{\rm fb, kin}^{\rm \,\,\,min}$. 
The simulation AqC5-FB1 adopts equation (\ref{FkEagle}), keeping both the aforementioned free 
parameters and $f_{\rm fb, kin}^{\rm \,\,\,max}=3.0$ and $f_{\rm fb, kin}^{\rm \,\,\,min}=0.3$, as
suggested in \citet[][see also Appendix \ref{appendixB} for further details]{schaye2015eagle}.

Note that, following the original scheme by \citet{DVS2012}, we
deliver the whole amount of SNII energy when a star is {\it spawned};
we do not use the continuous SFR provided by our SF model.
The reason for this choice will be detailed in Section \ref{FBtune}.
We verified that, in our implementation, the stochastic sampling of the
SNII deposited energy is accurate to within few \%, 
the ratio between the (cumulative) injected and expected SN feedback energy 
being $<4$ \% for all the simulations that adopt this galactic outflow model (including 
the lower resolution ones discussed in Appendix \ref{appendixB}).

\subsection{FB2: modified MUPPI kinetic feedback}
\label{sec:FB2}

In this model, we revised the kinetic stellar feedback scheme
presented in \citetalias{muppi2014}. 
Our original galactic outflow model was designed in order to produce outflows 
that are perpendicular to the forming disc, along the least-resistance direction. 
This is obtained by considering as eligible to receive energy only those wind particles 
lying in a cone centred on each star-forming particle. Moreover, the original model 
required the presence of a sufficient amount of gas mass to be accelerated
and produce an outflow; otherwise, the kinetic energy is considered to be lost. 
This may happen if no wind particles are present inside the cone.

Using such a scheme the chosen direction of least resistance is that
of the {\it star-forming} particle that gives energy to the wind one. 
However, also due to our choice of a new kernel, that adopts a larger number of 
neighbours, the least-resistance direction of the {\it wind} particle could be 
different from that of the star-forming one. For this reason we adopted a new scheme 
in which the least-resistance direction is that of the wind particle {\it receiving} energy. 

For this reason,
each star-forming particle spreads its
available amount of kinetic energy, $E_{\rm kin}=f_{\rm fb, \, kin} \cdot E_{\rm
  SN}$, among all wind particles within the smoothing length, with
kernel-weighted contributions (see below). Fig. \ref{fig:completeFB2orig}
illustrates this scheme: a wind particle gathers energy from all
star-forming particles of which it is a neighbour. The kinetic energy
of the receiving particle is then updated, the wind particle being
kicked in the direction of minus {\it its own} density gradient. Thus,
at variance with the original galactic outflow model, star-forming
particles give energy isotropically and not within a cone. In this way,
density gradients of star-forming particles are not used to give directionality 
to the outflows. Particles receiving energy use it to increase their velocity
along their least-resistance path. Energy contributions are weighted 
according to the distance that separates each wind particle that gathers 
energy and the considered energy donor star-forming particle. We 
note that this is at variance with the original galactic outflow model, where the distance 
between selected wind particles and the axis of the cone is considered, instead of the 
radial particle pair separation. 

\begin{figure}
\begin{center}
\includegraphics[trim=5.4cm 2.1cm 5.7cm 3.cm, clip, width=0.461\textwidth]{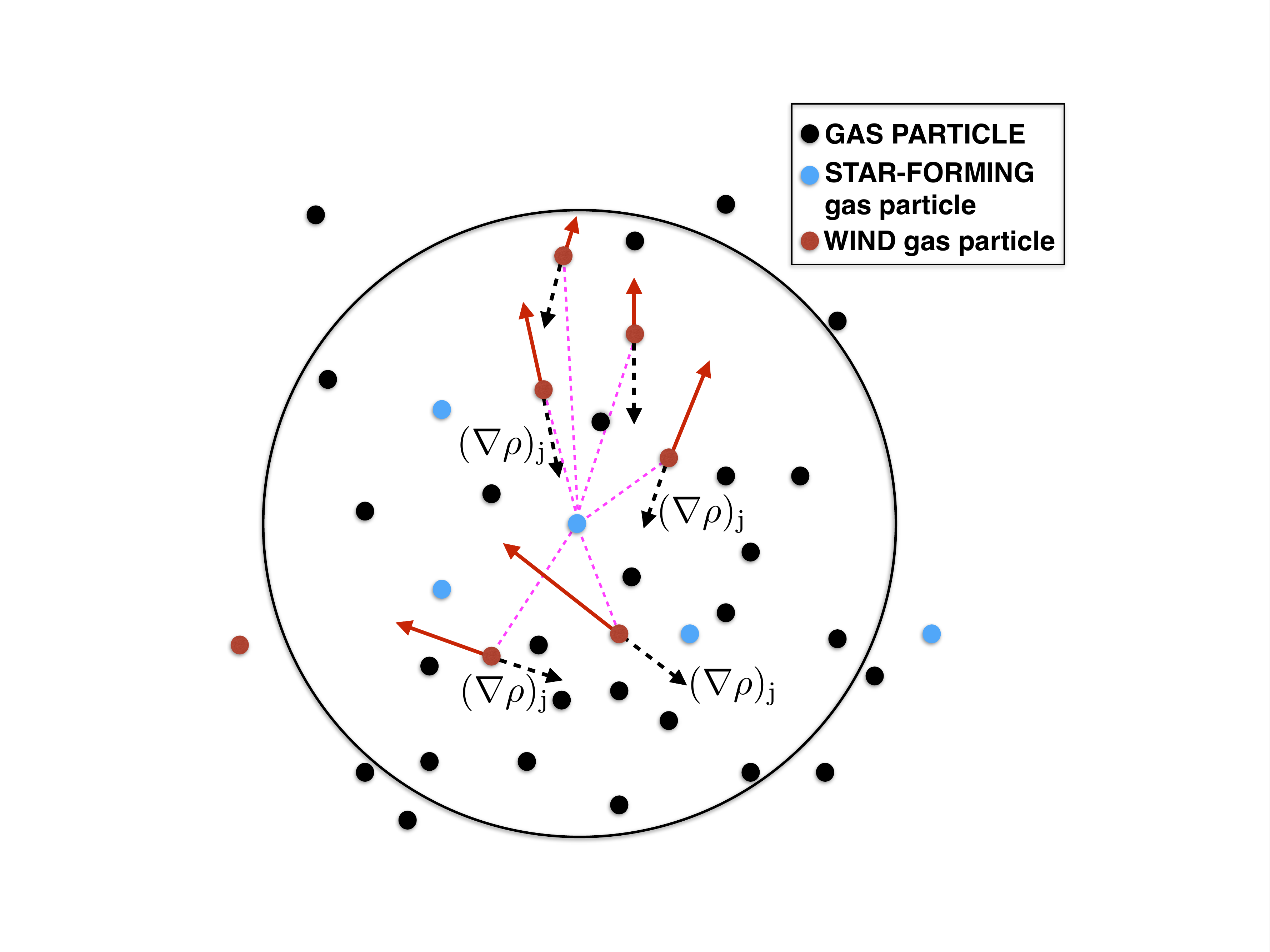}
\end{center}
\caption{Cartoon showing the FB2 galactic outflow model implemented within the MUPPI algorithm. 
At variance with the original implementation, a star-forming particle provides its feedback energy 
to all the wind particles that are located within the smoothing sphere. Each wind particle is
kicked in the direction of minus {\it its own} density gradient. Energy contributions are weighted 
according to the distance between wind particles and the energy donor star-forming particle.}
\label{fig:completeFB2orig}
\end{figure}

\subsection{FB3: fixing the mass loading factor}
\label{sec:FB3}

This model provides a new version of the original MUPPI kinetic
feedback model, that we described above, where we directly impose 
the mass load, instead of using as a free parameter the probability $P_{\rm kin}$ 
to promote gas particles to become wind particles. 

We aim at implementing a model similar to that of \citet{SpringelHernquist2003}. 
In their model equilibrium is considered to be always achieved: therefore, ISM 
properties and consequent SFR are evaluated in an instantaneous manner for each 
multiphase star-forming particle. Each star-forming particle has then a probability to 
become a wind particle.

In our model, a multiphase particle remains in its multiphase stage for a given time: 
this time is related to the dynamical time of the cold phase and computed when a 
sufficient amount of cold gas has been produced. Our model also takes into account  
the energy contribution from surrounding SN explosions, i.e. neighbouring star-forming 
particles contribute to the energy budget of each multiphase particle. 

We estimate ISM properties averaged over the lifetime that each multiphase
particle spends in its multiphase stage. We then use these properties to determine 
the probability for each multiphase particle of receiving kinetic energy and thus becoming 
a wind particle once it exits its multiphase stage.
For this reason, we have to sum the energy contributions from SN explosions provided by 
star-forming particles over the multiphase stage of each gas particle receiving the energy. 
Moreover, we also have to estimate the mass of gas that goes in outflow, and to compute 
time-averaged quantities.

In what follows, we always use the index $i$ when referring to star-forming, energy-giving particles, 
and the index $j$ for particles that receive that energy. 

Our aim is to produce an outflow with a fixed mass loading factor $\eta$. 
To obtain this, we relate the wind mass-loss rate $\dot{m}_{\rm w}$ to the SFR of each star-forming particle $i$:
\begin{equation}
\centering
\dot{m}_{\rm w}  \:=\:  \eta \, \dot{m}_{\ast} \,\,.
\label{eq:m02}
\end{equation}

In our FB3 model, particles that receive energy are the multiphase
neighbours of multiphase star-forming particles within their entire 
smoothing sphere. Each of these multiphase particles collects energy from all
its neighbouring multiphase star-forming particles and stores it
during its entire multiphase stage.  

The wind mass-loss rate $\dot{m}_{\rm w}$ can be expressed as 
the outflowing mass $\widetilde{m}_{\rm w}$ in a time interval (i.e., per time-step): 
\begin{equation}
\centering
\dot{m}_{\rm w}  \:=\: \frac{\widetilde{m}_{\rm w}}{\Delta t}   \,\,.
\label{eq:m1}
\end{equation}
Hence, for each star-forming particle $i$ one gets: 
\begin{equation}
\centering
\widetilde{m}_{\rm w, \, i, \, \Delta t}  \:=\: \eta \, \Delta t \, \dot{m}_{\rm  \ast, \, i} \,\,, 
\label{eq:m2}
\end{equation}
where $\dot{m}_{\rm \ast, \, i}$ is the SFR of that same particle.
Such a wind mass loss is taken out from the total gas mass contributed by all the SPH 
multiphase neighbours $k$\footnote{The index $k$ labels gas particles 
within the smoothing length of each star-forming particle $i$; note that each energy-receiving gas particle $j$
can have many star-forming particles inside its smoothing length.} of the particle $i$. 
This mass is computed as: 
\begin{equation}
\centering
m_{\rm tot, \, i, \, \Delta t}  \:=\: \sum_{\rm k} m_{\rm k} \,\,, 
\label{eq:p1}
\end{equation}
$m_{\rm k}$ being the mass of each energy-receiving gas particle
during the time-step interval $\Delta t$. Thus, each star-forming particle $i$ distributes 
energy to the multiphase receiving particles $k$; it also gives to each such particles the information 
on the required mass loading factor, and on the total gas mass to which the mass loss is referred. 

The total kinetic energy that a multiphase particle $j$ is provided
with in every time-step $\Delta t$ can be expressed as:
\begin{equation}
\centering
E_{\rm kin \, tot, \, j, \, \Delta t}  \:=\: \sum_{\rm i} f_{\rm fb, kin} \, \varepsilon_{\rm SN II} \, \frac{d m_{\rm \ast, \, i}}{dt} \, \Delta t \,\,, 
\label{eq:0}
\end{equation}
where $\varepsilon_{\rm SN II} \frac{d m_{\rm \ast, \, i}}{dt} \, \Delta t$ is the kinetic 
energy released per time-step by each neighbour star-forming particle whose 
stellar mass component is $m_{\rm \ast, \, i}$. 
The sum is over all neighbouring star-forming particles $i$. 
The corresponding total kinetic energy $E_{\rm kin \, tot, j}$ that is stored during the
entire multiphase stage of particle $j$ is therefore obtained by integrating equation 
($\ref{eq:0}$) over the time $t_{\rm MP, j}$ that the particle $j$ spends in its multiphase 
stage (i.e., summing over all the time-steps that make up the multiphase stage). 
This means that we have to sum all the energy contributions 
$E_{\rm kin \, tot, \, j, \, \Delta t}$ during the time interval $t_{\rm MP, j}$: 
\begin{equation}
\centering
E_{\rm kin \, tot, j}  \:=\: \sum_{\rm \Delta t} {E_{\rm kin \, tot, \, j, \, \Delta t} \,} \,\, \:=\: 
\sum_{\rm \Delta t } \sum_{\rm i} f_{\rm fb, kin} \, \varepsilon_{\rm SN II} \, \frac{d m_{\rm \ast, \, i}}{dt} \, \Delta t \,\,.
\label{eq:01}
\end{equation}
The rightmost term considers that each energy contribution $E_{\rm kin \, tot, \, j, \, \Delta t}$ has already 
been computed during each time-step $\Delta t$ in equation ($\ref{eq:0}$). 

The required outflowing mass per time-step, sampled by particle $j$, is the sum over all neighbouring 
star-forming particles within the smoothing length of $j$: 
\begin{equation}
\centering
\widetilde{m}_{\rm w, \, j, \, \Delta t}  \:=\:  \sum_{\rm i}  \widetilde{m}_{\rm w, \, i, \, \Delta t}   \:=\:  \eta \, \Delta t \, \sum_{\rm i} \dot{m}_{\ast \rm , \, i} \,. 
\label{eq:m3}
\end{equation}
Note that each star-forming particle $i$ gives a different contribution to particle $j$, being its 
SFR different. Therefore, the outflowing mass associated with each multiphase particle that 
experiences feedback can be obtained by summing all the contributions of equation ($\ref{eq:m3}$) over
the whole duration of the multiphase stage of particle $j$: 
\begin{equation}
\centering
\widetilde{m}_{\rm w, \, j}  \:=\: \sum_{\rm \Delta t} {\widetilde{m}_{\rm w, \, j, \, \Delta t} \,} \,\,
\:=\: \sum_{\rm \Delta t} \eta \, \Delta t \, \sum_{\rm i} \dot{m}_{\ast \rm , \, i} \,.
\label{eq:m4}
\end{equation}

We want to stochastically sample the outflowing mass $\widetilde{m}_{\rm w, \, j}$. Thus, 
when a multiphase particle exits the multiphase stage, this particle will have a probability:
\begin{equation}
\centering
p_{\rm j} \:=\: \frac{\widetilde{m}_{\rm w, \, j}}{m_{\rm j}}\cdot \frac{m_{\rm j}}{m_{\rm tot, \, j}}
\label{eq:1}
\end{equation}
of receiving the energy: 
\begin{equation}
\centering
E_{\rm rec, \, j} \:=\: E_{\rm kin \, tot, \, j} \cdot \frac{m_{\rm j}}{\widetilde{m}_{\rm w, \, j}} \,\,\,.
\label{eq:2}
\end{equation}

In equation ($\ref{eq:1}$), $m_{\rm tot, \, j}$ is the total mass of the multiphase gas 
that produces the outflow $\widetilde{m}_{\rm w, \, j}$. At each time-step, we have:
\begin{equation}
\centering
m_{\rm tot, \,j, \, \Delta t}  \:=\: \sum_{\rm i} m_{\rm tot, \, i, \, \Delta t} \,\,,
\label{eq:p2}
\end{equation}
where $m_{\rm tot, \, i, \, \Delta t}$ is computed according to equation ($\ref{eq:p1}$).
Since the mass $m_{\rm tot, \, j, \, \Delta t }$ can vary from time-step to time-step, 
we evaluate a time-averaged total mass of the multiphase gas that 
produces the outflow, for each particle $j$:
\begin{equation}
\centering
m_{\rm tot, \, j}  \:=\:  \frac{ \sum_{\rm \Delta t} (m_{\rm tot, \, j, \, \Delta t} \cdot {\rm \Delta t)}}{t_{\rm MP}} \,\,;
\label{eq:p3}
\end{equation}
here, the sum is over the time-steps that make up the multiphase stage whose total duration is $t_{\rm MP}$. 

In this way, each multiphase particle $j$ samples the SFR of its neighbours; this SFR is 
associated to the mass {\it of the star-forming particle} as far as the estimate of the mass loss is concerned, 
not to that of the receivers. 
At the same time, each multiphase particle collects the SNII energy output from
the same star-forming particles. Note that each star-forming particle
gives its {\it entire} energy budget to {\it all} the receivers. 
A particle ending its multiphase stage receives feedback energy 
and is uploaded into an outflow with its probability $p_{\rm j}$. 
The kinetic energy of the receiving multiphase particle is therefore updated, the particle being
kicked in the direction of minus its own density gradient.
Particles that experience feedback are hydrodynamically decoupled 
(for a time interval $t_{\rm wind}=15$ Myr) from the surrounding medium as soon as they gain
feedback energy. 

Energy conservation reads: 
\begin{equation}
\centering
E_{\rm tot} \:=\: \sum_{\rm j}  p_{\rm j}  E_{\rm rec, \, j} \:=\: 
\sum_{\rm j} m_{\rm j} \frac{E_{\rm kin \, tot, \, j}}{m_{\rm tot, \, j}}  \,\,,
\label{eq:enConserv}
\end{equation}
where the last equation has been obtained by plugging equations 
($\ref{eq:1}$) and ($\ref{eq:2}$) into the second term. 
Energy conservation is proven since the total feedback energy received by 
all the multiphase particles ending their multiphase stage amounts to the total energy 
budget provided by star-forming particles exploding as type-II SNe, as ensured 
by our stochastic sampling of the energy. 

We verified that the stochastic sampling correctly represents the
desired mass loading factor and energy deposition to within 5\% in
the considered cases. This figure can slightly vary with the parameters of the
model, as in our FB1 scheme; as for FB1 model, resolution 
leaves the accuracy of the stochastic sampling almost unaffected.

\begin{table}
\centering
\caption{Name and details of the simulations used in order to 
quantify the impact of the hydrodynamical scheme.
Column 1: label of the run.
Column 2: mass of DM particles. 
Column 3: initial mass of gas particles. 
Column 4: Plummer-equivalent softening length of the gravitational interaction. 
Column 5: hydrodynamic scheme.
Column 6: galactic outflow model.}
\renewcommand\tabcolsep{1.3mm}
\begin{tabular}{@{}lccccl@{}}
\hline
Simulation &  $M_{\rm DM}$ & $M_{\rm gas}$ & 
$\varepsilon_{\rm Pl}$ & Hydro & Stellar \\ 
 &   (h$^{-1}$M$_{\odot}$) &  (h$^{-1}$M$_{\odot}$) & 
 (h$^{-1}$kpc) & scheme  &  feedback \\ 
\hline
\hline
AqC5-newH &  $1.6 \cdot 10^6$ & $3.0 \cdot 10^5$ & 0.325 &  New & Original$^1$ \\  
\hline
AqC5-oldH &   $1.6 \cdot 10^6$ & $3.0 \cdot 10^5$ & 0.325 & Old & Original \\  
\hline
AqC6-newH &  $1.3 \cdot 10^7$ & $4.8 \cdot 10^6$ & 0.650 & New & Original \\  
\hline
AqC6-oldH &  $1.3 \cdot 10^7$ & $4.8 \cdot 10^6$ & 0.650 & Old & Original \\ 
\hline
\hline
 
\end{tabular}
\label{simmHC}
\rightline{$^1$ \citet{muppi2014}}
\end{table}

One important characteristic of this model is that the outflow
velocity is fixed once $\eta$ and $f_{\rm fb, kin}$ are fixed. In
fact, by equating the kinetic energy given to the particle to that
received from neighbouring type-II SNe, we find:
\begin{equation}
v_{\rm w} = \sqrt{{2 f_{\rm fb, kin}
    \varepsilon_{\rm SN II}}/{\eta}} 
\end{equation}
\noindent
where $\varepsilon_{\rm SN II}$ is the thermal energy per unit stellar
mass released by type-II SNe (that only depends on the chosen IMF).
For this reason, our model is similar to that of
\citet{SpringelHernquist2003}. The important change is that SNII
energy and SFRs are collected during the life of the portion of the ISM
that is sampled, taking into account the fact that changes in the
local hydrodynamics (e.g. mergers, shocks, and pressure waves) can
influence these values.

Also in this model, thermal energy is still distributed according to the original MUPPI scheme.

\section{The set of simulations}
\label{sec:simms}

We performed cosmological simulations with zoomed-in ICs
of an isolated halo of mass $M_{\rm halo, DM} \simeq 2 \cdot 10^{12}$
M$_{\odot}$, expected to host a disc galaxy resembling the Milky Way (MW), 
also because of its quiet
low-redshift merging history. The ICs used in
this work have been first introduced by \citet{Springel2008} and
then used, among others, in the Aquila comparison project
\citep{Scannapieco2012}. We refer to these ICs as AqC in the paper.
These ICs are available at different resolutions: here we consider an
intermediate-resolution case with a Plummer-equivalent softening
length for the gravitational interaction of 325 h$^{-1}$pc (AqC5),
and a lower resolution version (AqC6, with a maximum physical gravitational 
softening of 650 h$^{-1}$pc). Further details related to these ICs are given by
\citetalias{muppi2014}. The cosmological model is unchanged, thus we use a $\Lambda$ cold dark matter
cosmology, with $\Omega_{\rm m}=0.25$, $\Omega_{\rm \Lambda}=0.75$, 
$\Omega_{\rm baryon}=0.04$, and $H_{\rm 0}=73$ km s$^{-1}$ Mpc$^{-1}$.

Even if the mass of the AqC halo is similar to that of the MW, 
no attempts to mimic the accretion history of its dynamical environment were made. 
Thus, our simulations should not be taken as a model of our Galaxy. 

We list the series of simulations that we ran in order to quantify the
impact of the new hydrodynamical scheme in Table \ref{simmHC}. Each
simulation was given a name: the identifying convention encodes the
resolution of the simulation (AqC5 or AqC6) and the adopted
hydrodynamical scheme, where {\sl newH} refers to the improved SPH
implementation \citep{beck2015} and {\sl oldH} indicates the standard
SPH scheme. Mass resolutions (for both gas and DM particles) and
gravitational softenings are also provided in Table \ref{simmHC}. All
these simulations adopt the original \citetalias{muppi2014} galactic outflow model.

We ran simulations AqC5-newH and AqC6-newH adopting the advanced SPH
implementation and using the values of the relevant parameters of the
MUPPI model as listed in Table \ref{tab2}
(first and third rows). 

\begin{table*}
\centering
\begin{minipage}{\linewidth} %{175mm}
\caption{Relevant parameters of the sub-resolution model.
Column 1: label of the run.
Column 2: temperature of the cold phase.
Column 3: pressure at which the molecular fraction is $f_{\rm mol}=0.5$.
Column 4: half-opening angle of the cone, in degrees.
Column 5: maximum lifetime of a wind particle \citepalias[see also][]{muppi2014}.
Column 6: duration of a multiphase stage in dynamical times.
Column 7: number density threshold for multiphase particles.
Column 8: gas particle's probability of becoming a wind particle.
Columns 9 and 10: thermal and kinetic SN feedback energy efficiency, respectively.
Column 11: fraction of SN energy directly injected into the hot phase of the ISM.
Column 12: evaporation fraction. Column 13: SF efficiency, as a fraction of the molecular gas.}
\renewcommand\tabcolsep{2.901mm}
\begin{tabular}{@{}lcccccccccccc@{}}
\hline
Simulation & $T_{\rm c}$ & $P_{\rm 0}$ & $\theta$ & 
$t_{\rm wind}$ & $t_{\rm clock}/t_{\rm dyn}$ & $n_{\rm thresh}$ 
& $P_{\rm kin}$ & $f_{\rm fb, out}$ & $f_{\rm fb, kin}$ 
& $f_{\rm fb, local}$ & $f_{\rm ev}$ & $f_{\star}$ \\ 
 & $(K)$ & (k$_{\rm B}$ K cm$^{-3}$) & ($^{\circ}$) 
& (Myr) &   & (cm$^{-3}$) 
&  &  &  
&  &  &  \\ 
\hline
\hline
AqC5-newH & 300 & 2 $\cdot 10^4$ & 30 & 20$-t_{\rm clock}$ & 1 & 0.01 & 0.05 & 0.2 & 0.7 
& 0.02 & 0.1 & 0.02 \\  
\hline
AqC5-oldH & 300 & 2 $\cdot 10^4$ & 60 & 30$-t_{\rm clock}$ & 1 & 0.01 & 0.03 & 0.2 & 0.5 
& 0.02 & 0.1 & 0.02 \\  
\hline
AqC6-newH & 300 & 2 $\cdot 10^4$ & 30 & 20$-t_{\rm clock}$ & 1 & 0.01 & 0.03 & 0.2 & 0.8 
& 0.02 & 0.1 & 0.02 \\  
\hline
AqC6-oldH & 300 & 2 $\cdot 10^4$ & 60 & 30$-t_{\rm clock}$ & 1 & 0.01 & 0.03 & 0.2 & 0.5 
& 0.02 & 0.1 & 0.02 \\  
\hline
\hline

\end{tabular}
\label{tab2}
\end{minipage}
\end{table*}

AqC5-oldH and AqC6-oldH are the reference simulations with the old
hydrodynamical scheme; these simulations marginally differ from the runs that 
have been presented and described by \citetalias{muppi2014}, as described below. Here, we adopt 
sets of stellar yields that are newer with respect to those used by \citetalias{muppi2014}; they
are taken from \citet{Karakas2010} for AGB 
stars, from \citet{Thielemann2003} and \citet{ChieffiLimongi2004} for
type-Ia and type-II SNe, respectively. After this change, the parameter 
$f_{\rm fb, kin}$ describing the fraction of SN energy directly injected into 
the ISM had to be slightly fine-tuned again (from $0.6$ to $0.5$, see Table \ref{tab2}). 
Such a change modifies only barely the outcome of the original 
simulations. All the parameters of the sub-resolution model
that have been chosen to carry out AqC5-oldH and AqC6-oldH simulations
are provided in Table \ref{tab2} (second and fourth rows).

Table \ref{simmFC} lists the simulations that we carried out in
order to investigate the effect of wind modelling on final properties
of the simulated disc galaxy. In this table we summarize the
hydrodynamical scheme adopted and the type of galactic outflow model
implemented within the MUPPI sub-resolution prescriptions for cooling
and SF. 

\begin{table}
\centering
\caption{Name and main features of the simulations performed in order 
to investigate the effect of the adopted galactic outflow model on final 
properties of the simulated galaxy.
Column 1: label of the run.
Column 2: hydrodynamic scheme.
Column 3: galactic outflow model (see Section \ref{sec:confronto} for details).
All these runs adopt the softening length and masses of DM and gas particles 
of the simulation AqC5-newH.}
\renewcommand\tabcolsep{3.6mm}
\begin{tabular}{@{}lcr@{}}
\hline
Simulation & Hydro scheme & Stellar feedback \\ 
\hline
\hline
AqC5-FB1 &  New & First outflow model$^2$ (\ref{sec:DvS})\\  
\hline
AqC5-FB2 &  New & Second new outflow model (\ref{sec:FB2})\\  
\hline
AqC5-FB3 &  New & Third new outflow model (\ref{sec:FB3})\\  
\hline
\hline
 
\end{tabular}
\label{simmFC}
\rightline{$^2$ \citet{DVS2012}}
\end{table}

Simulations AqC5-FB1, AqC5-FB2, and AqC5-FB3 all adopt the new
hydrodynamical scheme and all have the same gravitational softening
and mass resolutions of AqC5-newH. We will compare these simulations
to AqC5-newH, that has been performed with the \citetalias{muppi2014} original outflow
model, in order to quantify the variation in the final results when
modifying the model of stellar feedback.

\section{Results}
\label{sec:bestGal}

In this section we present results from the simulations listed in Section 
\ref{sec:simms}, that led to the formation of disc galaxies. After discussing
the relevance of introducing a switch to suppress AC in 
star-forming gas particles (Section \ref{sec:sfr}), we show the properties 
of two disc galaxies simulated by adopting the new and the old hydrodynamic 
schemes, respectively, and we focus on the differences between the two SPH 
implementations (Section $\ref{NewHydroOldFb}$). In Section 
$\ref{NewHydroNewFb}$ we show results for different versions of the model 
of galactic outflows, all implemented within the new SPH scheme.

\subsection{The importance of the AC switch}
\label{sec:sfr}

\begin{figure*}
\newcommand{\captionfonts}{\small}
\begin{minipage}{\linewidth}%{1.01\linewidth}
\centering
\includegraphics[trim=0.6cm 1.0cm 0.25cm 0.05cm, clip, width=0.49\textwidth]{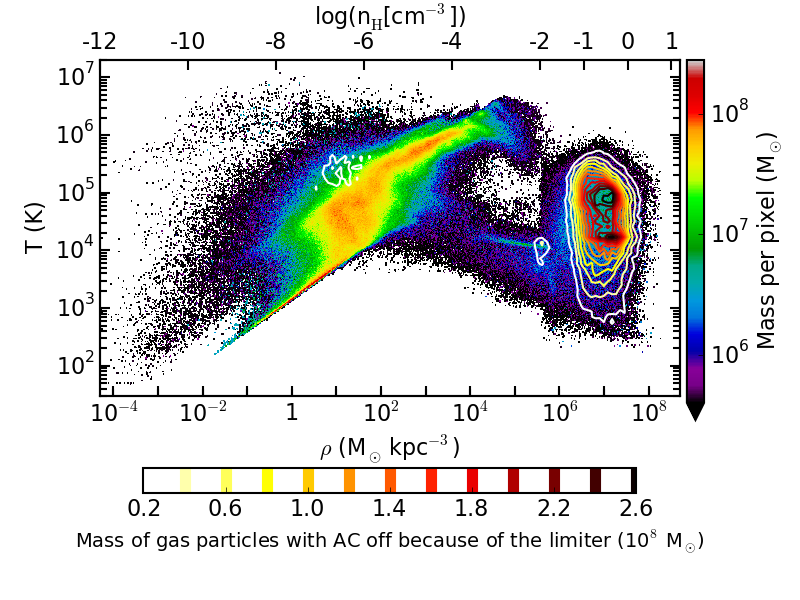} 
\includegraphics[trim=0.6cm 1.0cm 0.25cm 0.05cm, clip, width=0.49\textwidth]{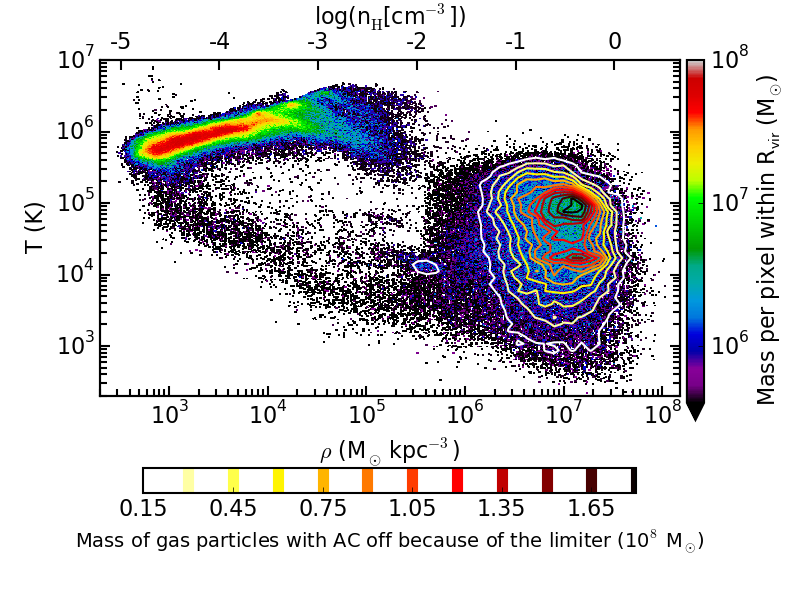} 
\end{minipage} 
\caption{{\sl Left-hand panel:} phase diagram of all the gas particles (within
  the entire Lagrangian region) of the AqC5-newH galaxy simulation. {\sl
    Right-hand panel:} phase diagram of gas particles within the virial
  radius for the AqC5-newH run. In both panels, contours depict the mass  
  distribution of gas particles with the AC switched off because of
  the effect of the new AC limiter, within the entire Lagrangian region (left-hand panel) 
  and within the virial radius (right-hand panel). Phase diagrams are shown at
  redshift $z=0$. 
  Colour encodes the gas mass per density-temperature bin (colour scale on the right of each panel). 
  Colours of contours are encoded as shown in the bottom colour scales. Colour scales differ for the two panels.}
\label{AqC5_newHydro_ACoff_PDs} 
\end{figure*}

In order to assess the effect of the AC switch, we first investigate
the properties of the gas particles with AC turned off. Fig.
\ref{AqC5_newHydro_ACoff_PDs} shows the density-temperature phase
diagram of gas particles for the AqC5-newH simulation (see also Section
\ref{NewHydroOldFb}) at redshift $z=0$. The left-hand panel is for all the
gas particles of the simulation, while the right-hand panel shows the phase
diagram of the gas particles located within the virial radius\footnote{We 
consider virial quantities as those calculated in a sphere that is centred 
on the minimum of the gravitational potential of the halo and that encloses 
an overdensity of 200 times the {\sl critical} density at present time.} 
of the galaxy ($R_{\rm vir}\simeq 240$ kpc).

In both panels, contours show the mass distribution of gas
particles with AC switched off as a consequence of the implementation
of our AC limiter. In fact, contours encircle gas particles that have
just exited their multiphase stage and still keep their AC turned
off, and gas particles that exited their last multiphase lapse few
time-steps ago but whose temperature is not in the range 
(see Section \ref{sec:AC}) allowed to artificially conduct again. 
We note that particles encircled by contours in Fig.
\ref{AqC5_newHydro_ACoff_PDs} are not all the particles that cannot
artificially conduct: multiphase and wind particles have AC switched off, too.

As expected, the majority of non-multiphase particles having AC
switched off are located in high-density regions, where also
multiphase particles lie. This makes it clear that the switch is acting
to avoid too fast a smearing of the thermodynamical properties of the
gas that here is mainly determined by the effect of our sub-grid model
rather than by hydrodynamics. It is also apparent that AC 
is normally acting in all the remaining gas particles
within the virial radius.

\begin{figure*}
\newcommand{\captionfonts}{\small}
\begin{minipage}{\linewidth}%{1.0\linewidth}
\centering
\includegraphics[trim=0.5cm 0.02cm 1.9cm 1.25cm, clip, width=0.49\textwidth]{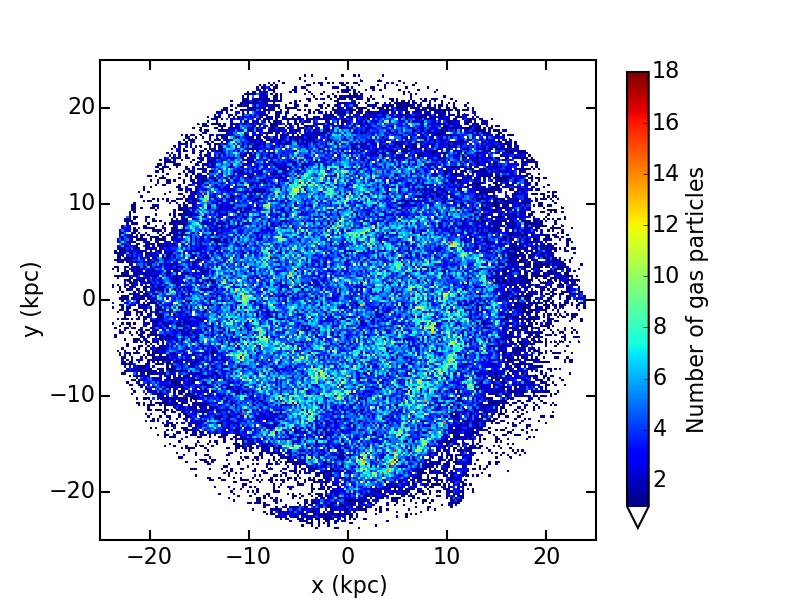} 
\includegraphics[trim=0.3cm 0.02cm 1.9cm 1.25cm, clip, width=0.49\textwidth]{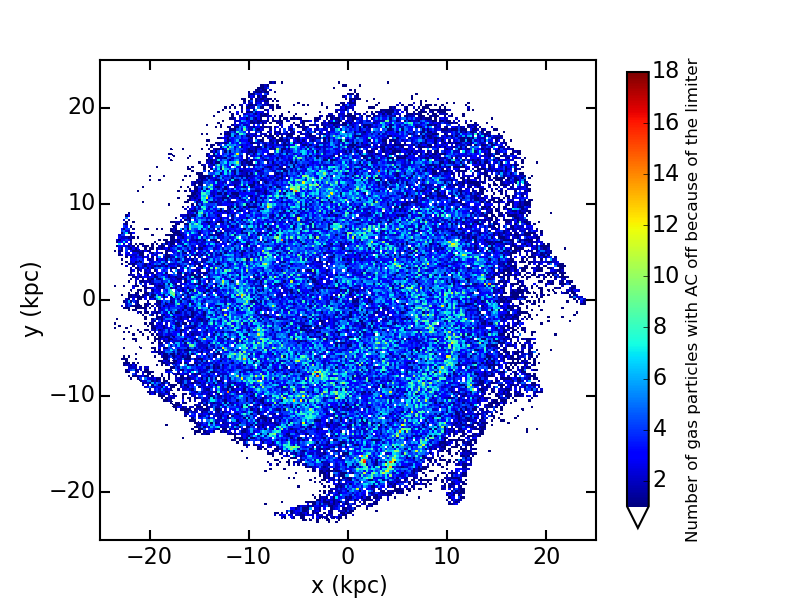} 
\includegraphics[trim=0.5cm 0.02cm 1.9cm 1.25cm, clip, width=0.49\textwidth]{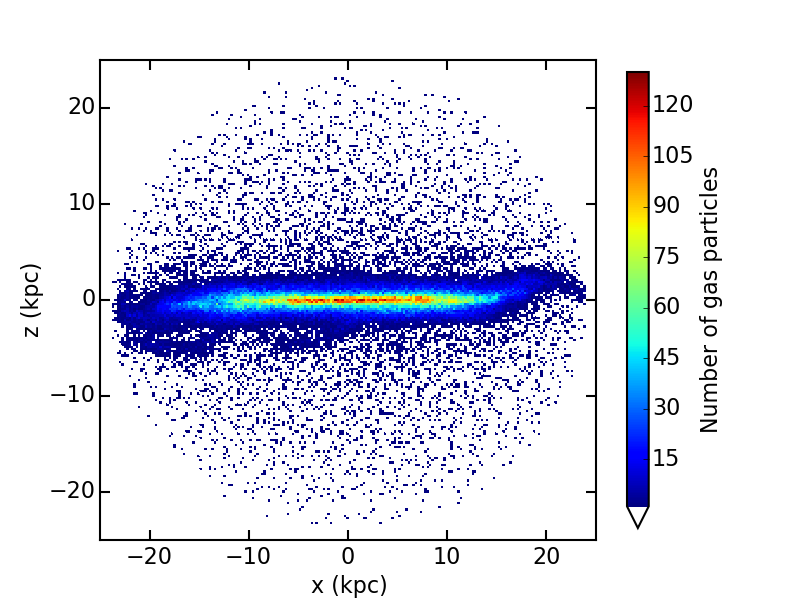} 
\includegraphics[trim=0.3cm 0.02cm 1.9cm 1.25cm, clip, width=0.49\textwidth]{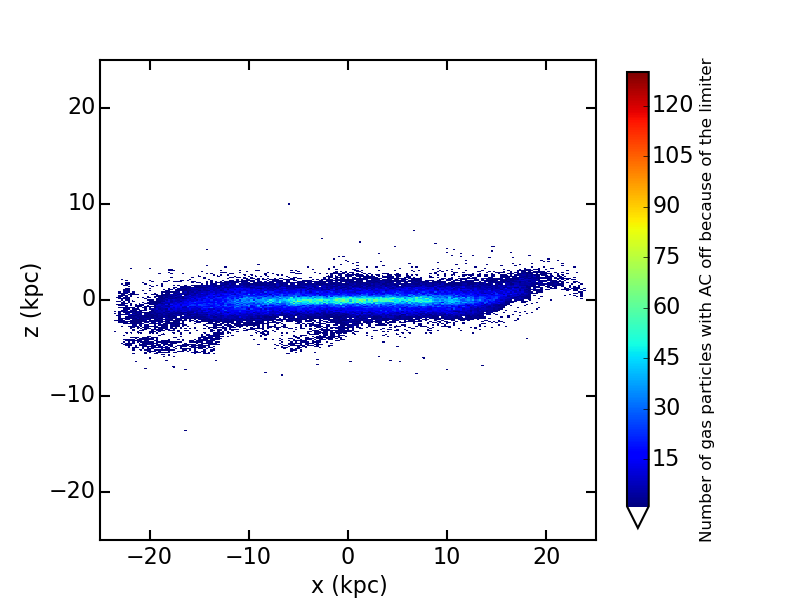} 
\end{minipage} 
\caption{{\sl Top panels:} face-on distribution of all the gas
  particles located within the galactic radius for the AqC5-newH
  galaxy simulation (left); face-on distribution of gas particles
  within $R_{\rm gal}$ with the AC switched off as a 
  consequence of the implementation of our AC limiter, in the 
  same run (right). {\sl Bottom panels:} edge-on distribution of 
  all the gas particles located within $R_{\rm gal}$ for the 
  AqC5-newH simulation (left); edge-on distribution of gas particles 
  within $R_{\rm gal}$ with the AC switched off because of the
  AC limiter (right). Plots are shown at redshift $z=0$; the
  colour encodes the numerical density of the particles per bin.}
\label{AqC5_newHydro_ACoff_mapXyz} 
\end{figure*}

\begin{figure*}
\newcommand{\captionfonts}{\small}
\vspace{-1.1ex}
\includegraphics[trim=0.0cm 0.0cm 0.0cm 0.0cm, clip, width=0.325\textwidth]{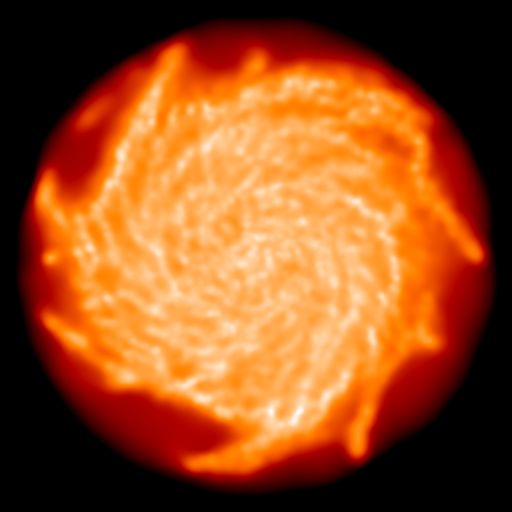}
\hspace{-0.6ex}
\includegraphics[trim=0.0cm 0.0cm 0.0cm 0.0cm, clip, width=0.325\textwidth]{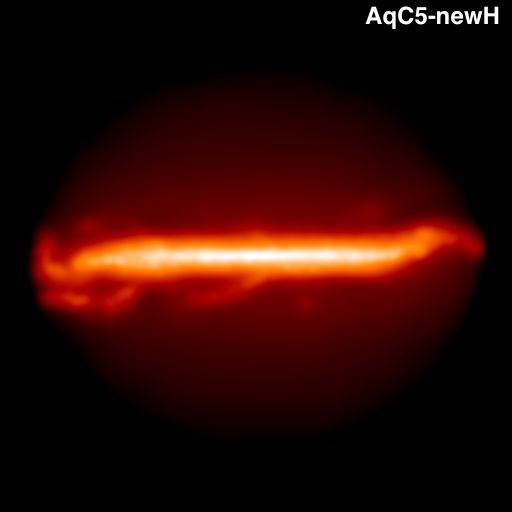}\\
\vspace{0.3ex}
\includegraphics[trim=0.0cm 0.0cm 0.0cm 0.0cm, clip, width=0.325\textwidth]{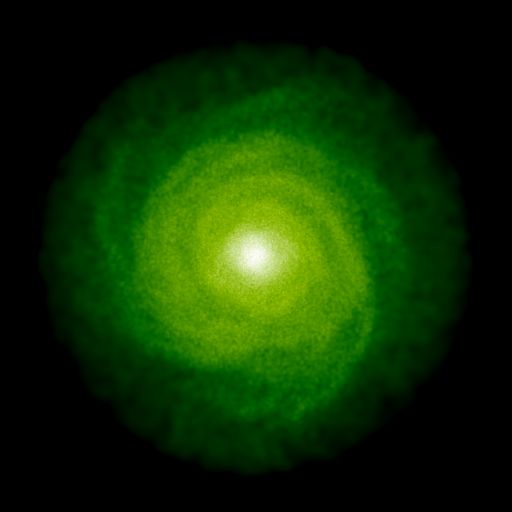} 
\hspace{-0.6ex}
\includegraphics[trim=0.0cm 0.0cm 0.0cm 0.0cm, clip, width=0.325\textwidth]{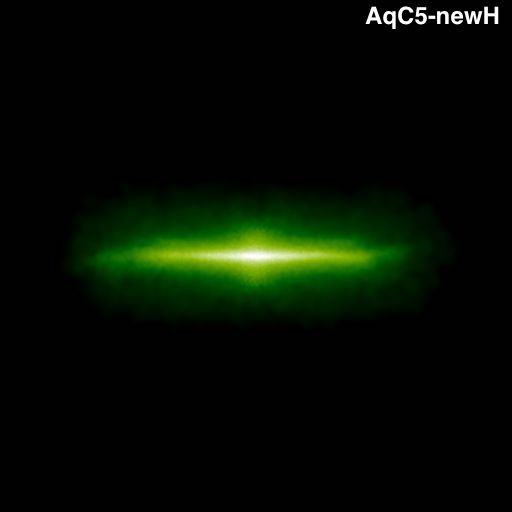}\\ 
\vspace{1.1ex}
\includegraphics[trim=0.0cm 0.0cm 0.0cm 0.0cm, clip, width=0.325\textwidth]{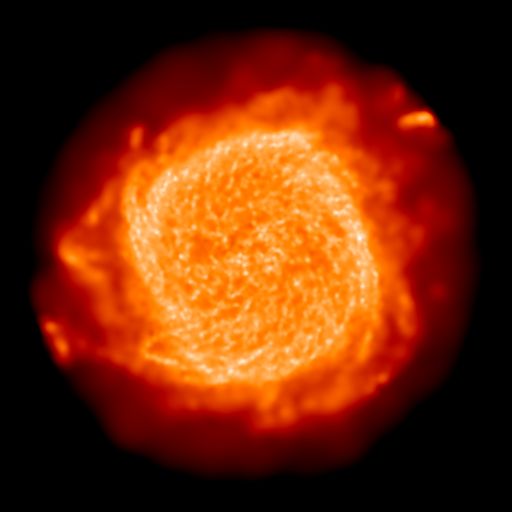}
\hspace{-0.6ex}
\includegraphics[trim=0.0cm 0.0cm 0.0cm 0.0cm, clip, width=0.325\textwidth]{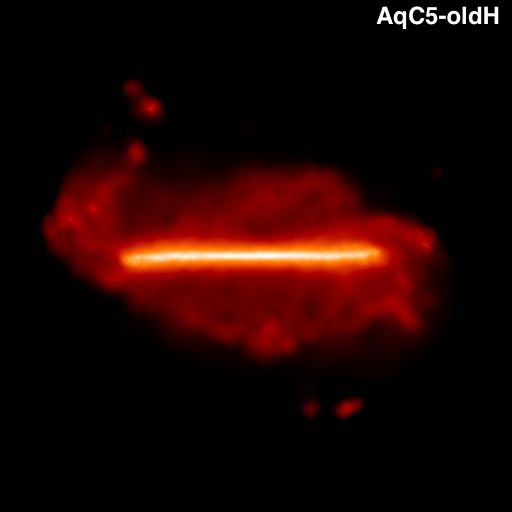}\\
\vspace{0.3ex}
\includegraphics[trim=0.0cm 0.0cm 0.0cm 0.0cm, clip, width=0.325\textwidth]{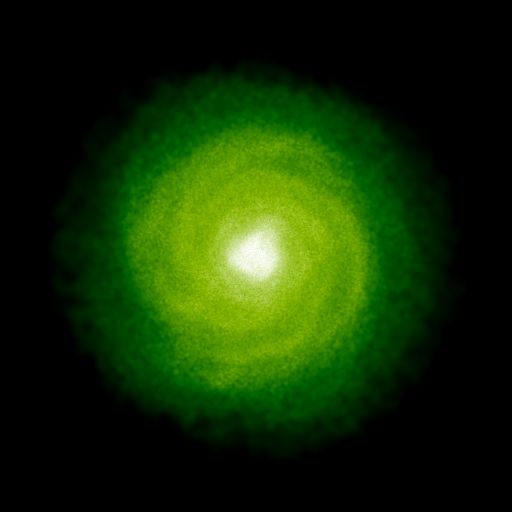} 
\hspace{-0.6ex}
\includegraphics[trim=0.0cm 0.0cm 0.0cm 0.0cm, clip, width=0.325\textwidth]{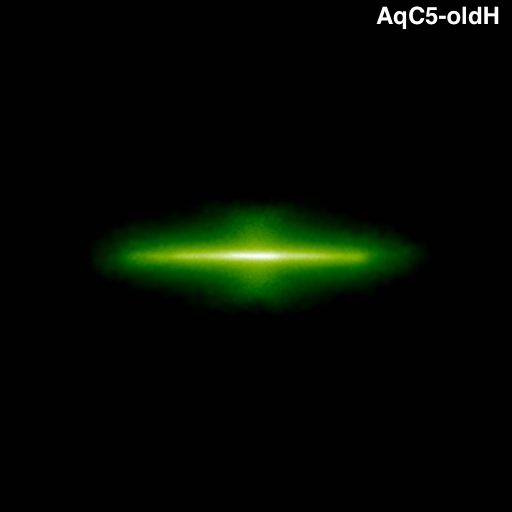} 

\vspace{-0.9ex}
\caption{{\sl Top four panels:} projected gas (first and second panels) and star (third and fourth) density for the AqC5-newH simulation. {\sl Bottom four panels:} projected gas (fifth and sixth panels) and star (bottom) density for AqC5-oldH. Left- and right-hand panels show face-on and edge-on densities, respectively. 
All the maps are shown at redshift $z=0$. The size of the box is 55 kpc.} 
\label{AqC5_newHydro} 
\end{figure*}

We also examine the position and the distribution of gas particles
with AC turned off as a consequence of the implementation of our AC
limiter in the AqC5-newH simulation at $z=0$. Fig.
\ref{AqC5_newHydro_ACoff_mapXyz} shows, in the left column, face-on
(top) and edge-on (bottom) distribution of gas particles within the
galactic radius\footnote{We define here the galactic radius as one-tenth 
  of the virial radius, i.e. $R_{\rm gal}= 0.1 R_{\rm vir}$. We choose 
  this radius $R_{\rm gal}$ in order to identify and select the region of the 
  computational domain that is dominated by the central galaxy.}, 
$R_{\rm gal} \simeq 24$ kpc; the right column shows the location of gas 
particles within $R_{\rm gal}$ with AC switched off due to the effect 
of the switch, for the same simulation. 
In the four panels of Fig. \ref{AqC5_newHydro_ACoff_mapXyz} we rotated 
the reference system in order to have the $z$-axis aligned with the angular 
momentum of stars and multiphase gas particles within 8 kpc from the 
location of the minimum of the gravitational potential (the same procedure 
has been adopted in Figs \ref{AqC5_newHydro} and \ref{AqC6_newHydro}). 
The origin is set at the centre of the galaxy, that is defined as the centre of mass 
of stars and multiphase gas particles within 8 kpc from the location of the minimum 
of the gravitational potential.

Particles with AC switched off mainly trace the densest and the
innermost regions of the simulated disc galaxy. Thus, Figs 
\ref{AqC5_newHydro_ACoff_PDs} and \ref{AqC5_newHydro_ACoff_mapXyz} 
show how AC is switched off for particles that are sampling the ISM, described 
by sub-grid physics, whose evolution strongly affects the gas thermodynamics 
and dominates over the resolved hydrodynamics. 
In particular, both the edge-on view of the AqC5-newH galaxy (bottom-right panel 
of Fig. \ref{AqC5_newHydro_ACoff_mapXyz}) and Fig. \ref{AqC5_newHydro_ACoff_PDs} 
(combined with information provided by Fig. \ref{AqC5_newHydro_MP_PD}) 
show how AC is instead normally acting outside the galaxy.

\subsection{Effect of the new SPH scheme}
\label{NewHydroOldFb}
In this section, we focus our attention on the results from AqC5 simulations,
just stressing few differences with respect to the low-resolution case
where needed. In order to investigate the significant role
played by the advanced hydrodynamic scheme in the formation and
evolution of a disc galaxy, we performed the comparison at both
available resolutions, i.e. between runs AqC5-oldH and AqC5-newH, and
AqC6-oldH and AqC6-newH (see Table \ref{simmHC}). A more detailed study
of the effect of the resolution is performed in Appendix
\ref{appendixA}.

Fig. \ref{AqC5_newHydro} introduces galaxies AqC5-newH and AqC5-oldH. 
The first four panels show face-on and edge-on projections of gas (top panels) 
and stellar (third and fourth ones) density for AqC5-newH at $z=0$. 
The four lower panels present face-on and edge-on projections of gas (fifth and 
sixth panels) and stellar (bottom ones) density for AqC5-oldH, at $z=0$, for comparison. 

The galaxy AqC5-newH has a limited bulge and a dominant disc,
with a clear spiral pattern both in the gaseous and in the stellar component.
The gaseous disc is more extended than the stellar one (see also the 
middle-left panel of Fig. \ref{AqC5_HydroComparison3} and discussion below).

\begin{figure}
\newcommand{\captionfonts}{\small}
\hspace{-1.65ex}
\centering
\includegraphics[trim=0.4cm 0.4cm 0.35cm 0.0cm, clip, width=0.488\textwidth]{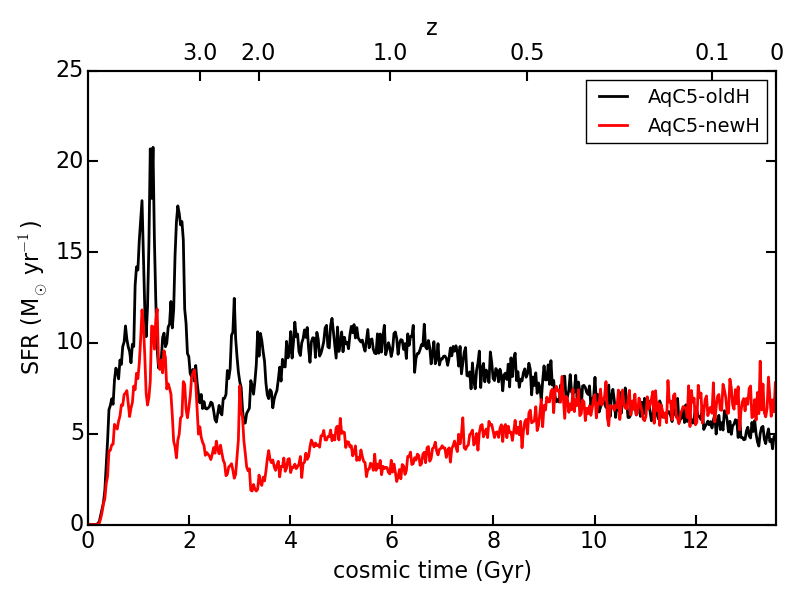} 
\caption{SFR as a function of cosmic time for
  the AqC5 simulations with the old (black) and the new (red)
  hydrodynamic scheme.}
\label{AqC5_HydroComparison1} 
\end{figure}

\begin{figure}
\newcommand{\captionfonts}{\small}
\hspace{-1.65ex}
\centering
\includegraphics[trim=0.4cm 0.4cm 0.35cm 0.0cm, clip, width=0.488\textwidth]{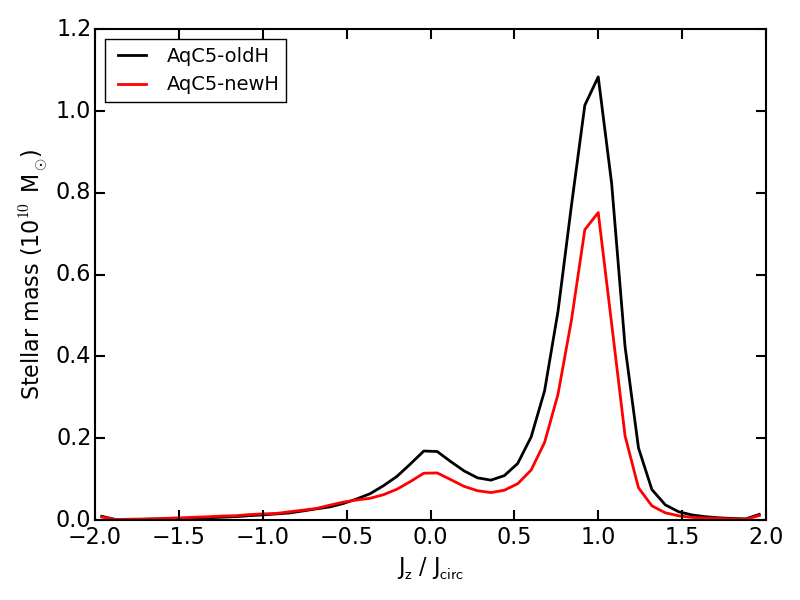} 
\caption{Stellar mass as a function of the circularity of stellar 
  orbits at $z=0$. The red curve refers to the AqC5-newH and the black
  curve shows the result from the AqC5-oldH simulation. The heights
  of the peaks located at $J_{\rm z}/J_{\rm circ}=0$ and at $J_{\rm z}/J_{\rm
    circ}=1$ describe the relative contributions to the total stellar
  mass of the galaxy from the bulge and the disc,
  respectively. We find $B/T=0.30$ and 0.23 for AqC5-newH and AqC5-oldH, respectively.}
\label{AqC5_HydroComparison2} 
\end{figure}

Fig. \ref{AqC5_HydroComparison1} shows the SFR as a function of the
cosmic time for AqC5-newH and AqC5-oldH simulations. Here, and in the
rest of this work, we only show the SFR evaluated for all stars that at $z=0$ are 
located within $R_{\rm gal}$. In all the figures presented and discussed in
this section results of AqC5-newH are shown in red, while those
referring to AqC5-oldH are in black. The burst of SF occurring at
redshift $z>3$ builds up the bulge component \citepalias[see][]{muppi2014} in the two 
simulated galaxies. High redshift ($z>3$) SFR is reduced when 
the new flavour of SPH is considered: as a consequence, the bulge 
component of AqC5-newH is slightly less massive (see Fig. \ref{AqC5_HydroComparison2}, 
described below). Below $z=2.5$, the SFR evolves differently
in the two simulations. AqC5-newH has a lower SFR till $z=0.5$, when
the two SFRs become again comparable. 

\begin{figure*}
\newcommand{\captionfonts}{\small}
\begin{minipage}{\linewidth}%{1.0\linewidth}
\centering
\includegraphics[trim=0.4cm 0.4cm 0.35cm 0.0cm, clip, width=0.49\textwidth]{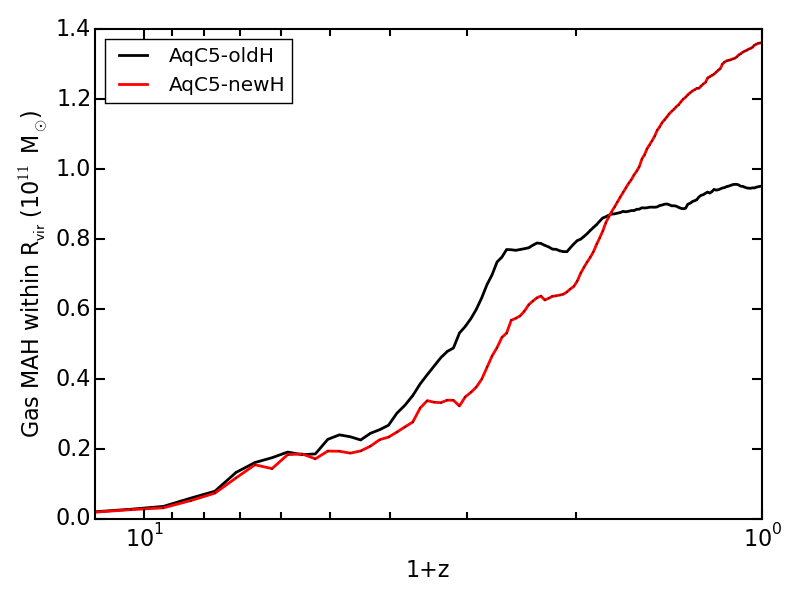} 
\includegraphics[trim=0.4cm 0.4cm 0.35cm 0.0cm, clip, width=0.49\textwidth]{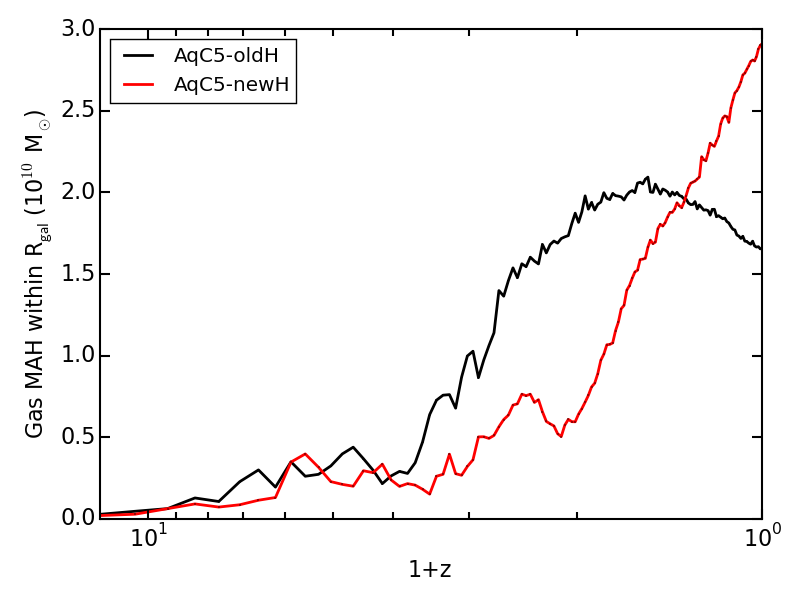} 
\end{minipage} 
\caption{{\sl Left-hand panel:} MAH, i.e. the
  redshift evolution of the gas mass, within the virial
  radius. AqC5-oldH is shown in black, while the AqC5-newH simulation
  in red. {\sl Right-hand panel:} MAH for gas within the galactic radius.}
\label{AqC5_HydroComparison4} 
\end{figure*}

One of the reasons of the discrepant SFR between AqC5-oldH and AqC5-newH 
for redshift that spans the range $2.5>z>0.5$ lies in the presence of the 
time-step limiting particle wake-up scheme. This feature of the improved SPH 
promotes the accurate treatment of feedback \citep{Durier2012,Schaller2015,beck2015}. 
When the wake-up scheme is used, energy deposited by wind particles in the surrounding 
medium when they hydrodynamically recouple with the ambient gas is more accurately 
distributed, especially to cold inflowing gas. This slows down the gas accretion 
on to the galaxy (or even on to the halo) and reduces the amount of fuel that is 
available for the disc growth (see also Fig. \ref{AqC5_HydroComparison4}, and discussion below).
We note that is an interaction of the outflow model with the wake-up feature of our
new hydrodynamical scheme, and depends not only on the sub-grid model parameters
but also on the details of the wake-up scheme. Here, we decided to keep the wake-up 
scheme and its parameter ($f_{\rm w}$, see Section \ref{sec:beck}) fixed. We use indeed 
the wake-up scheme as tested in \citet{beck2015}, where a number of hydrodynamical 
only tests are shown. A deeper investigation of the interaction between the wake-up and 
the outflows will be the subject of future studies.

Besides the smaller amount of the accreted gas in AqC5-newH, inflowing gas is characterized 
by a lower (radial component of the) velocity than in AqC5-oldH (for $2.5>z>1$): as a result, 
the SFR drops. Moreover, gas expelled by high-redshift outflows falls back at later times in the 
AqC5-newH: the SFR starts to gently rise for $z<1$, and then keeps roughly constant below 
$z<0.5$, while the SFR of AqC5-oldH is moderately declining. The final SFRs slightly overestimate 
that observed in typical disc galaxies \citep[see, e.g.,][for our Galaxy]{snaith2014}. 
The different SF history is the result of the lower baryon conversion efficiency of AqC5-newH 
(see Fig. \ref{AqC5_HydroComparison5}, right-hand panel, described below).

We then analyse the distribution of stellar circularities in order to
perform a kinematic decomposition of the simulated galaxies, by
separating the dispersion supported component (bulge) from the
rotationally supported one (disc). Fig. \ref{AqC5_HydroComparison2}
shows the distribution of stellar mass as a function of the circularity of star
particles\footnote{Numbers of DM, gas, and star particles within 
   $R_{\rm vir}$ at $z=0$ are 710150, 362648 and 692125 for AqC5-newH, respectively, 
   and 691717, 252660 and 895657 for AqC5-oldH.} 
at $z=0$ for the two simulations. We consider star particles within $R_{\rm gal}$. 
Virial radii of the two galaxies are $R_{\rm vir}=238.64$ kpc 
(AqC5-oldH) and $R_{\rm vir}=240.15$ kpc (AqC5-newH). The circularity of a stellar orbit is 
characterized by the ratio $J_{\rm z}$/$J_{\rm circ}$, i.e. the specific angular momentum in the 
direction perpendicular to the disc over the specific angular momentum of a reference 
circular orbit. This latter quantity is defined according to \citet{Scannapieco2009} as 
$J_{\rm circ}=r \cdot v_{\rm c}(r) = r (G M(<r)/r)^{1/2}$, $v_{\rm c}(r)$ being the circular velocity of
each star at the distance $r$ from the galaxy centre.

The bulk of the stellar mass is rotationally supported, thus highlighting that the disc 
represents the dominant component. The relative contribution to the total galaxy stellar 
mass from the bulge and the disc components can indeed be appreciated by weighting 
the heights of the peaks located at $J_{\rm z}/J_{\rm circ}=0$ and at $J_{\rm z}/J_{\rm circ}=1$, 
respectively. The total stellar mass of AqC5-newH ($4.74 \cdot 10^{10}$ M$_{\odot}$, 
within $R_{\rm gal}$) is lower than that of AqC5-oldH ($7.22 \cdot 10^{10}$ M$_{\odot}$); 
both simulated galaxies are clearly disc-dominated. Assuming that the counter-rotating stars 
are symmetrically distributed with zero average circularity, we estimate the bulge mass as 
twice the mass of counter-rotating stars. 
We can thus calculate the bulge over total stellar mass ratio within $R_{\rm gal}$. We obtain 
$B/T=0.30$ and $B/T=0.23$ for AqC5-newH and AqC5-oldH, respectively.  
The mass of the bulge component in AqC5-newH is marginally reduced with respect to AqC5-oldH 
(see Fig. \ref{AqC5_HydroComparison1} and discussion above); 
the stellar mass in the disc component is decreased as well in AqC5-newH. Gas mass 
within $R_{\rm gal}$ for AqC5-newH is $2.91 \cdot 10^{10}$ M$_{\odot}$, the cold ($T<10^5$ K) gas 
mass being $2.54 \cdot 10^{10}$ M$_{\odot}$; gas mass within $R_{\rm gal}$ for AqC5-oldH is 
$1.66 \cdot 10^{10}$ M$_{\odot}$, the amount of cold gas being $1.55 \cdot 10^{10}$ M$_{\odot}$.

\begin{figure*}
\newcommand{\captionfonts}{\small}
\begin{minipage}{\linewidth}%{1.01\linewidth}
\centering
\includegraphics[trim=0.4cm 0.3cm 0.65cm 1.5cm, clip, width=0.49\textwidth]{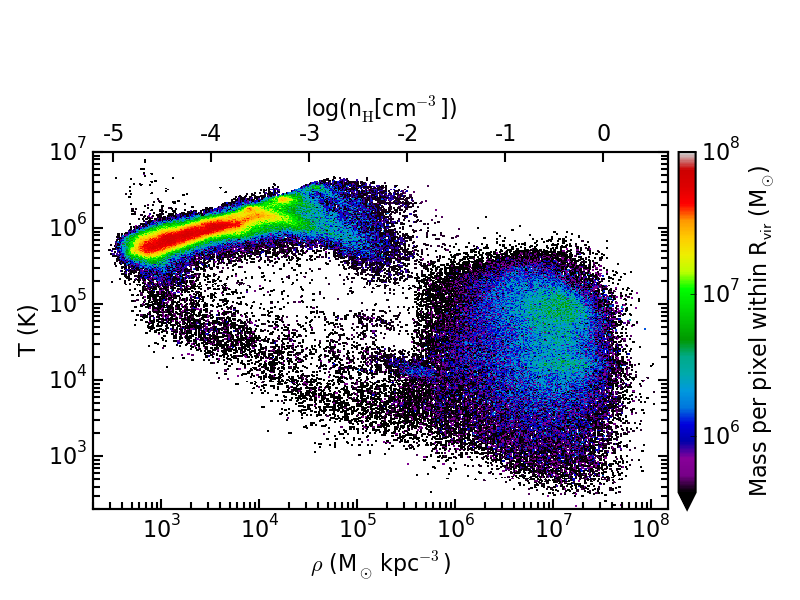} 
\includegraphics[trim=0.4cm 0.3cm 0.65cm 1.5cm, clip, width=0.49\textwidth]{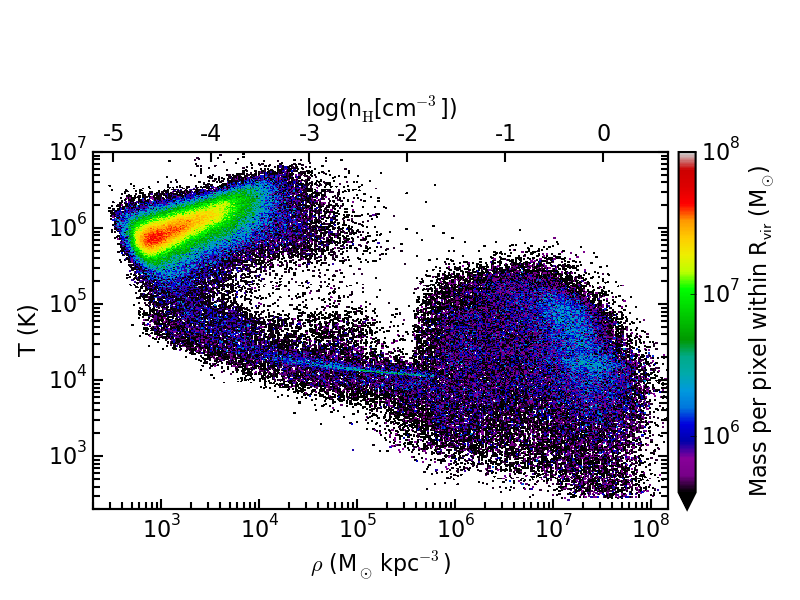} 
\end{minipage} 
\caption{Phase diagrams of gas particles within the virial radius for the simulations AqC5-newH
({\sl left-hand panel}) and AqC5-oldH ({\sl right-hand panel}). Colour encodes the mass of gas particles per 
density-temperature bin. The colour scale is the same for both panels.}
\label{AqC5_HydroComparison7} 
\end{figure*}

Fig. \ref{AqC5_HydroComparison4} describes the mass accretion
history (MAH), i.e. the redshift evolution of the gas mass, within the
virial radius (left-hand panel) and the galactic radius (right-hand panel). 
Gas accretion within $R_{\rm vir}$ is comparable in the two reference
simulations up to $z\simeq4$; below this redshift, the gas accretion
is delayed when the new SPH implementation is adopted and then, after
redshift $z=1$, continues rising down to $z=0$. On the other hand, the old
hydrodynamic scheme reduces the gas mass accretion rate below $z\simeq2$,
and this is much more evident when the gas accretion within $R_{\rm gal}$ is 
analysed, where it is stopped below $z\simeq1$. This different gas accretion 
pattern is one of the reasons of the differences that we found between the 
two AqC5 simulations. The new hydrodynamic scheme is responsible for the 
delay of the gas inflow, by slowing down the gas fall-back within the virial 
and the galactic radii. Moreover, low-redshift ($z<1$) gas accretion is faster in 
AqC5-newH and still ongoing at $z=0$: as a consequence, the stellar disc forms 
later and it is still growing at redshift $z=0$, when fresh gas keeps accreting on the 
galaxy, sustaining the SF.

Fig. \ref{AqC5_HydroComparison7} shows the density-temperature
phase diagram of all the gas particles within the virial radius for the 
AqC5-newH simulation (left-hand panel) and for the run AqC5-oldH
(right-hand one). The mass distribution is shown at $z=0$ and both single-phase 
and multiphase particles are taken into account. As for the temperature, 
it is the temperature of single-phase particles and the mass-averaged 
temperature for the multiphase ones. 
The presence of AC affects the gas particle distribution in the phase 
diagram: cooling properties are modified, phase-mixing is promoted. 
When more diffuse gas is considered, i.e. $\rho < 3 \cdot 10^4$ M$_\odot$ 
kpc$^{-3}$, we find that the gas is kept hotter in the AqC5-newH 
simulation, with a lot of gas at temperatures $3 \cdot 10^5 < T < 4 \cdot 10^6$ K.
At intermediate densities, i.e. $3 \cdot 10^4 < \rho < 4 \cdot 10^5$ M$_\odot$ 
kpc$^{-3}$, and for $T > 3 \cdot 10^5$ K we notice the presence of 
a large number of gas particles that have been recently heated because of 
stellar feedback in the AqC5-newH simulation, while in the same density 
and temperature regime there is a lack of these particles in the AqC5-oldH. 
This feature confirms the importance of the time-step limiting particle 
wake-up scheme in the improved SPH, as discussed above.
At higher densities, above the multiphase threshold, AqC5-newH is 
characterized by a larger amount of multiphase and star-forming particles, 
as a consequence of the more active SFR at $z=0$.

From these phase diagrams we conclude that the interplay between 
hydrodynamical scheme and sub-resolution model plays an important role 
in determining the thermodynamical properties of gas particles and the
resulting history of SF and feedback. 

Fig. \ref{AqC5_HydroComparison3} shows radial profiles at $z=0$ of
some interesting physical properties of the two galaxies obtained with
the new and the original implementation of SPH. The top-left panel
describes rotation curves for AqC5-oldH (black) and AqC5-newH (red).
In this panel, the solid curves describe the circular velocity due to
the total mass inside a given radius, while dashed, dotted and
dot-dashed curves show the different contribution of DM, stars and gas
to the total curve, respectively. Both galaxies, and AqC5-newH in particular, 
are characterized by total rotation curves that exhibit a flat behaviour at 
large radii and that are not centrally peaked, thus pointing to 
galaxies with a limited bulge and a dominant disc component. Differences 
between contributions of considered components can be understood by 
analysing density radial profiles.

\begin{figure*}
\newcommand{\captionfonts}{\small}
\hspace{-1.6ex}
\begin{minipage}{\linewidth}%{1.0\linewidth}
\centering
\includegraphics[trim=0.4cm 0.1cm 0.35cm 0.1cm, clip, width=0.49\textwidth]{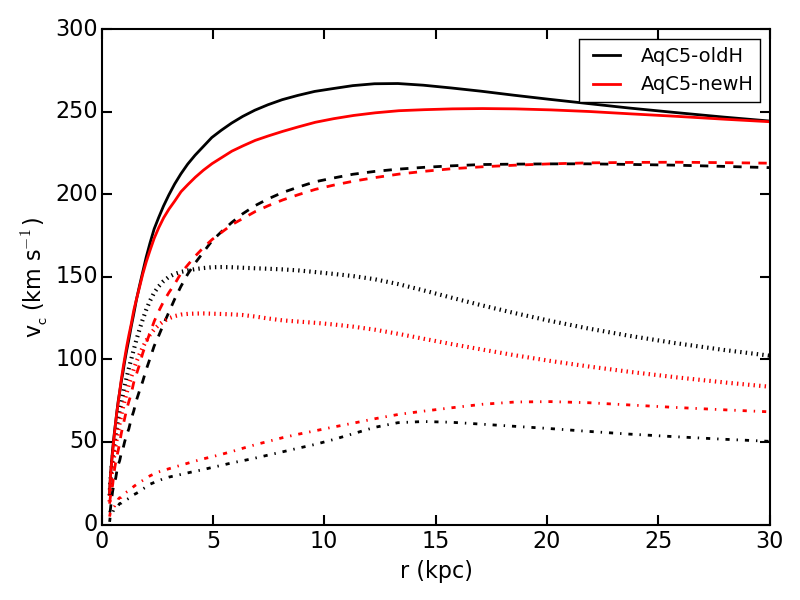} 
\includegraphics[trim=0.4cm 0.25cm 0.25cm 0.1cm, clip, width=0.488\textwidth]{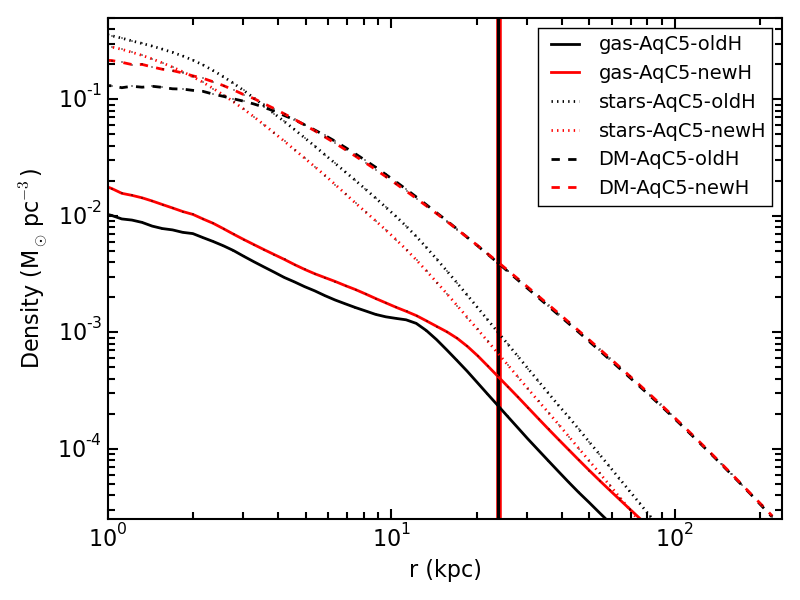}
\includegraphics[trim=0.4cm 0.2cm 0.35cm 0.1cm, clip, width=0.49\textwidth]{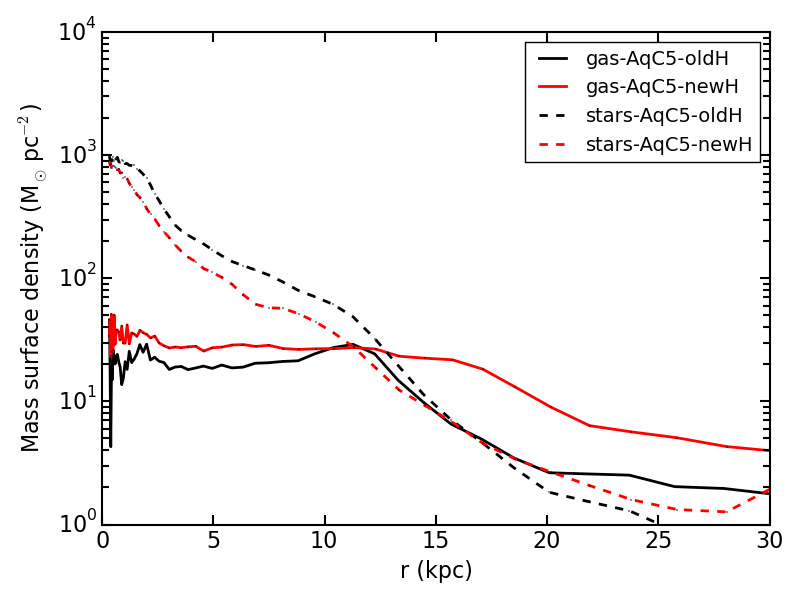} 
\includegraphics[trim=0.4cm 0.2cm 0.35cm 0.1cm, clip, width=0.49\textwidth]{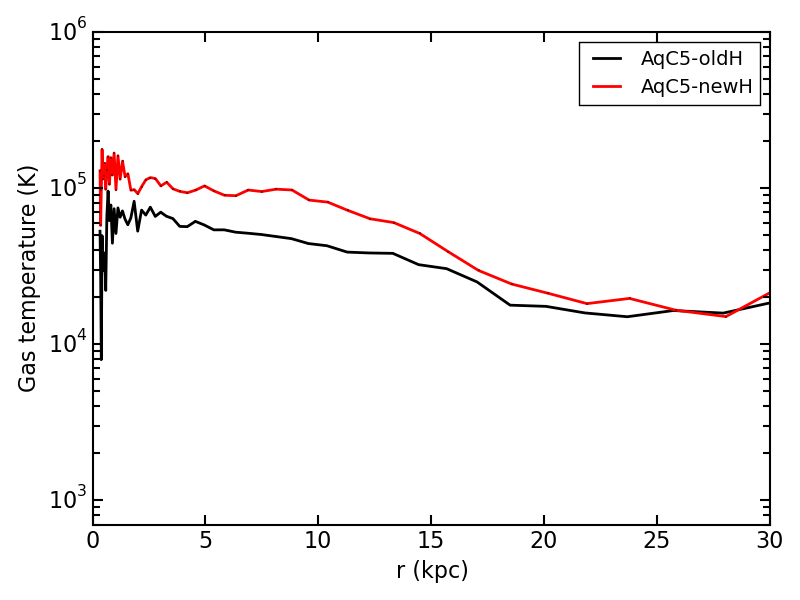} 
\includegraphics[trim=0.4cm 0.2cm 0.35cm 0.cm, clip, width=0.49\textwidth]{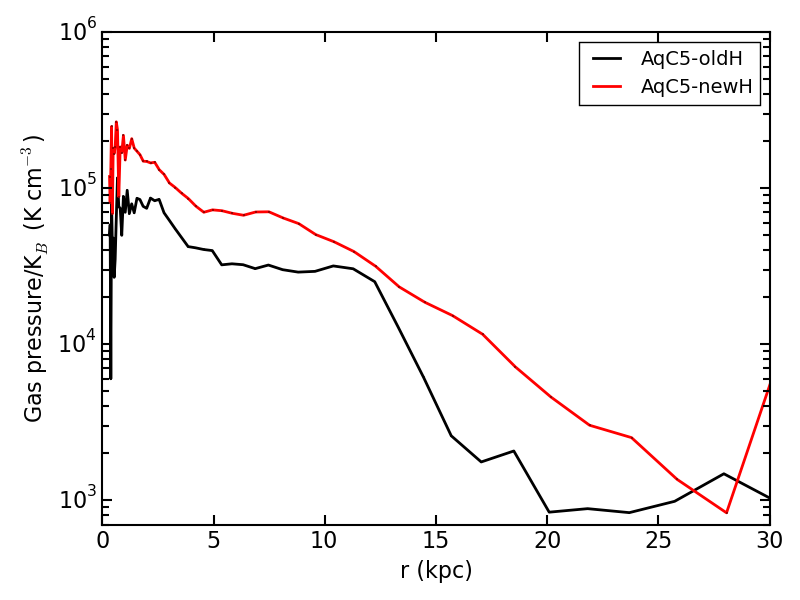} 
\includegraphics[trim=0.4cm 0.2cm 0.35cm 0.cm, clip, width=0.49\textwidth]{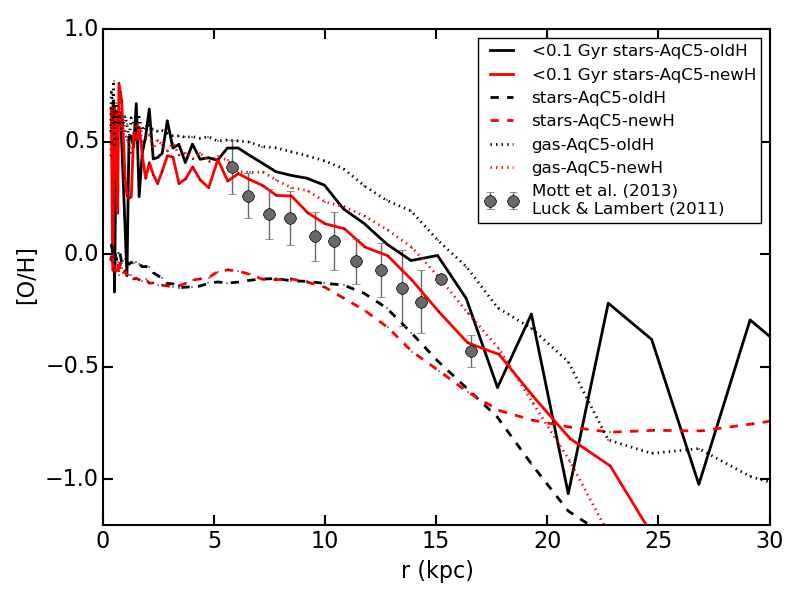} 
\end{minipage} 

\caption{{\sl Upper panels:} left: rotation curves for the AqC5
  simulations with the old (black) and the new (red) hydrodynamic
  scheme. The solid curve describes the circular velocity due to the
  total mass inside a given radius. The different contribution of DM
  (dashed line), stars (dotted), and gas (dot-dashed) to the total
  curve are shown.  Right: density radial profiles of gas (solid
  curve), stars (dotted), and DM (dashed) for the AqC5-newH and
  AqC5-oldH simulations. Black and red vertical solid lines mark the galactic 
  radii of AqC5-oldH and AqC5-newH, respectively, the $x$-axis extending to 
  the virial radius of the second one.  
  {\sl Middle panels:} left: surface density
  profiles. Solid curves describe the contribution of gas and dashed
  lines are the one of the stars. Right: gas temperature radial
  profiles.  {\sl Bottom panels:} left: gas pressure radial profiles.
  Right: radial profiles of oxygen abundance of gas (dotted curves), 
  stars (dashed), and young stars (solid, see the text)
  for AqC5-newH (red) and AqC5-oldH (black). 
  Data (grey filled circles) show the radial abundance gradient for oxygen from 
  observations of Cepheids \citep{Luck2011}, according to the division 
  in bins performed by \citet{Mott2013}. 
  Error bars represent the standard deviation in each bin. In
  all the panels, results for the simulation AqC5-oldH are shown in
  black and for AqC5-newH in red. All the profiles are analysed at
  $z=0$.}
\label{AqC5_HydroComparison3} 
\end{figure*}

The top-right panel of Fig. \ref{AqC5_HydroComparison3} shows the density 
radial profiles of gas (solid), stars (dotted) and DM (dashed) in the reference 
simulations, out to $R_{\rm vir}=240.15$ kpc, the virial radius of AqC5-newH. 
Solid vertical red and black lines pinpoint the location of $R_{\rm gal}$ of AqC5-newH 
and AqC5-oldH, respectively. 
DM profiles only differ in the innermost regions, i.e. $r<4$ kpc. 
Such a discrepancy is related to the different high-$z$ ($z>3$) SFR between the 
two simulated galaxies (see Fig. \ref{AqC5_HydroComparison1}). The lower 
high-$z$ SFR in AqC5-newH produces a stellar feedback associated to the SF 
burst that is not as strong as in AqC5-oldH: as a consequence, the DM 
component in AqC5-newH experiences a gentler adiabatic expansion, the 
feedback induced gas displacement being not as striking as in AqC5-oldH, 
at that epoch. This results in a higher central DM volume density in AqC5-newH.
Gas and stellar density profiles have similar shapes but different
normalizations. A higher gas content characterizes the AqC5-newH
simulation, while the contribution of stars is more important in
AqC5-oldH, as already seen in the circularity histograms (and 
quantified with gas and stellar masses above). 

The middle-left panel of Fig. \ref{AqC5_HydroComparison3} shows
surface density profiles. Solid and dashed curves refer to gas and to
stars, respectively. The comparison between the gas contribution in the two 
simulations further shows that AqC5-newH has a higher gas content. 
Moreover, the size of the gas disc is more extended and the
gas mass surface density is larger in the innermost regions.
The higher gas content in AqC5-newH at $z=0$ is due to the fact that
we have a larger amount of gas that is infalling towards the galaxy at
$z \textless 1$ with the new hydrodynamic implementation (see Fig. 
\ref{AqC5_HydroComparison4}), thus being consistent with the increasing 
low-$z$ SFR shown in Fig. \ref{AqC5_HydroComparison1}.
Stellar surface density profile is higher for AqC5-oldH, and this is in 
agreement with the fact that high-redshift expelled gas fell back on this galaxy at 
earlier times and produced more stars, while this process is still ongoing in AqC5-newH. 

The middle-right and bottom-left panels of Fig. \ref{AqC5_HydroComparison3} 
show the radial profiles of gas temperature and pressure, respectively. 
Discs obtained with the improved SPH formalism are hotter and more pressurized 
in the innermost regions. The higher pressure in AqC5-newH with respect to AqC5-oldH 
is a direct consequence of the higher gas surface density in the galaxy simulated 
with the new hydrodynamic scheme (see discussion above).
Temperature and pressure profiles can be further illustrated by considering the 
phase diagrams of gas particles shown in Fig. \ref{AqC5_HydroComparison7} 
and discussed above: we indeed see that in AqC5-newH there is a larger number 
of particles having a higher temperature, when both the low- 
($\rho \la 5 \cdot 10^5$ M$_\odot$ kpc$^{-3}$) 
and the high-density ($\rho \ga 5 \cdot 10^5$ M$_\odot$ kpc$^{-3}$) regimes 
are considered. This holds in particular for multi-phase particles, thus probing 
the higher temperature and pressure in the innermost regions of AqC5-newH.

\begin{figure*}
\newcommand{\captionfonts}{\small}
\begin{minipage}{\linewidth}%{1.0\linewidth}
\centering
\includegraphics[trim=0.4cm 0.4cm 0.35cm 0.0cm, clip, width=0.49\textwidth]{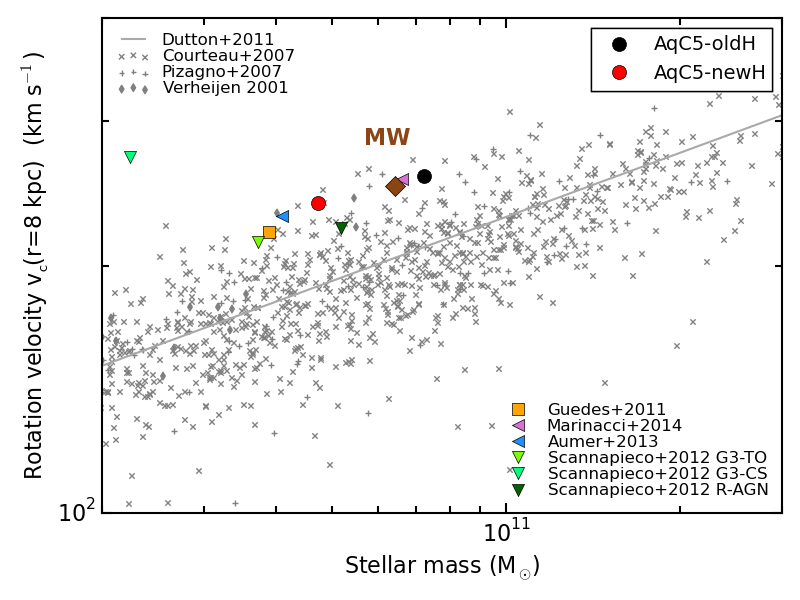}
%tf_AqC5_new_HC.png} 
\includegraphics[trim=0.4cm 0.4cm 0.35cm 0.0cm, clip, width=0.49\textwidth]{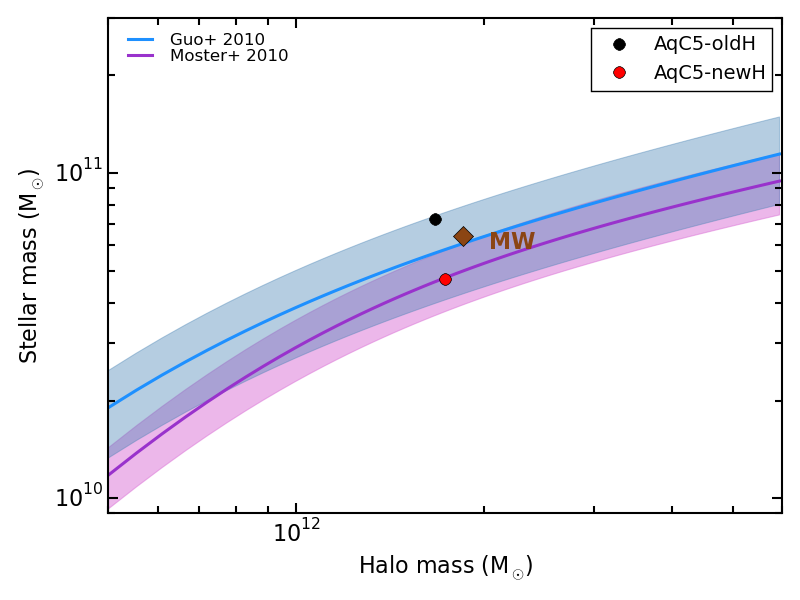}
%bce_AqC5_new_HC.png} 
\end{minipage} 
\caption{ {\sl Left-hand panel:} Tully-Fisher relation for the AqC5-newH (red) and
AqC5-oldH (black) simulations. We use the circular velocity at a
distance of $8$ kpc from the galaxy centre as a function of the
galaxy stellar mass within the galactic radius. Grey crosses, diamonds,
and plus symbols are observations from \citet{cou2007},
\citet{ver2001}, and \citet{piz2007}, respectively, the solid line
providing the best fit \citep{Dutton2011}. Other coloured symbols
refer to the following simulations of late-type galaxies: the purple
triangle is the AqC5 from \citet{Marinacci2014}, the blue triangle is
the AqC5 from \citet{Aumer2013}, the orange filled square shows Eris
simulation from \citet{Guedes2011}, and green triangles are AqC5 galaxies
from the Aquila comparison project \citep[][simulations G3-TO (light
  green), G3-CS (medium green) and R-AGN (dark
  green)]{Scannapieco2012}. {\sl Right-hand panel:} Baryon conversion
efficiency of the two disc galaxies at $z=0$. Blue solid line shows
the stellar-to-halo mass relation derived by \citet{Guo2010}, with the 
shaded blue region representing an envelope of 0.2 dex around it. 
Purple solid curve describes the fit proposed by \citet{Moster2010}, 
with 1$\sigma$ uncertainty (shaded envelope) on the normalization. 
Brown diamonds show where the MW is located in the Tully-Fisher 
relation and in the stellar-to-halo mass relation plots. We include our Galaxy for completeness.
Properties of MW according to \citet{mcMillan2011}, \citet{bhatta2014}, and \citet{Zaritsky2017}.
}
\label{AqC5_HydroComparison5} 
\end{figure*}

The last panel of Fig. \ref{AqC5_HydroComparison3} describes metallicity radial 
profiles of the two simulations. 
We compare our simulated metallicity profiles with high-quality data 
measured for stars in our Galaxy. We note that the current simulations are not
aimed at modelling the MW; the goal of the comparison is to understand
if simulated results are in broad agreement with observations of a
typical disc galaxy like our MW.  
We consider radial profiles of oxygen abundance
(that is an accurate and reliable tracer of the total metallicity) in gas and stars 
for AqC5-newH (red) and AqC5-oldH (black). Data (grey filled circles) show the radial abundance 
gradient for oxygen from observations of Cepheids \citep{Luck2011}, according to the division 
in bins as a function of the Galactocentric distance performed by \citet{Mott2013}. Error 
bars represent the standard deviation computed in each bin, whose width is $1$ kpc. Dotted and 
dashed curves depict the oxygen abundance radial profile from all the gas and star particles within $R_{\rm gal}$ 
at $z=0$, respectively. We also consider oxygen abundance radial profiles (solid curves) by limiting the 
considered stars to those that have been formed after $z \la 0.007$, i.e. less than $\sim 100$ Myr ago. 
Cepheids are indeed young stars \citep[estimated age of $\la 100$ Myr,][]{Bono2005}: solid curves allow 
a fair comparison with observations (we will therefore focus on them in the following analysis). 
We find a good agreement with observations: the slope of the radial profile predicted by 
considering only young stars in AqC5-newH, for distances from the galaxy centre spanning the 
range $5 \la r \la 15$ kpc, well conforms to the trend suggested by observations. 
The ability to recover well the slope predicted by observations 
highlights the effectiveness of our sub-resolution model to properly describe processes that affect the 
spread and the recycling of metals. We also note that there is a flattening of profiles in 
simulations for distances $r \la 5$ kpc from the galaxy centre. 
The normalization of the oxygen abundance profile of young stars in AqC5-newH is consistent with  
observations within the error bars (despite a slight overestimate by $\sim 0.05$ dex, while the overestimate 
is about $\sim 0.1$ dex when AqC5-oldH is considered). 
However, a different normalization of the abundance profiles in simulations with respect to the observed 
ones is expected, as the normalization of the profiles in simulations is affected by many aspects 
(e.g. uncertainties related to the yields and IMF). On the other hand, their slope can be directly 
compared to that of the observed abundance gradients.

The left-hand panel of Fig. \ref{AqC5_HydroComparison5} displays the
stellar Tully-Fisher relation for the AqC5-newH (red) and AqC5-oldH
(black) simulations. We measure the circular velocity at a distance
of $8$ kpc from the galaxy centre and consider the galaxy stellar
mass within the galactic radius. We also plot observational data
\citep{cou2007, ver2001, piz2007}, the best fit to them
\citep{Dutton2011} and results from the following simulations of
late-type galaxies: the purple triangle is the AqC5 from
\citet{Marinacci2014}, the blue triangle is the AqC5 from
\citet{Aumer2013}, and the orange filled square shows Eris simulation from
\citet{Guedes2011}. Green triangles are AqC5 galaxies from the Aquila
comparison project \citep[][simulations G3-TO (light green), G3-CS
(medium green), and R-AGN (dark green)]{Scannapieco2012}. 
The brown diamond shows where the MW is placed in the Tully-Fisher relation diagram. 
We do not intend to claim that AqC is a model of the MW, our Galaxy being included only for 
completeness. As for the properties of the MW, we adopted the following values: 
$6.43 \cdot 10^{10}$ M$_{\odot}$ is the total stellar mass, 
$250$ km s$^{-1}$ is the circular velocity (evaluated at $8$ kpc, to be consistent with our analysis), and 
$1.85 \cdot 10^{12}$ M$_{\odot}$ is the halo mass 
\citep{mcMillan2011, bhatta2014, Zaritsky2017}.  
We note that
our galaxies are located in the upper edge of the region traced by
observational results, thus highlighting a predicted circular velocity
slightly higher than the one expected from observations at the
considered stellar mass. As discussed in \citetalias{muppi2014}, this appears to be a
feature shared among several similar simulations of late-type
galaxies. ICs likely play a role in the aforementioned shift, rather than resolution. 

The right-hand panel of Fig. \ref{AqC5_HydroComparison5} shows the baryon
conversion efficiency of the two disc galaxies at $z=0$. Blue solid
line shows the stellar-to-halo mass relation derived by \citet{Guo2010}, 
with the shaded blue region representing an interval of $0.2$ dex around it 
\citep[as done in][and in \citetalias{muppi2014}]{Marinacci2014}. Purple curve describes 
the fit (solid line) proposed by \citet{Moster2010}, with 1-$\sigma$ uncertainty 
(shaded envelope) on the normalization. 
The simulation performed with the 
new SPH implementation is characterized by a lower baryon conversion
efficiency, that is in good agreement with both the predictions by 
\citet{Moster2010} and \citet{Guo2010}. As discussed above, this reflects on the lower 
stellar mass of the new galaxy.

As for the comparison between AqC5-oldH and AqC5-newH simulations, 
the key differences can be interpreted as an effect of the different MAH: 
gas accretion within the virial and galactic radii is delayed in AqC5-newH. 
In this simulation we still have fresh gas that is infalling towards the galaxy 
centre at $z<1$. In general, we are able to obtain a disc-dominated 
galaxy with both the old and the new hydrodynamic schemes. The 
free parameters of our sub-grid model had to be re-tuned 
when the improved SPH implementation has been adopted.

\subsection{Changing the galactic outflow model}
\label{NewHydroNewFb}

\subsubsection{Tuning the models}
\label{FBtune}

In this section, we compare the results obtained from the three implementations 
of galactic outflows described in Section \ref{sec:confronto}, and compare them 
to the original one presented by \citetalias{muppi2014}. We remind that all the differences we present 
in this section are only due to the change in the outflow model, being the ICs, the 
code, the SPH implementation and the MUPPI SF model exactly the same in all cases.
Note that we kept fixed the scheme for the {\it thermal} energy release of MUPPI. 
All the changes only concern the distribution of the fraction of SN energy that is provided 
in {\it kinetic} form in the original model. 

\begin{figure}
\newcommand{\captionfonts}{\small}
\hspace{-1.65ex}
\centering
\includegraphics[trim=0.4cm 0.4cm 0.35cm 0.2cm, clip, width=0.486\textwidth]{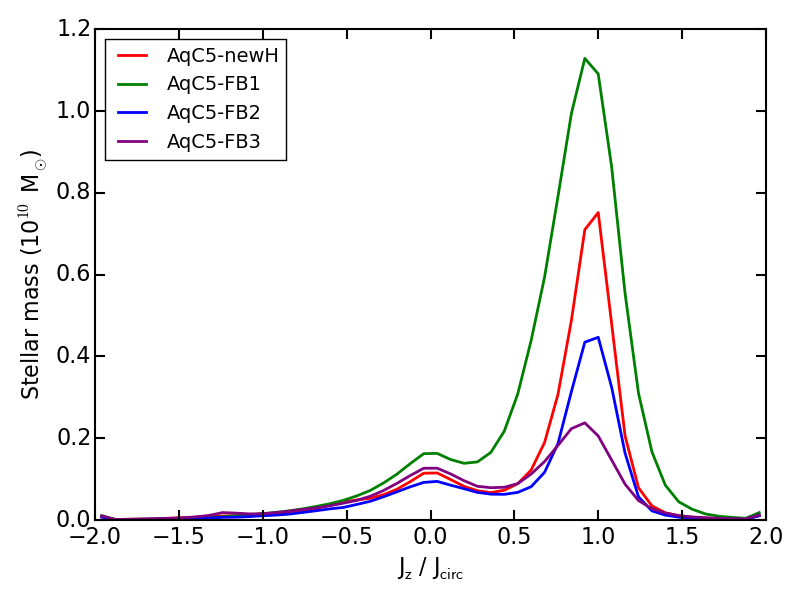} 
\caption{Stellar mass as a function of the circularity of stellar
  orbits at $z=0$. The red curve refers to the AqC5-newH run, the
  green one to AqC5-FB1, the blue curve identifies AqC5-FB2 and the
  purple one AqC5-FB3. The values of the $B/T$ ratios are: 0.30, 0.19, 0.35,
  and 0.54, respectively. }
\label{AqC5_FBall_jcirc} 
\end{figure}

\begin{figure}
\newcommand{\captionfonts}{\small}
\hspace{-1.65ex}
\centering
\includegraphics[trim=0.4cm 0.4cm 0.35cm 0.2cm, clip, width=0.488\textwidth]{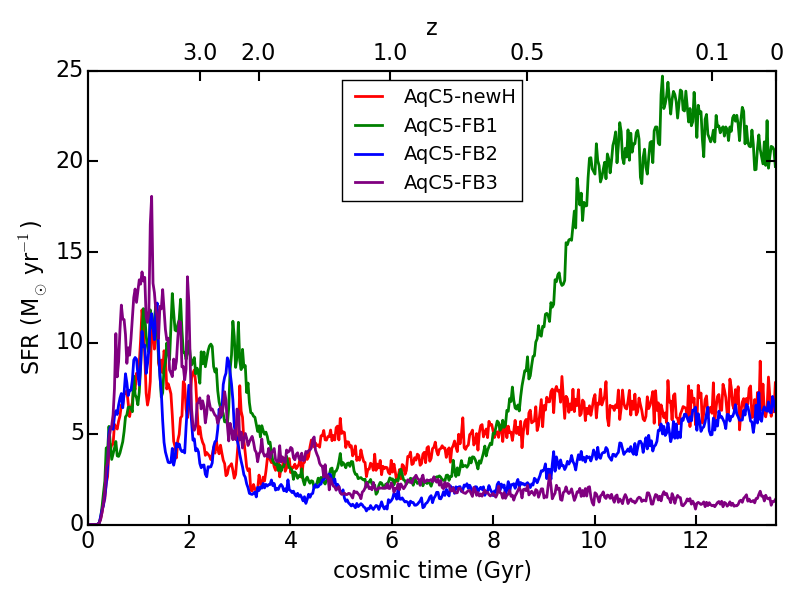} 
\caption{SFRs for the AqC5-FB1, AqC5-FB2, and AqC5-FB3
  simulations, compared with that of AqC5-newH. Colours are as in
  Fig. \ref{AqC5_FBall_jcirc}. Results refer to $z=0$.}
\label{AqC5_FBall_sfr} 
\end{figure}

For each of the three outflow models, we carried out a detailed exploration of 
the corresponding parameter space. To select our preferred set of parameters, we 
mainly focused on those simulations that produce the most prominent stellar disc 
component, as evaluated by visual inspection and by comparing the circularity histograms, 
as those shown in Fig. \ref{AqC5_FBall_jcirc}. We also analysed the galactic SFRs, 
as those shown in Fig. \ref{AqC5_FBall_sfr}, trying to minimize the high-redshift SF burst, 
in order to prevent a too large amount of late gas infall and, consequently, a too high low-redshift SFR. 
We did not succeed in simultaneously calibrating other
  properties of our simulated galaxies, such as the $B/T$ ratio or the total stellar mass
  (or one between the disc or the bulge stellar mass).
  However, as we will show below, our main point is that
  galaxy properties are deeply linked to the selected outflow
  model: it is very difficult to obtain the same set of properties by simply tuning 
  the model parameters. 

We performed an accurate exploration of the models' parameter space mainly using 
the low-resolution AqC6 ICs. When needed, we slightly re-tuned some of the parameters at the higher 
resolution of the AqC5 ICs. In this section, we only show results for our AqC5 best cases. 
The procedure for parameter tuning is described in Appendix \ref{appendixB}.

\begin{figure*}
\newcommand{\captionfonts}{\small}
\begin{minipage}{\linewidth}%{1.0\linewidth}
\centering
\includegraphics[trim=0.4cm 0.4cm 0.35cm 0.0cm, clip, width=0.49\textwidth]{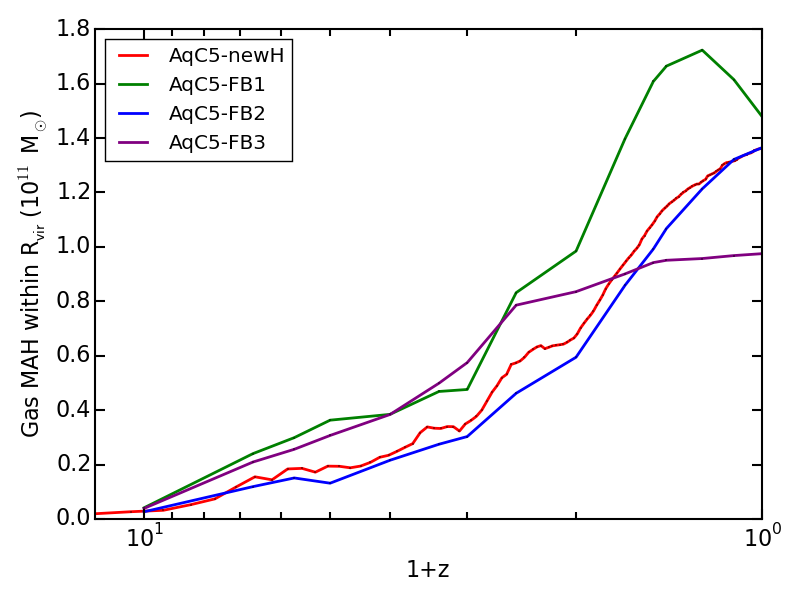} 
\includegraphics[trim=0.4cm 0.4cm 0.35cm 0.0cm, clip, width=0.49\textwidth]{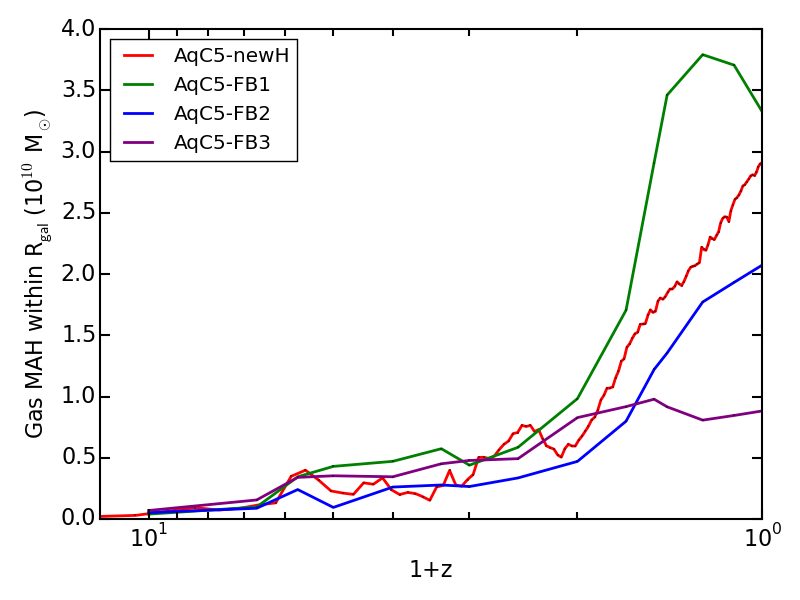} 
\end{minipage} 
\caption{Gas MAH for AqC5-FB1, AqC5-FB2, and
  AqC5-FB3, compared with the result of AqC5-newH. Colours are as in
  Fig. \ref{AqC5_FBall_jcirc}. Left-hand panel shows the gas MAH within the
  virial radius, and right-hand panel within the galactic radius.}
\label{AqC5_FBall_mah} 
\end{figure*}

Our FB1 model is an implementation of the model originally introduced
by \cite{DVS2012}, in the context of our MUPPI SF and ISM model. Here,
the relevant parameters are the temperature jump $\Delta T$, and the
fraction of SNII energy available to power the outflows, $f_{\rm fb, kin}$, 
i.e. the feedback efficiency. We adopt a temperature jump that depends on 
the density of gas particles that are eligible to produce feedback, as detailed 
in Section \ref{sec:DvS} 
(see also Appendix \ref{appendixB} for results of a simulation where we fix the temperature 
jump to the value of $T=10^{7.5}$ K). 
As far as the feedback efficiency is concerned, we adopted 
the density- and metallicity-dependent feedback efficiency used in the reference Eagle 
simulation \citep{schaye2015eagle}, as discussed in Section \ref{sec:DvS}. 
Such a choice successfully produces a disc-dominated galaxy at redshift $z=0$, 
even if this galaxy is characterized by high low-redshift gas MAH 
and resulting SFR (see figs \ref{AqC5_FBall_mah}, \ref{AqC5_FBall_sfr}, and 
discussion in Section \ref{sec:FBcomp}). 
Nonetheless, we prefer to consider the model adopting this feedback efficiency (equation \ref{FkEagle})
as reference FB1 simulation: we then discuss in Appendix \ref{appendixB} how different choices for 
$f_{\rm fb, kin}$ can reduce the amount of gas available for the low-$z$ phase of disc formation,
thus regulating the SFR at $z=0$. 

Energy provided by SF is distributed according to \citet{DVS2012}: first, we calculate 
the SNII energy release of the whole stellar population. Then, when a star is spawned, 
we use it to heat the surrounding gas. In principle, such mechanism of delivering energy contrasts 
with the continuous SFR of MUPPI model. In our sub-grid model, energy provided by SNe 
is indeed computed by using the energy contributed by the virtual stellar component in 
multiphase gas particles in each time-step, regardless of the moment in which the 
spawning of a star particle occurs. This results in a gradual way of distributing and releasing
energy. On the other hand, if the whole amount of energy is released impulsively when a star 
is really spawned, the feedback turns out to be more effective. The gradual way of distributing 
energy does not allow us to produce disc-dominated galaxies when the FB1 galactic outflow 
model is adopted, since the stellar feedback is not effective enough to prevent an excessive 
high-redshift SF burst. This set of choices allows us to produce an AqC5-FB1 galaxy with a 
dominant disc component. 

The FB2 scheme is a modification of the original MUPPI outflow model. In the process of 
tuning this model's parameters, we had to reduce the value of the SNII energy faction down 
to $f_{\rm fb, kin}=0.12$ and the probability of a gas particle to become a wind one to $P=0.03$, 
in order to account for the fact that this scheme does not limit the search of available wind 
particles to a cone but uses the whole smoothing sphere.

Finally, results for our FB3 scheme are shown with parameters $\eta=3.0$ and 
$f_{\rm fb, kin}=2.0$. In Appendix \ref{appendixB}, we discuss our choice of the parameter 
$f_{\rm fb, kin}$, that is the parameter to which our model is most sensitive. There, we also 
refer to \citet{SpringelHernquist2003}, whose model this galactic outflow scheme is inspired. 
Here, we note that the higher is the value of the mass loading factor (values of $\eta$ ranging 
between $0.5$ and $8$), the narrower and more reduced is the high-redshift SF burst, 
thus shortening the phase of bulge formation and limiting bulge size accordingly. 
Observations of low-redshift kpc-scale neutral gas outflows suggest a value for the 
mass loading factor that slightly exceeds unity \citep{Cazzoli2014}.

%%%%%%%%%%%%%%%%%%%%%%%% JCIRC di FB(all)

%%%%%%%%%%%%%%%%%%%%%%%% SFR di FB(all)

%%%%%%%%%%%%%%%%%%%%%%%% MAH di FB(all)

%%%%%%%%%%%%%%%%%%%%%%%% VROT di FB(all)
\begin{figure}
\newcommand{\captionfonts}{\small}
\hspace{-1.65ex}
\centering
\includegraphics[trim=0.4cm 0.4cm 0.35cm 0.0cm, clip, width=0.488\textwidth]{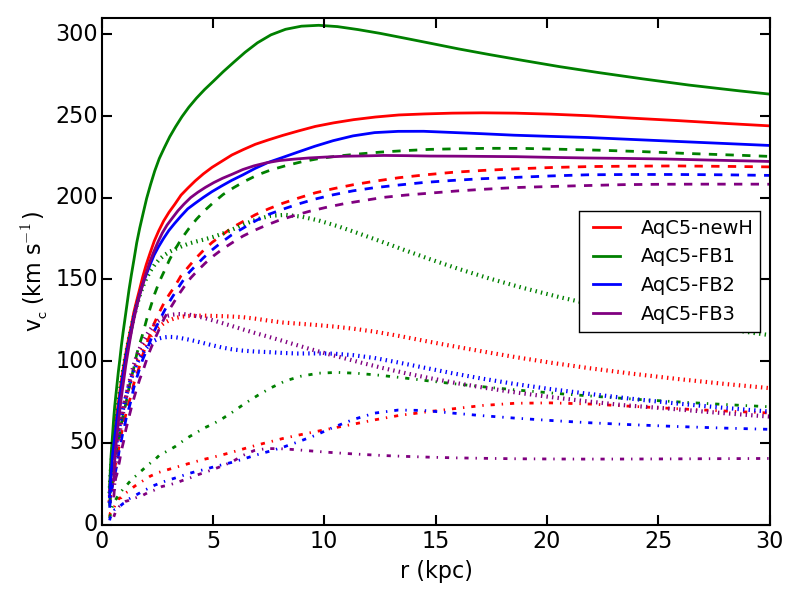} 
\caption{Rotation curves at $z=0$ for AqC5-FB1, AqC5-FB2, and AqC5-FB3,
  compared with that of AqC5-newH. Colours as in
  Fig. \ref{AqC5_FBall_jcirc}. The solid curves describe the circular
  velocity due to the total mass inside a given radius. The different
  contribution of DM (dashed line), stars (dotted) and gas
  (dot-dashed) to the total curve are shown.}
\label{AqC5_FBall_vrot} 
\end{figure}

\subsubsection{Comparing results}
\label{sec:FBcomp}

In this section, we discuss results obtained using our four galactic outflow 
models, the original one and those described in Section \ref{sec:confronto}. 
We focus on the following properties of the simulated galaxies: circularity of 
stellar orbits, SFRs, gas MAHs, rotation curves, Tully-Fisher relation and baryon 
conversion efficiencies, metallicity, and surface density profiles.

In Fig. \ref{AqC5_FBall_jcirc}, we show the circularities for all of our models, 
at redshift $z=0$. In all cases, we obtain rotationally supported galaxies. However, 
the bulge over total mass ratio changes significantly with the outflow model, with 
$B/T=0.30, 0.19, 0.35$, and $0.54$ for the original model, FB1, FB2 and FB3, 
respectively\footnote{We remind that our estimate 
  for $B/T$ ratio could also include satellites, stellar streams, contribution from 
  bars, etc, within $R_{\rm gal}$. Therefore, our $B/T$ values 
  should not be directly compared with observational photometric estimates.}. 
The total stellar mass of the galaxy (the integral of the circularity histogram) 
does change, too. Total stellar masses within $R_{\rm gal}$ for AqC5-newH, AqC5-FB1, 
AqC5-FB2, and AqC5-FB3 are 
$4.74 \cdot 10^{10}$ M$_{\odot}$, 
$9.34 \cdot 10^{10}$ M$_{\odot}$, 
$3.29 \cdot 10^{10}$ M$_{\odot}$, 
and $2.94 \cdot 10^{10}$ M$_{\odot}$, respectively. 
Galactic radii are $24.01$ kpc for AqC5-newH, 
$24.76$ kpc for AqC5-FB1, 
$23.90$ kpc for AqC5-FB2, 
and $23.58$ kpc for AqC5-FB3.

Fig. \ref{AqC5_FBall_sfr} shows the SFRs obtained for our four implementations 
of the outflow model, as a function of the cosmic time. SFRs are computed by considering 
all stars within $R_{\rm gal}$ at $z=0$.
The SFR of all models is comparable at high ($z>3$) redshifts. Bulges of simulated 
galaxies form at these epochs. At later times, during the disc formation phase, the 
behaviour of our outflow models differs. 
FB1 scheme has an SFR that is slightly higher than the other models between $3>z>2$, 
thus producing a further growth of the bulge, that is the most massive one. The SFR of this model 
has then a drop; at $z\simeq 0.7$ it starts to rapidly increase again, and it reaches the 
roughly constant value of $\simeq 20$ M$_\odot$ yr$^{-1}$ from $z\simeq 0.35$ on, 
i.e. over the last $\sim 4$ Gyr. 
\citet{schaye2015eagle} predict the SFR to be at most $\simeq 8$ M$_{\odot}$ yr$^{-1}$
for a galaxy with a stellar mass slightly lower than $8 \cdot 10^{10}$ M$_{\odot}$ at $z=0.1$ (this value 
corresponds to the upper edge of the 1$-\sigma$ uncertainty of their fig. 11). 
Our galaxy (whose stellar mass is $7.39 \cdot 10^{10}$ M$_{\odot}$ at $z=0.1$) 
has therefore an SFR in excess of that predicted by \citet{schaye2015eagle} by a factor of $2.5$.
However, unlike in \citet{schaye2015eagle}, our simulations do not include SMBHs 
(supermassive black holes) and the ensuing AGN (active galactic nucleus) feedback: 
AGN feedback is indeed expected to play a role in reducing low-redshift SFR (in Appendix \ref{appendixB}, 
we show how we regulated the low-$z$ SFR with a different choice of $f_{\rm fb, kin}$). 
Fig. \ref{AqC5_FBall_sfr} highlights how different prescriptions for the SF impact on final results, 
even when the same galactic outflow model is adopted. 
Both FB2 and FB3 schemes do show a drop in the SFR below $z=2$ with respect to AqC5-newH. 
In the FB2 case, SFR rises again after $z\simeq 0.5$, while it remains below 
$\sim 2$ M$_{\odot}$ yr$^{-1}$ for FB3. This is clearly at the
origin of the higher $B/T$ ratios produced by these two schemes: the early-forming 
bulge component is similar to that of AqC5-newH, but the disc formation phase 
is delayed or suppressed.

%%%%%%%%%%%%%%%%%%%%%%%% Surface densities e Metallicità di FB(all)
\begin{figure*}
\newcommand{\captionfonts}{\small}
\begin{minipage}{\linewidth}%{1.0\linewidth}
\centering
\includegraphics[trim=0.4cm 0.4cm 0.35cm 0.1cm, clip, width=0.489\textwidth]{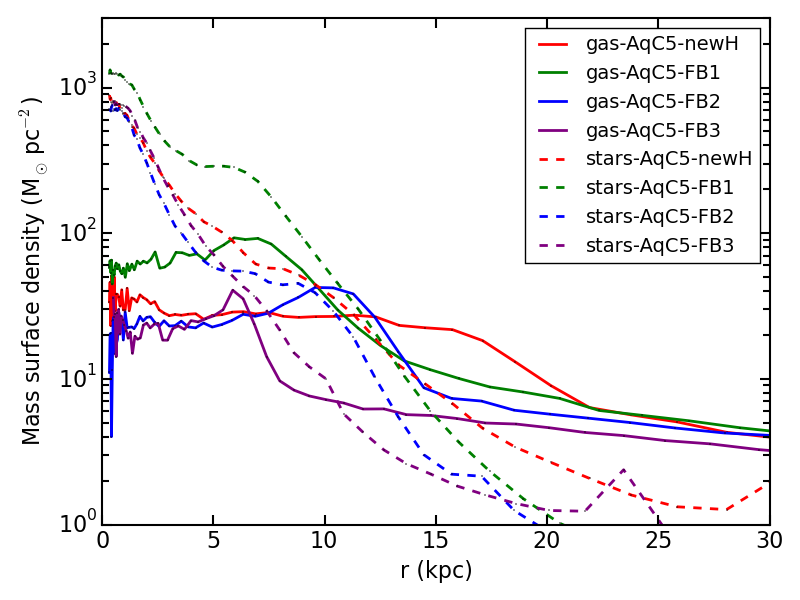} 
\includegraphics[trim=0.4cm 0.4cm 0.3cm 0.1cm, clip, width=0.496\textwidth]{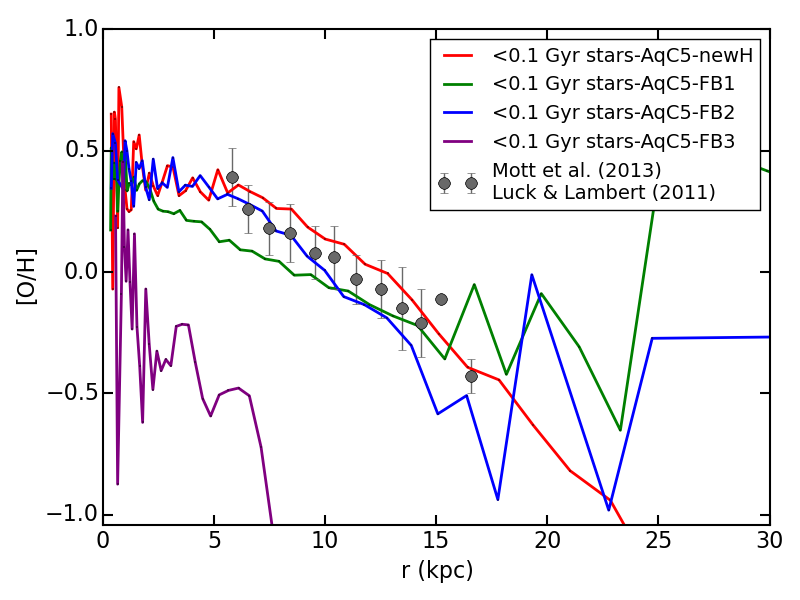}
\end{minipage} 
\caption{{\sl Left-hand panel:} surface density profiles for simulations AqC5-newH (red), AqC5-FB1 (green), 
  AqC5-FB2 (blue) and AqC5-FB3 (purple). Solid curves describe the contribution of gas and 
  dashed lines the one of the stars. {\sl Right-hand panel:} radial profiles of oxygen abundance in 
  young stars (see the text) for the four simulations. Data (grey filled circles) show the radial abundance gradient 
  for oxygen from observations of Cepheids \citep{Luck2011}, according to the division 
  in bins as a function of the Galactocentric distance performed by \citet{Mott2013}. Error 
  bars represent the standard deviation computed in each bin. All the profiles are analysed at $z=0$.}
\label{AqC5_FBall_met} 
\end{figure*}

The reason for the different behaviour of the SFR among models is further illustrated 
by Fig. \ref{AqC5_FBall_mah}. 
Here, we show the gas MAH on to the main progenitor of the galaxy as a function of the 
redshift. We have already shown in Fig. \ref{AqC5_HydroComparison4} a comparison 
of the gas MAH using the old and the new SPH implementation, and found a delay in the 
gas fall-back when we use the new scheme with respect to the old one. Here, the qualitative
behaviour of the MAH is similar for all of our four galactic outflow models for $z \ga 2$, 
both within the virial (left-hand panel) and the galactic (right-hand panel) radii. 
FB1 has the highest gas accretion rate from $z \simeq 1.5$ on. 
This feature highlights that this galactic outflow model is not as effective as the 
others in preventing gas accretion and explains the highest low-redshift SFR and 
final stellar mass of the galaxy simulated with this scheme. 
FB2 gas MAH closely follows that of our original scheme at the virial radius, but the gas accretion rate within 
$R_{\rm gal}$ at low redshift (below $z\sim2$) is significantly lower. We interpret this as the evidence that this 
kind of feedback is more effective at low redshift, generating strong outflows 
that stop gas from fuelling the disc growth. FB3 is even more effective in expelling gas from 
the galaxy, and at $z \la 1$ also from the halo. In fact, it has the lowest SFR, the smallest 
disc component and, as a consequence, the highest value of the $B/T$ ratio\footnote{We 
   show, for each galactic outflow model, the simulation that produces the best 
   results in terms of circularity histogram and SFR. Any shortfalls of each model 
   is therefore a direct result of how the outflow is described and cannot be  
   ascribed to an unreliable tuning of the parameters.}.

In Fig. \ref{AqC5_FBall_vrot}, we show the circular velocity profiles for all the schemes. 
AqC5-newH, FB2, and FB3 models have profiles that are flat at large radii, thus confirming their 
ability to prevent the formation of a strong baryonic concentration at the centre. 
FB1 scheme produces a galaxy whose stellar component of the rotation curve 
has the most pronounced peak at small radii ($r \la 5$ kpc): this is related to the larger mass 
of the bulge that is produced in this case. The morphology of this galaxy, with prominent stellar spiral arms, 
originates the peak of the velocity profile of the FB1 model, at $r \simeq 8$ kpc (see 
the stellar density profile in Fig. \ref{AqC5_FBall_met}, too). 
The normalization of the circular velocity profiles straightforwardly follows the 
different stellar (see above) and gas (see below) masses of the four simulations.

The left-hand panel of Fig. \ref{AqC5_FBall_met} shows surface density profiles of the 
four simulated galaxies. Dashed and solid curves refer to stars and to gas, respectively. 
Total gas masses within $R_{\rm gal}$ for AqC5-newH, AqC5-FB1, AqC5-FB2, and AqC5-FB3 are 
$2.91 \cdot 10^{10}$ M$_{\odot}$, 
$3.33 \cdot 10^{10}$ M$_{\odot}$, 
$2.07 \cdot 10^{10}$ M$_{\odot}$, 
and $8.82 \cdot 10^{9}$ M$_{\odot}$, respectively. 
Total gas masses within $R_{\rm vir}$ for AqC5-newH, AqC5-FB1, AqC5-FB2, and AqC5-FB3 are 
$1.36 \cdot 10^{11}$ M$_{\odot}$, 
$1.48 \cdot 10^{11}$ M$_{\odot}$, 
$1.36 \cdot 10^{11}$ M$_{\odot}$, 
and $9.75 \cdot 10^{10}$ M$_{\odot}$, respectively. 
Gas masses directly come from the different gas MAHs within $R_{\rm vir}$ and $R_{\rm gal}$ 
(see Fig. \ref{AqC5_FBall_mah}).
AqC5-newH has both the most extended gas disc and the stellar component 
that extends farthest from the galaxy centre. 
AqC5-FB1 has the largest gas mass surface density in the innermost regions.  
Gas surface density profiles of AqC5-FB1, AqC5-FB2 and AqC5-FB3 exhibit a bump 
at the edge of the gas disc (at distances of $7$, $10$ and $6$ kpc, respectively). 
In particular for AqC5-FB1 and AqC5-FB2, it is possible to note a corresponding excess 
in the stellar surface density. This is the effect of high angular momentum gas infalling 
within $R_{\rm gal}$ at low redshift. The shallower decrease in the stellar surface density profiles is a 
consequence of the enhancement of the SF due to the recently accreted gas. The  
infalling gas is material ejected from the forming halo -- due to the high-redshift ($z \ga 3$) 
SF burst that formed the bulge -- with a low angular momentum. After acquiring angular 
momentum from halo gas \citep{Brook2012,Ubler2014,Teklu2015,Genel2015}, the expelled material
fell back, in a way regulated by the stellar feedback. As for AqC5-FB3, the gas fall-back 
occurred at earlier times than in AqC5-FB1 and AqC5-FB2 (see Fig. \ref{AqC5_FBall_mah} 
and discussion above), where instead the process is still ongoing.

The right-hand panel of Fig. \ref{AqC5_FBall_met} shows the metallicity radial profiles of the four 
simulations, where we consider radial profiles of oxygen abundance in stars (within $R_{\rm gal}$) 
that are younger than $\sim 100$ Myr, in order to make a fair comparison with observations (as 
discussed in Section \ref{NewHydroOldFb}). Data (grey filled circles) are the same as in the last 
panel of Fig. \ref{AqC5_HydroComparison3}. 
Negative radial abundance gradients are recovered for all the models, in line with observational 
results. The agreement between oxygen abundance radial profiles predicted by different simulations 
and observations demonstrates the effectiveness of a sub-resolution model (and in particular of a 
galactic outflow model) to properly describe the chemical enrichment of the galaxy at different positions.
We see that the galactic outflow model of AqC5-newH and FB2 well reproduce the slope of observed 
radial abundance profiles.
The slope of the profile in AqC5-FB1 is slightly flatter than in observations. 
We note that the normalization of radial abundance profiles in simulations is a delicate 
aspect (as discussed in Section \ref{NewHydroOldFb}); nonetheless, the normalization of the oxygen 
abundance profile in both the AqC5-newH and AqC5-FB2 simulations is consistent with the observed 
abundance gradient within the error bars. The metallicity radial profile predicted by the AqC5-FB1 simulation 
shows an oxygen abundance lower than observations in the inner regions ($r \la 10$ kpc). The 
metal content of young stars throughout the galaxy is more homogeneous: this galactic outflow model is 
able to promote a more efficient spread of metals from stars to the surrounding gas. 
As for the FB3 model, it does not succeed in reproducing observations. The AqC5-FB3 galaxy has 
a limited stellar disc (see Fig. \ref{AqC5_FBall_met}, left-hand panel), it is characterized by 
a low SFR (Fig. \ref{AqC5_FBall_sfr}) and a reduced gas MAH (Fig. \ref{AqC5_FBall_mah}) below $z \sim 1$: 
as a consequence, a lower amount of metals has been produced in a confined region (subsequently 
generated stars being richer in heavy metals than previous ones).  
As for the relative difference between slopes and normalizations of profiles in simulations 
AqC5-newH and AqC5-FB2, metals are less locked into stars in AqC5-newH with respect 
to AqC5-FB2. 

\begin{figure}
\newcommand{\captionfonts}{\small}
\hspace{-1.65ex}
\centering
\includegraphics[trim=0.5cm 0.5cm 0.2cm 0.0cm, clip, width=0.4855\textwidth]{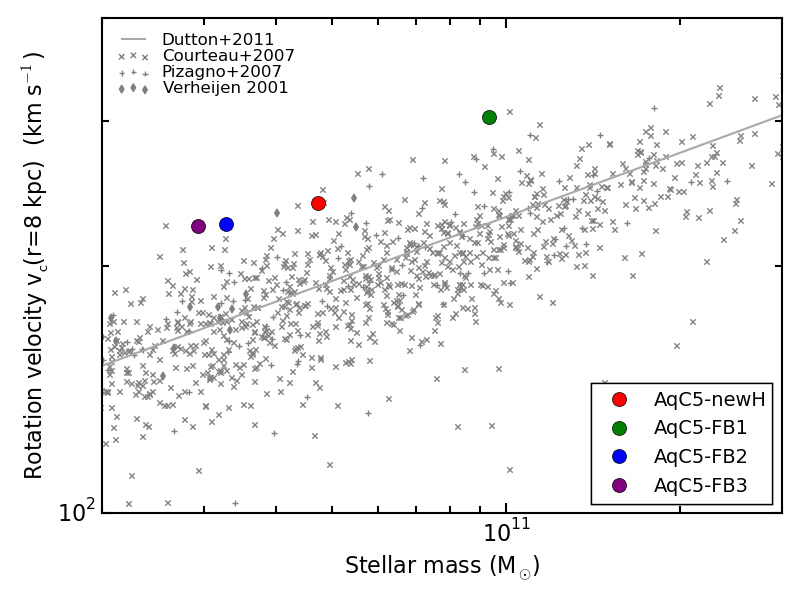} 
\caption{Tully-Fisher relation at $z=0$ for AqC5-FB1, AqC5-FB2, and
  AqC5-FB3, compared with that of AqC5-newH. Colours are the same used in
  Fig. \ref{AqC5_FBall_jcirc}. We use the circular velocity at a distance of 
  $8$ kpc from the galaxy centre as a function of the galaxy stellar mass 
  within the galactic radius. Grey symbols are as in Fig. \ref{AqC5_HydroComparison5}.}
\label{AqC5_FBall_tf} 
\end{figure}

Fig. \ref{AqC5_FBall_tf} shows the position of the four simulated galaxies in the 
Tully-Fisher diagram. Also in this case, as already discussed in Section \ref{NewHydroOldFb}, 
all the simulated galaxies lie close to the upper limit of the region allowed by observations. 
AqC5-FB1 galaxy lies just outside the region identified by the scatter of the observed data: 
this is a consequence of its peaked rotation curve (Fig. \ref{AqC5_FBall_vrot}). 
Rotation velocity scales with the stellar mass of the simulated galaxies, thus the position in the diagram of 
our four galaxies just relates with their stellar mass.

%%%%%%%%%%%%%%%%%%%%%%%% TF di FB(all)

%%%%%%%%%%%%%%%%%%%%%%%% BCE di FB(all)
\begin{figure}
\newcommand{\captionfonts}{\small}
\centering
\hspace{-1.85ex}
\includegraphics[trim=0.44cm 0.5cm 0.cm 0.0cm, clip, width=0.488\textwidth]{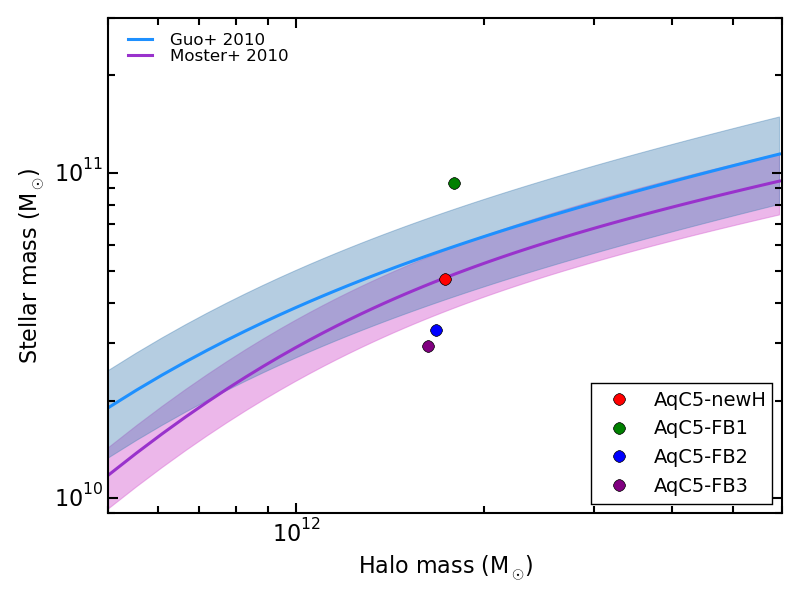} 
\caption{Baryon conversion efficiencies at $z=0$ for AqC5-FB1,
  AqC5-FB2, and AqC5-FB3, compared with that of AqC5-newH. Colours are as
  in Fig. \ref{AqC5_FBall_jcirc}. Blue solid line shows the stellar-to-halo 
  mass relation derived by \citet{Guo2010}, with the shaded blue region representing 
  an envelope of 0.2 dex around it. Purple solid curve describes the fit proposed by
  \citet{Moster2010}, with 1$\sigma$ uncertainty (shaded envelope) on the normalization.}
\label{AqC5_FBall_bce} 
\end{figure}

Finally, in Fig. \ref{AqC5_FBall_bce}, we show the position of our simulations in the 
baryon conversion efficiency plot. 
FB1 is characterized by a baryon conversion efficiency that slightly exceeds the uncertainty 
contour of the prediction by Guo et al. (2010) and that overestimates the one by Moster et al. (2010). 
FB2 shows a baryon conversion efficiency lower than the expectations 
from \citet{Guo2010} and \citet{Moster2010}, and also FB3 underproduces stars. 
We note that the lower the baryon conversion efficiency of simulated galaxies is, 
the larger their $B/T$ (see Fig. \ref{AqC5_FBall_jcirc}).

A general result of this section is that the observational properties of the 
simulated galaxies are sensitive to the details of the different outflow models, 
even when they are implemented within the same SF scheme. 
Some of these properties, for instance
the bulge mass and (to a lower extent) the radial dependence of the rotation velocity (at large radii)  
are relatively stable against the considered implementations of galactic outflow model. 
On the other hand, the prominence of the disc component is highly sensitive to the
implemented model and, more specifically, to the timing of the gas fall-back and 
the mass of the gas involved. In fact, a feedback model that is successful in 
producing a disc-dominated galaxy should be able to regulate the pristine SF, 
which is responsible for the creation of the bulge component, and, at the same time, 
not to allow a too late gas fall-back, that would turn into a too high SFR at $z=0$. 
The results of our simulations confirm that this represents a non-trivial requirement 
on galactic outflow models to be implemented in cosmological simulations of galaxy formation.

\section{Summary and Conclusions}
\label{sec:conclusions}
In this paper, we presented a suite of cosmological simulations of a 
Milky-Way sized halo ($M_{\rm halo, DM} \simeq 2 \cdot 10^{12}$
M$_{\odot}$), namely Aquila C5 \citep{Springel2008}, using the 
TreePM+SPH GADGET3 code. 
As a first aim of our analysis, we coupled a sub-resolution star formation (SF) 
and stellar feedback model, called MUPPI \citep[][M15]{muppi2014}, with an 
improved implementation of SPH that includes a high-order interpolating
kernel, an improved treatment of artificial viscosity (AV) and an artificial 
conduction (AC) term that promotes mixing at gas interfaces \citep{beck2015}. 
As a second goal, we implemented three different models for galactic outflows, 
besides the original model presented by \citetalias{muppi2014}, and used them to re-simulate the 
same initial conditions (ICs) with the same SF prescription. For each scheme, we 
tuned the relevant model parameters by carrying out an exploration of their
respective parameter space (see Appendix \ref{appendixB}). To choose
the reference values of the parameters, we used the criterion based on
maximizing the stellar disc component, both by visual inspection
and by plotting the circularity histograms. Our simulations do not include AGN feedback. 

The key features of our galactic outflow models are the following:
\begin{itemize}
\item In our original scheme, each particle ending its multiphase
  stage has a probability of receiving kinetic energy from neighbour 
  star-forming particles, providing that it is inside a cone centred on the
  star-forming particles and directed towards the least-resistance direction. 

\item FB1 is a stochastic thermal feedback model based on \citet{DVS2012}. 
  It is here implemented within our sub-resolution model for the ISM and SF,
          different from the original one by \citet{DVS2012}.

\item FB2 is a kinetic stellar feedback scheme, where the ISM is isotropically 
  provided with energy that triggers galactic outflows.

\item FB3 is a feedback model where we directly impose the mass loading factor 
  and stochastically sample the outflowing mass. SN energy and SFRs are collected 
  during the multiphase stage of gas particles, thus accounting for the evolution of the 
  ISM. This scheme is inspired by \citet{SpringelHernquist2003}.
\end{itemize}

In order to successfully introduce the improved SPH implementation in cosmological 
simulations adopting the sub-resolution model MUPPI, we switched off AC for both 
multiphase and wind particles, and implemented a switch that allows former multiphase 
particles to artificially conduct whenever some conditions on their temperature are met. 
In fact, the physical state of the ISM is determined by continuous energy injections due 
to the sub-grid model. This results in a particle-by-particle difference in temperature and 
density. AC acts in mixing internal energy, contrasting the effect of the sub-grid model. 
The new switch suppresses AC for those particles that are sampling the non-uniform thermal 
structure of the galaxy, where the sub-resolution physics accounts for the evolution of the ISM. 

Our main results can be summarized as follows.
\begin{itemize}

\item MUPPI is able to produce a disc galaxy using the AqC5 ICs, 
  also when coupled with the improved SPH scheme. The simulated 
  galaxy has a flat rotation velocity profile (flatter than the one of the galaxy simulated 
  with the previous SPH implementation), a low value of the $B/T$ ratio, 
  extended gaseous and stellar discs, baryon conversion efficiency, and 
  position on the Tully-Fisher plot in broad agreement with expectations. 
  However, we note that significant differences arise with respect to the 
  results obtained by adopting the standard SPH scheme. Using our new 
  SPH solver introduces a delay in the gas fall-back that causes the stellar 
  disc to form later, and the simulated galaxy to have a relatively higher low-redshift SFR.

\item Changing the outflow model does affect the properties of the
  simulated galaxies at redshift $z=0$. Some of them are relatively stable, 
  e.g. the high-redshift ($z \ga 3$) SFR which determines the present-day stellar mass of the
  bulge. However, the majority of the properties, in particular the prominence of the disc component, 
  shows a sensitive dependence on the employed outflow model.

\item FB1 model generates the galaxy with the largest stellar mass, a
  strong disc component, high SFR at redshift below $z \simeq 0.7$. 
  FB2 model has a smaller stellar mass at $z=0$ with respect to
  our reference outflow scheme, and shows a drop in SFR between
  $z=2$ and $z=0.3$. FB3 model strongly suppresses SF after $z=2$, 
  producing in turn the lowest value of the $B/T$ ratio.

\item The different behaviour of our outflow models can be understood
  in terms of gas mass accretion history (MAH). FB1 almost always has the highest 
  gas MAH; FB2 shows an MAH similar to that of our reference model within the
  virial radius, but lower within the galactic radius; on the other hand,
  FB3 is too effective in quenching the gas fuelling of the disc. In this latter outflow 
  model, the outflowing gas is kept inside the halo, but away from the galactic radius.

\end{itemize}

Our analysis demonstrates and quantifies the strong interplay between details of
the hydrodynamic scheme and the sub-resolution model describing SF and feedback. 
The sub-resolution model plays a fundamental role in determining the properties of 
simulated galaxies and the exact values of free parameters entering into this model 
can significantly change with the hydrodynamic solver. 
The hydrodynamic scheme remarkably affects properties of simulated galaxies 
(stellar mass is the quantity that is affected the most) and highly regulates gas inflow 
and outflow across cosmic time. Moreover, the higher the resolution is, the more 
sensitive to the accuracy of the hydrodynamic solver properties of simulated galaxies 
are (see Appendix \ref{appendixA}).

Our results further show that the sub-resolution prescriptions adopted to
generate galactic outflows are the main shaping factor of the stellar
disc component at low redshift (when the hydrodynamic scheme is held fixed). 
In particular, the timing of the gas fall-back after its expulsion from the forming 
halo at high redshift is of paramount importance, together with the ability of a 
given model to tune or even quench the cosmological infall of gas from the
large-scale environment. For this reason, a detailed comparison between 
simulation results and observations of the properties of the circumgalactic
medium at different redshifts should provide invaluable information on
the outflow/inflow gas properties and, therefore, on the history of feedback.
Another interesting direction of investigation will be to compare the properties of
simulated galaxy populations (instead of those of a single galaxy),
obtained by adopting one or more of our outflow models, with observations.

\section*{Acknowledgments}
We thank the anonymous referee for a prompt, careful and constructive report.  
We are greatly indebted to Volker Springel for giving us access to the
developer version of the GADGET3 code. We acknowledge Francesca Matteucci 
and Emanuele Spitoni for useful discussions and for providing observational data 
of radial abundance gradient for oxygen.
This work received financial support from the PRIN-MIUR 201278X4FL Grant, 
the PRIN-INAF 2012 \textquotedblleft The Universe in a Box: Multi-scale Simulations of 
Cosmic Structures\textquotedblright $\:$ Grant, the INFN INDARK Grant, and \textquotedblleft 
Consorzio per la Fisica\textquotedblright $\:$ of Trieste. AMB acknowledges support by the DFG 
cluster of excellence \textquotedblleft Universe\textquotedblright, the DFG research unit 1254 
and the Leibniz-Rechenzentrum with computing time assigned to the project \textquotedblleft 
pr92ju\textquotedblright. Simulations were carried out using Galileo at CINECA (Italy), with 
CPU time assigned through Italian Super-Computing Resource Allocation (ISCRA) proposals 
and an agreement with the University of
Trieste, PICO at CINECA through our expression of interest, and ULISSE at SISSA.

%%%%%%%%%%%%%%%%%%%%%%%%%%%%%%%%%%%%%%%%%%%%%%%%%%
%%%%%%%%%%%%%%%%%%%%%%%%%%%%%%%%%%%%%%%%%%%%%%%%%%



\bibliographystyle{mnras} 
\bibliography{cool_ref}

%%%%%%%%%%%%%%%%%%%%%%%%%%%%%%%%%%%%%%%%%%%%%%%%%%
%%%%%%%%%%%%%%%%%%%%%%%%%%%%%%%%%%%%%%%%%%%%%%%%%%

%%%%%%%%%%%%%%%%% APPENDICES %%%%%%%%%%%%%%%%%%%%%
%%%%%%%%%%%%%%%%%%%%%%%%%%%%%%%%%%%%%%%%%%%%%%%%%%

\appendix
\section{Effect of the resolution}
\label{appendixA}

In this appendix, we discuss how our main results change when we
decrease the mass resolution by a factor of $\sim10$ (see Table
\ref{simmHC}).

Table \ref{simmAPP} lists the set of simulations that we analyse in
Appendices \ref{appendixA} and \ref{appendixB}. We do not report again
here simulations already introduced in Tables \ref{simmHC} and
\ref{simmFC}. The simulation AqC6-newH that has been introduced in
Tables \ref{simmHC} and \ref{tab2} is now labelled as AqC6-newH\_p003.
The simulation AqC5-FB3 that had been introduced in Tables
\ref{simmFC} is now labelled as AqC5-FB3\_fk2.0.

\begin{table}
\centering
\caption{Name and details of the simulations discussed in Appendices \ref{appendixA} and \ref{appendixB}.
Column 1: label of the run.
Column 2: parameter of interest for the comparison investigated here. 
Column 3: hydrodynamic scheme.
Column 4: galactic outflow model.}
\begin{tabular}{@{}lccr@{}}
\hline
Simulation &   
Parameter & Outflow & Hydro \\ 
 &    
 $f_{\rm fb, kin}$ or $P_{\rm kin}$  & model  &  scheme \\ 
\hline
\hline
AqC6-newH\_p003 &   $P_{\rm kin}$=0.03 &  M15$^1$ & New \\  
\hline
AqC6-newH\_p005 &    $P_{\rm kin}$=0.05 & M15 & New \\  
\hline
%new
AqC5-FB1$f$2.0-1.0 &   Equation (\ref{eqTheta}) with & FB1 & New \\  
  &   ${f_{\rm fb, kin}}^{\rm max}_{\rm min}={}^{2.0}_{1.0}$ &   &   \\
\hline
%new
AqC6-FB1$f$Eagle &   Equation (7) of & FB1 & New \\ 
  & \citet{schaye2015eagle} & &  \\ 
  &  for $f_{\rm fb, kin}$ (see the text) & &  \\ 
\hline
%new
AqC6-FB1- $\Delta$T$_{\rm fix}$ &   $f_{\rm fb, kin}$: equation (7) of & FB1 & New \\ 
  & \citet{schaye2015eagle}; & &  \\ 
  & $\Delta T_{\rm fix}=10^{7.5}$ K & &  \\ 
\hline
%new
AqC6-FB1$f$1.0 &   $f_{\rm fb, kin}$=1.0 & FB1 & New \\ 
\hline
AqC5-FB3\_fk2.0 &   $f_{\rm fb, kin}$=2.0 & FB3 & New \\
\hline
AqC5-FB3\_fk0.5 &   $f_{\rm fb, kin}$=0.5 & FB3 & New \\
\hline
AqC5-FB3\_fk0.25 &   $f_{\rm fb, kin}$=0.25 & FB3 & New \\
\hline
\hline
 
\end{tabular}
\label{simmAPP}
\rightline{$^1$ \citet{muppi2014}}
\end{table}

\begin{figure*}
\newcommand{\captionfonts}{\small}
\vspace{-1.1ex}
\includegraphics[trim=0.0cm 0.0cm 0.0cm 0.0cm, clip, width=0.325\textwidth]{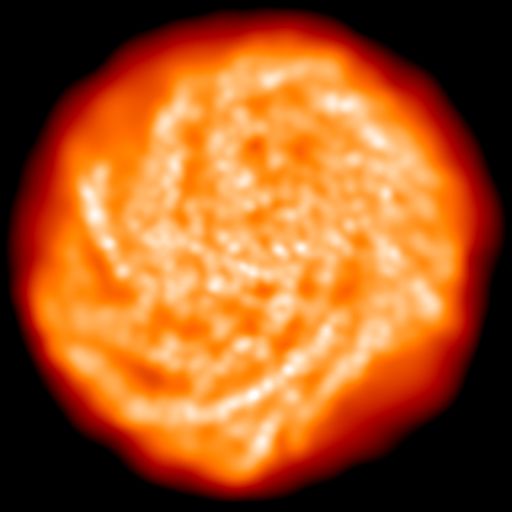}
\hspace{-0.6ex}
\includegraphics[trim=0.0cm 0.0cm 0.0cm 0.0cm, clip, width=0.325\textwidth]{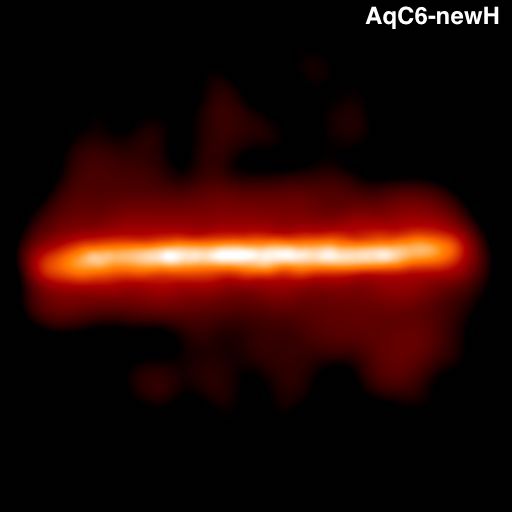}\\
\vspace{0.3ex}
\includegraphics[trim=0.0cm 0.0cm 0.0cm 0.0cm, clip, width=0.325\textwidth]{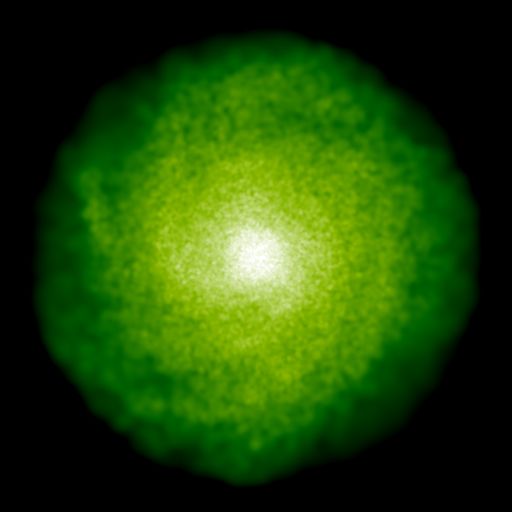} 
\hspace{-0.6ex}
\includegraphics[trim=0.0cm 0.0cm 0.0cm 0.0cm, clip, width=0.325\textwidth]{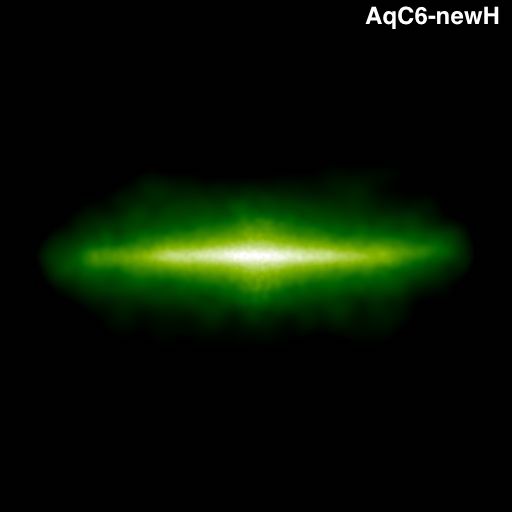}\\ 
\vspace{1.1ex}
\includegraphics[trim=0.0cm 0.0cm 0.0cm 0.0cm, clip, width=0.325\textwidth]{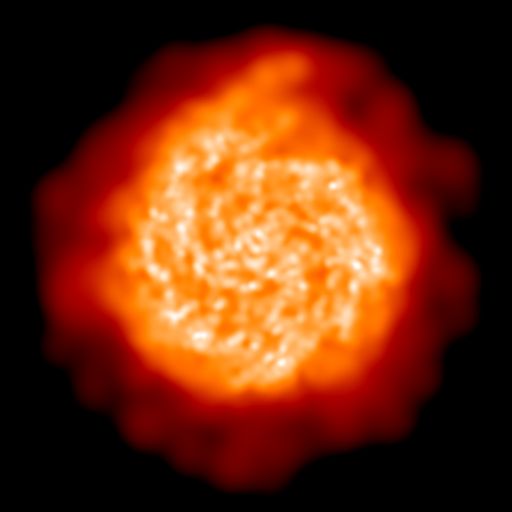}
\hspace{-0.6ex}
\includegraphics[trim=0.0cm 0.0cm 0.0cm 0.0cm, clip, width=0.325\textwidth]{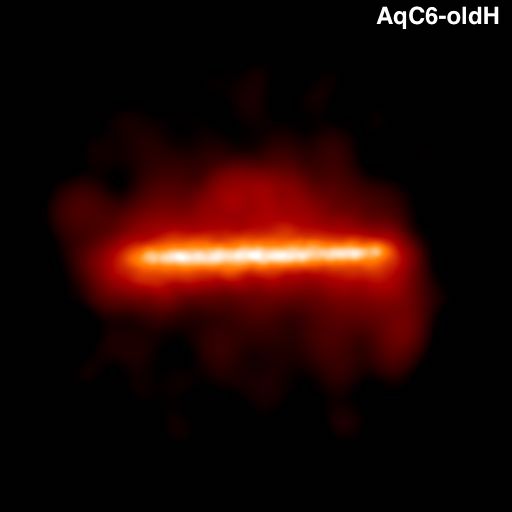}\\
\vspace{0.3ex}
\includegraphics[trim=0.0cm 0.0cm 0.0cm 0.0cm, clip, width=0.325\textwidth]{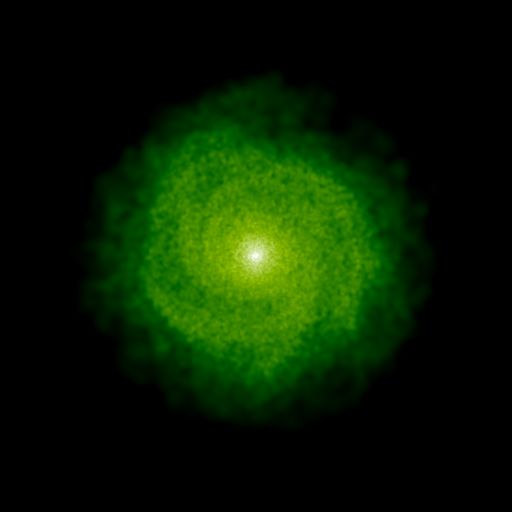} 
\hspace{-0.6ex}
\includegraphics[trim=0.0cm 0.0cm 0.0cm 0.0cm, clip, width=0.325\textwidth]{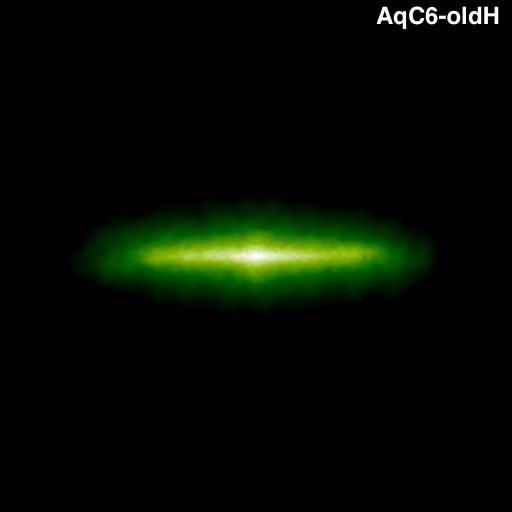} 

\vspace{-0.9ex}
\caption{{\sl Top four panels:} projected gas (first and second panels) and star (third and fourth) density for the AqC6-newH\_p003 simulation. {\sl Bottom four panels:} projected gas (fifth and sixth) and star (bottom) density for AqC6-oldH. Left- and right-hand panels show face-on and edge-on densities, respectively. 
All the maps are shown at redshift $z=0$. The size of the box is 55 kpc.} 
\label{AqC6_newHydro} 
\end{figure*}

\begin{figure} \newcommand{\captionfonts}{\small} \hspace{-1.65ex}
\centering 
\includegraphics[trim=0.4cm 0.4cm 0.35cm 0.4cm, clip, width=0.488\textwidth]{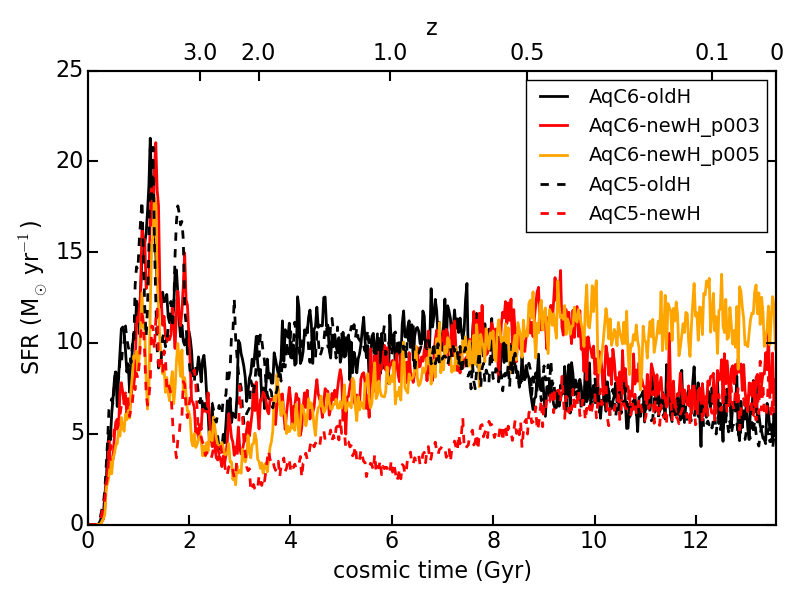}
  \caption{Evolution of the SFR for simulations
    AqC6-oldH (black solid line), AqC6-newH\_p003 (red solid line),
    and AqC6-newH\_p005 (orange solid line). Dashed lines refer to
    higher resolution simulations: AqC5-oldH (black) and AqC5-newH
    (red).}
\label{AppA_sfr} 
\end{figure}

\begin{figure}
\newcommand{\captionfonts}{\small}
\hspace{-1.65ex}
\centering
\includegraphics[trim=0.4cm 0.4cm 0.35cm 0.2cm, clip, width=0.488\textwidth]{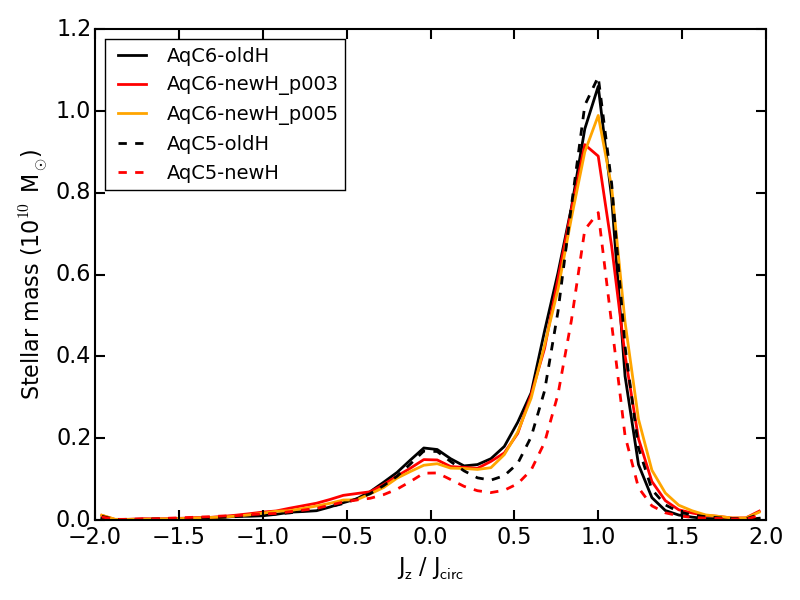} 
\caption{Stellar mass as a function of the circularity of stellar
  orbits at $z=0$ for the following simulations: AqC6-oldH (black
  solid), AqC6-newH\_p003 (red solid), AqC6-newH\_p005 (orange solid),
  AqC5-oldH (black dashed) and AqC5-newH (red dashed).}
\label{AppA_jcirc} 
\end{figure}

\begin{figure} \newcommand{\captionfonts}{\small} \hspace{-1.65ex}
\centering 
\includegraphics[trim=0.4cm 0.4cm 0.35cm 0.1cm, clip, width=0.488\textwidth]{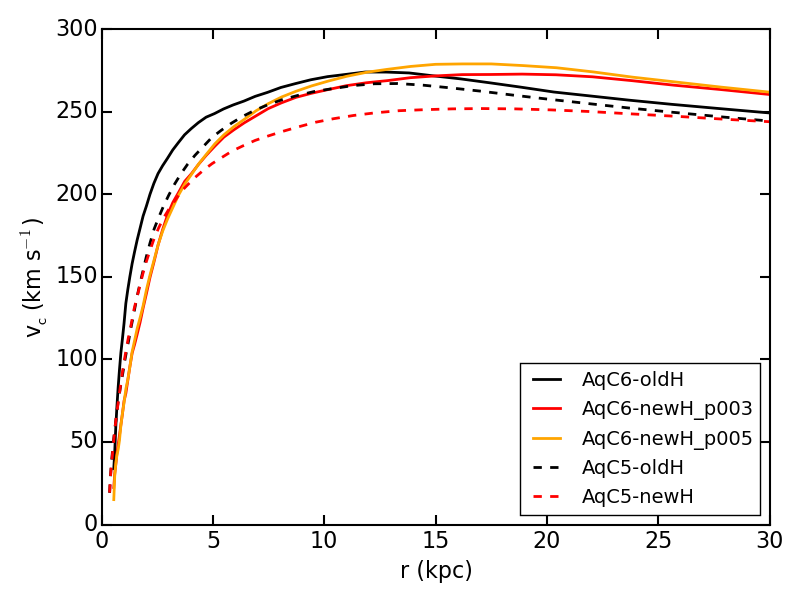}
  \caption{Rotation curves for simulations AqC6-oldH (black solid
    line), AqC6-newH\_p003 (red solid line), and AqC6-newH\_p005
    (orange solid line). Dashed lines refer to higher resolution
    simulations: AqC5-oldH (black) and AqC5-newH (red). All the curves
    describe the circular velocity due to the total mass inside a
    given radius.}
\label{AppA_vrot} 
\end{figure}

\begin{figure}
\newcommand{\captionfonts}{\small}
\hspace{-1.65ex}
\centering
\includegraphics[trim=0.4cm 0.4cm 0.35cm 0.1cm, clip, width=0.488\textwidth]{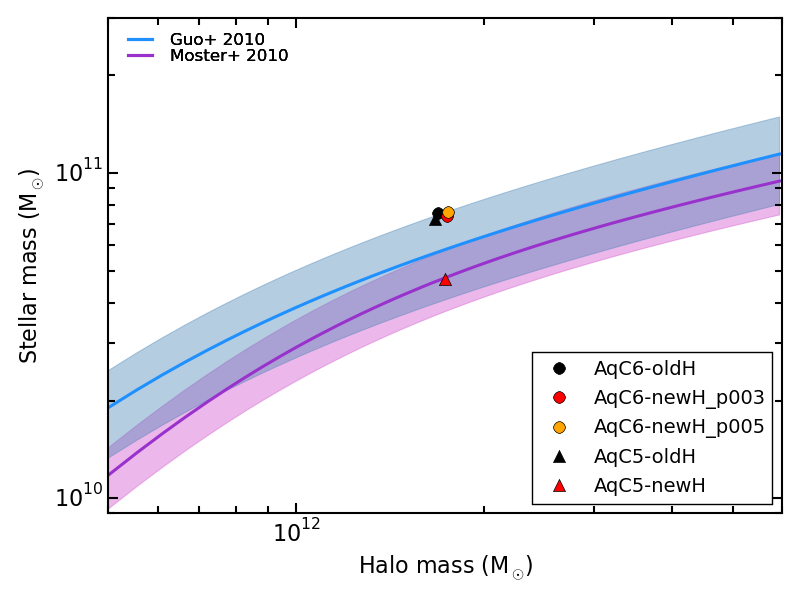} 
\caption{Comparison between the baryon conversion efficiency from observations and from
  the simulated galaxies at $z=0$. Curves show the stellar-to-halo mass relations derived by 
  \citet[][blue]{Guo2010} and by \citet[][purple]{Moster2010}, as in the right-hand panel of Figs
  \ref{AqC5_HydroComparison5} and \ref{AqC5_FBall_bce}.
  Symbols refer to simulations: AqC6-oldH (black filled circle),
  AqC6-newH\_p003 (red filled circle), AqC6-newH\_p005 (orange filled
  circle), AqC5-oldH (black filled triangle) and AqC5-newH (red filled
  triangle), respectively.}
\label{AppA_bce} 
\end{figure}

We now focus on simulations AqC6-oldH, AqC6-newH\_p003,
AqC6-newH\_p005, and we contrast their results with those of AqC5-oldH
and AqC5-newH that we have analysed in Section \ref{NewHydroOldFb}.

The four upper panels of Fig. \ref{AqC6_newHydro} show face-on and edge-on 
projections of gas (top panels) and stellar (third and fourth ones) density for the
AqC6-newH\_p003 galaxy simulation, at $z=0$. The four lower panels depict face-on 
and edge-on projections of gas (fifth and sixth panels) and stellar (bottom ones) density 
for AqC6-oldH, at $z=0$. 
The reference system has been rotated as described for Fig. 
\ref{AqC5_newHydro_ACoff_mapXyz}. 
AqC6-newH\_p003 is a clearly disc-dominated galaxy, and a spiral pattern is evident, 
especially in the gas component. The gaseous disc of the galaxy produced by adopting 
the improved hydrodynamic scheme is more extended than the stellar one at this resolution, too.
Continuing to focus on the galaxy simulated with the new SPH, it is possible to notice how the 
density in the outer part of the gaseous halo is no longer resolved above a given height 
on the disc plane (see top-right panel) at this resolution, with respect to Fig. \ref{AqC5_newHydro}.
Moreover, both the gaseous and the stellar disc of AqC6-newH\_p003 are more extended 
than the analogous ones of AqC6-oldH when this lower resolution is considered, too.

Figs \ref{AppA_sfr}, \ref{AppA_jcirc}, \ref{AppA_vrot}, and
\ref{AppA_bce} show the impact of the hydrodynamic solver on the
results of AqC6 simulations and how our main results are sensitive to
the change of resolution. For the sake of completeness, besides
AqC6-newH\_p003, we also show results of AqC6-newH\_p005, where we
have set the probability $P_{\rm kin}=0.05$ (see Table \ref{simmAPP})
that is the same parameter value adopted for the reference higher
resolution simulation AqC5-newH.

Fig. \ref{AppA_sfr} shows SFRs as a function of the cosmic time (and
of the redshift, too). SFRs of all the analysed simulations are
comparable for $z>3$. Then, if we focus on the AqC6 simulations, we
see that the impact of the new hydrodynamic solver makes SFRs of
AqC6-newH\_p003 and AqC6-newH\_p005 be lower than that of AqC6-oldH
between $3 \la z \la 0.7$. The main reason of this discrepancy lies in the presence 
of the time-step limiting particle wake-up scheme, as discussed for AqC5 runs in 
Section \ref{NewHydroOldFb}. AqC6-newH runs then overpredict the SFR
until the cosmic time reaches $\sim10$ Gyr; from that moment on, the SFR
of AqC6-newH\_p003 starts to gently decline and matches the SFR of
AqC6-oldH, while AqC6-newH\_p005 keeps constantly larger than the
prediction of AqC6-oldH (this is the actual reason why we prefer
AqC6-newH\_p003 rather than AqC6-newH\_p005, though comparable). 
SFRs of AqC5-oldH and AqC6-oldH are relatively insensitive to mass
resolution \citepalias[see also][]{muppi2014}; on the other hand, the variation of
resolution is quite remarkable between AqC5-newH and AqC6-newH\_p003
(and AqC6-newH\_p005).

Fig. \ref{AppA_jcirc} describes the distribution of stellar mass as
a function of the circularity of stellar orbits. All the galaxies are
disc-dominated, with the hydrodynamic solver only slightly affecting
the amount of stellar mass in both the bulge and the disc component.
However, the different stellar mass in the disc component of AqC6-oldH
and AqC6-newH\_p003 (or AqC6-newH\_p005, similarly) is not as
pronounced as in the comparison between simulations AqC5-oldH and
AqC5-newH. The effect of the resolution is much more apparent when
using the new SPH scheme: a complete agreement between 
AqC6-newH\_p003 and AqC5-newH is not reached even when the 
weak convergence is considered \citep{schaye2015eagle}. 
On the other hand, AqC6-oldH and AqC5-oldH 
show strong convergence with numerical resolution. 
$B/T$ ratios for AqC6-oldH, AqC6-newH\_p003, for AqC6-newH\_p005, 
AqC5-oldH, and AqC5-newH are:
0.23, 0.25, 0.22, 0.23, and 0.30, respectively.

Fig. \ref{AppA_vrot} shows results for the circular velocity
profiles. Quite remarkably, none of the analysed
profiles is centrally peaked, thus pinpointing that the adopted
galactic outflow model is effective in preventing the formation of a
too much centrally concentrated galaxy at both the resolutions.
The different normalizations highlight the differences in the total
stellar and gas mass that we obtain for the different tests. 
Values of the total stellar mass within the galactic radius $R_{\rm gal}$ are: 
$7.57 \cdot 10^{10}$ M$_{\odot}$ ($R_{\rm gal}=24.00$ kpc) for AqC6-oldH, 
$7.38 \cdot 10^{10}$ M$_{\odot}$ ($R_{\rm gal}=24.19$ kpc) for AqC6-newH\_p003, 
$7.61 \cdot 10^{10}$ M$_{\odot}$ ($R_{\rm gal}=24.24$ kpc) for AqC6-newH\_p005, 
$7.22 \cdot 10^{10}$ M$_{\odot}$ ($R_{\rm gal}=23.86$ kpc) for AqC5-oldH, 
and $4.74 \cdot 10^{10}$ M$_{\odot}$ ($R_{\rm gal}=24.01$ kpc) for AqC5-newH. 
Values of the gas mass within the galactic radius are: 
$1.85 \cdot 10^{10}$ M$_{\odot}$ for AqC6-oldH, 
$4.47 \cdot 10^{10}$ M$_{\odot}$ for AqC6-newH\_p003, 
$4.37 \cdot 10^{10}$ M$_{\odot}$ for AqC6-newH\_p005, 
$1.66 \cdot 10^{10}$ M$_{\odot}$ for AqC5-oldH, 
and $2.91 \cdot 10^{10}$ M$_{\odot}$ for AqC5-newH, at $z=0$. 
We note that the difference in total stellar and gas mass between
AqC5-oldH and AqC6-oldH are not as prominent as in the case AqC5-newH
and AqC6-newH\_p003 (or AqC6-newH\_p005, similarly).

Fig. \ref{AppA_bce} shows the position of our simulations in the
baryon conversion efficiency plot. 
We find that AqC5-newH has a lower baryon conversion efficiency than AqC5-oldH, 
while the baryon conversion efficiency is comparable when AqC6 simulations are analysed.  
Moreover, the lower the resolution is, the higher the stellar content of the simulated galaxy: 
this is independent of the adopted SPH scheme. However, such an effect is much 
more evident when the new SPH scheme is considered.
For completeness, total baryonic masses within $R_{\rm vir}$ 
for the AqC6 runs are:
$1.89 \cdot 10^{11}$ M$_{\odot}$ for AqC6-oldH, 
$1.99 \cdot 10^{11}$ M$_{\odot}$ for AqC6-newH\_p003, 
and $2.06 \cdot 10^{11}$ M$_{\odot}$ for AqC6-newH\_p005, at $z=0$.  

Our general result is that properties of galaxies simulated at higher resolution 
are significantly more sensitive to the accuracy of the hydrodynamic solver.

\section{Tuning the parameters of galactic outflow models}
\label{appendixB}

\begin{figure}
\newcommand{\captionfonts}{\small}
\hspace{-1.65ex}
\centering
\includegraphics[trim=0.4cm 0.4cm 0.35cm 0.4cm, clip, width=0.488\textwidth]{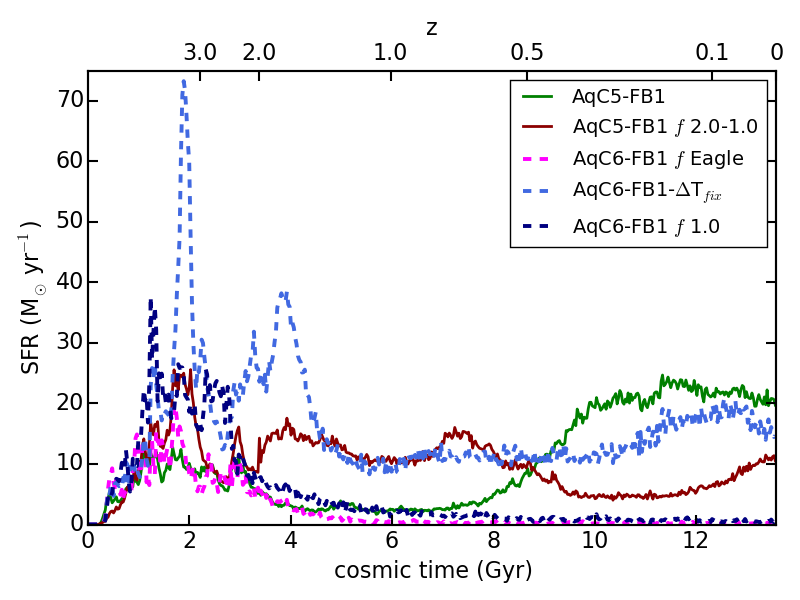} 
\caption{Evolution of the SFR for different simulations 
that adopt the FB1 outflow model.} 
\label{AppB_sfrFB1} 
\end{figure}

\begin{figure}
\newcommand{\captionfonts}{\small}
\hspace{-1.65ex}
\centering
\includegraphics[trim=0.4cm 0.4cm 0.35cm 0.2cm, clip, width=0.488\textwidth]{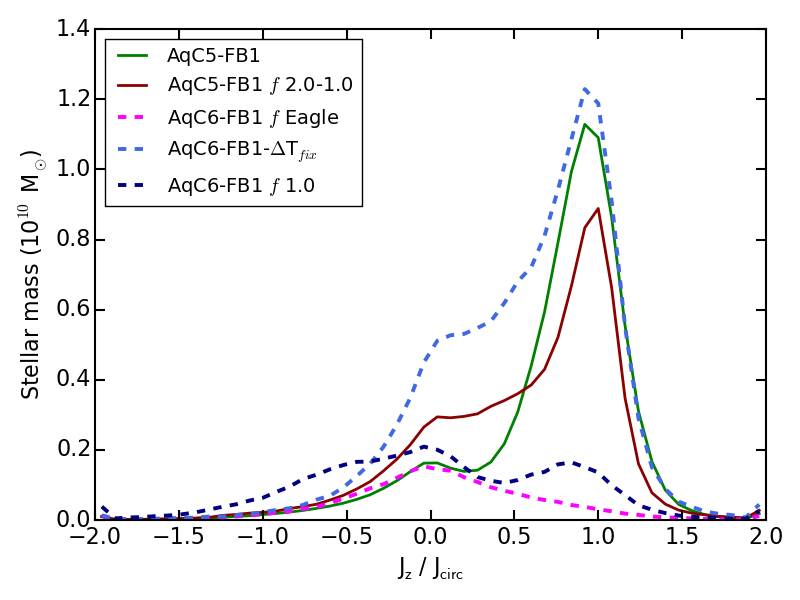} 
\caption{Stellar mass as a function of the circularity of stellar
  orbits at $z=0$ for different simulations adopting the FB1 galactic
  outflow model.}
\label{AppB_jcircFB1} 
\end{figure}

\begin{figure}
\newcommand{\captionfonts}{\small}
\hspace{-1.65ex}
\centering
\includegraphics[trim=0.4cm 0.4cm 0.35cm 0.4cm, clip, width=0.488\textwidth]{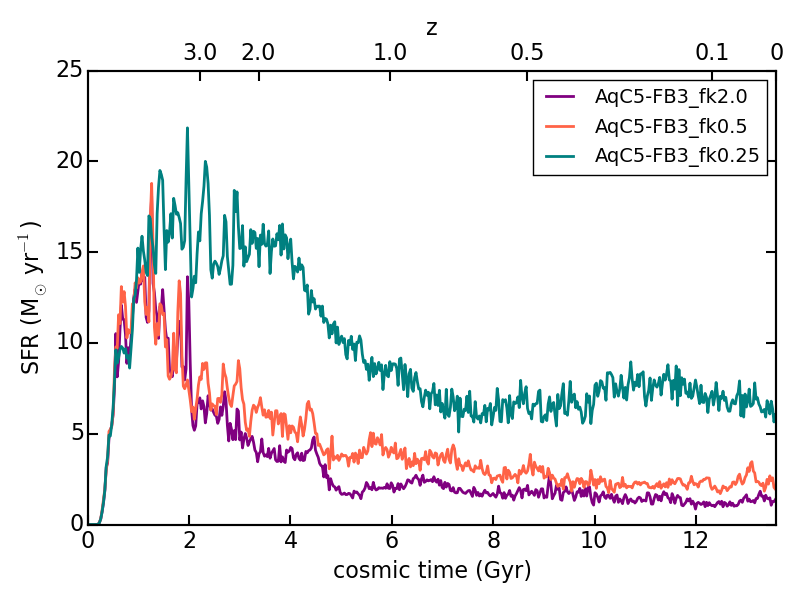} 
\caption{Evolution of the SFR for different simulations 
that adopt the FB3 outflow model.} 
\label{AppB_sfrFB3} 
\end{figure}

\begin{figure}
\newcommand{\captionfonts}{\small}
\hspace{-1.65ex}
\centering
\includegraphics[trim=0.4cm 0.4cm 0.35cm 0.2cm, clip, width=0.488\textwidth]{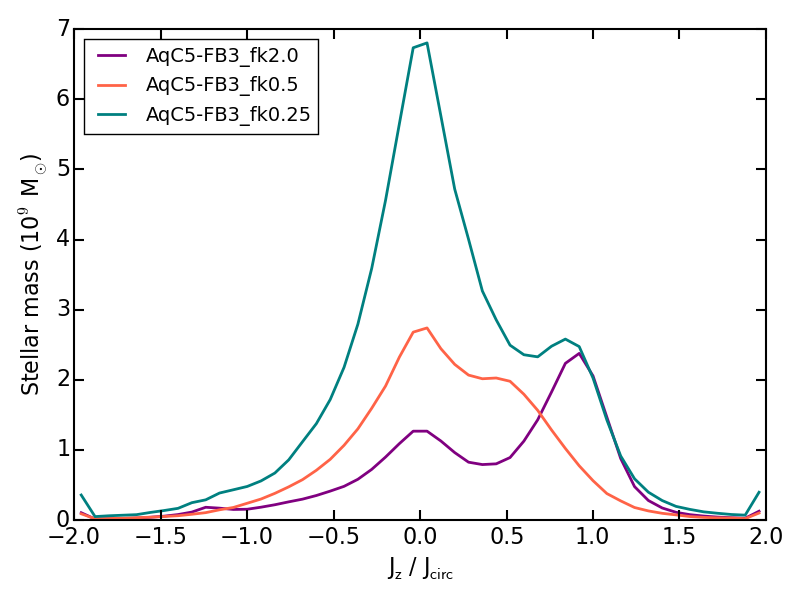} 
\caption{Stellar mass as a function of the circularity of stellar
  orbits at $z=0$ for different simulations adopting the FB3 galactic
  outflow model.}
\label{AppB_jcircFB3} 
\end{figure}

Feedback efficiency $f_{\rm fb, kin}$ is the most delicate parameter of galactic outflow
models. We report in this appendix our tests aimed at calibrating an optimal value of 
$f_{\rm fb, kin}$ and its impact on the resulting properties of the simulated galaxies. 
In particular, we concentrate here on the effect of feedback efficiency in the FB1 and 
FB3 outflow models.

Simulations adopting the galactic outflow model FB1 (Section
\ref{sec:DvS}) that we investigate in this appendix are: 
AqC5-FB1,
AqC5-FB1$f$2.0-1.0, 
AqC6-FB1$f$Eagle, 
AqC6-FB1-$\Delta$T$_{\rm fix}$, 
and AqC6-FB1$f$1.0 (see Table \ref{simmAPP}). 
Results are described in Figs \ref{AppB_sfrFB1} and \ref{AppB_jcircFB1}, where 
we show the SFR evolution and the circularity diagram for the set of FB1 runs.
Our fiducial simulation AqC5-FB1 adopts equation (\ref{FkEagle}), i.e. the density- and 
metallicity-dependent feedback efficiency used in the reference Eagle 
simulation \citep{schaye2015eagle}, as discussed in Section \ref{sec:DvS}. 
The simulation AqC5-FB1$f$2.0-1.0 employs a redshift-dependent feedback 
efficiency (see below), with $f_{\rm fb, kin}$ spanning values between 
$f_{\rm fb, kin}^{\rm \,\,\,max}=2.0$ and $f_{\rm fb, kin}^{\rm \,\,\,min}=1.0$, 
where higher efficiency is adopted at higher redshift. 
The analytical expression for the redshift-dependent feedback efficiency reads:
\begin{equation}
f_{\rm fb, kin}(z) = f_{\rm fb, kin}^{\rm \,\,\,min} + (f_{\rm fb, kin}^{\rm \,\,\,max} 
- f_{\rm fb, kin}^{\rm \,\,\,min} ) \cdot \Theta( z - 2 )    \,\,\,,
\label{eqTheta}
\end{equation}
\noindent
where $\Theta(z)$ is the Heaviside step function\footnote{We also 
	investigated the effect of a smoother transition for the
	redshift evolution of the feedback efficiency. We verified
	that, by adopting the following analytical expression:
	\begin{equation}
               f_{\rm fb, kin}(z) = f_{\rm fb, kin}^{\rm \,\,\,min} + f_{\rm fb, kin}^{\rm \,\,\,max} \cdot \biggl( 0.5 \cdot \text{erf} \biggl( \frac{ z - 2 }{0.5} \biggr) + 0.5 \biggr)   \,\,\,,
	\end{equation}
	where $\text{erf}(z)$ is the error function, instead of using equation
	(\ref{eqTheta}), the SFR and the final outcome of simulations with model
	FB1 are left almost unaffected.
}.

Equation (\ref{eqTheta}) for the feedback efficiency has been introduced according to 
the phenomenological approach \citep[following][]{schaye2015eagle} of using a higher 
feedback efficiency at high redshift. The above expression
for $f_{\rm fb, kin}$ is simpler than equation (\ref{FkEagle}), and has been adopted to understand how 
different values for high-$z$ and low-$z$ feedback efficiency impact on SFR. By tuning the parameters 
$f_{\rm fb, kin}^{\rm \,\,\,max}$ and $f_{\rm fb, kin}^{\rm \,\,\,min}$, it is possible to regulate high-$z$ and low-$z$ SFR. 

\noindent
Redshift $z=2$ has been identified as the approximate epoch at which the bulge formation phase ends 
and the SF starts building up the disc component. 
After an exploration of the parameter space, we found that $f_{\rm fb, kin}^{\rm \,\,\,max}=2.0$ 
and $f_{\rm fb, kin}^{\rm \,\,\,min}=1.0$ are the most suitable parameters in reproducing a typical
late-type galaxy SF history for our AqC5 simulations adopting the FB1 model. In particular, with such 
a choice, we are able to control low-redshift SF (see Fig. \ref{AppB_sfrFB1}): ongoing SFR 
in the simulation AqC5-FB1$f$2.0-1.0 has been reduced by half with respect to AqC5-FB1. 

Figs \ref{AppB_sfrFB1} and \ref{AppB_jcircFB1} describe the evolution of SFR and 
the distribution of the stellar mass as a function of the circularity of stellar orbits for the 
simulations adopting the galactic outflow model FB1, respectively. In these figures, we also 
include results from AqC6 resolution. 

AqC6-FB1$f$Eagle is a simulation analogous to AqC5-FB1. At variance with the higher resolution simulation, 
AqC6-FB1$f$Eagle does not succeed to form a convincing disc galaxy, and it 
produces a spheroidal (see Fig. \ref{AppB_jcircFB1}).
AqC6-FB1-$\Delta$T$_{\rm fix}$ adopts the same feedback efficiency as the AqC6-FB1$f$Eagle 
(i.e. equation \ref{FkEagle}): in this simulation, we fix the temperature jump to the reference value of 
$\Delta T=10^{7.5}$ K (see Section \ref{sec:DvS}). We find that this model is not able to prevent 
an excess of high-redshift SF, thus ending up with a disc galaxy having an unrealistically 
massive bulge. 

The simulation AqC6-FB1$f$1.0 is analogous to AqC6-FB1$f$Eagle, but it adopts a constant 
$f_{\rm fb, kin}=1.0$. In this case, an SF burst at high redshift takes place (see Fig. \ref{AppB_sfrFB1}). 
As a consequence, a prominent bulge is produced, with a resulting too small stellar disc due to the 
drastically reduced late-time SF. 
By adopting just one value for the feedback efficiency throughout the simulation, we do not succeed 
in producing a disc-dominated galaxy with this model (see Fig. \ref{AppB_jcircFB1}). 
If the parameter $f_{\rm fb, kin}$ is high enough to quench the high-redshift SF, then it depletes 
the amount of gas available for a subsequent phase of disc formation at low $z$. 
This is in agreement with \citet{Crain2015}, who find that disc galaxies are characterized by a compact 
bulge and a too low low-redshift SFR if a constant stellar feedback efficiency is adopted. 

Figs \ref{AppB_sfrFB1} and \ref{AppB_jcircFB1} also show that this galactic outflow model is 
resolution dependent (by considering AqC5-FB1 and AqC6-FB1$f$Eagle). 
AqC5 simulations have an initial mass of gas particles that is $\sim 4.4$ times smaller than the initial 
mass of baryonic particles in the reference Eagle simulation \citep{schaye2015eagle}, while 
AqC6 simulations have gas particles with a mass $\sim 3.7$ times larger than that of particles 
of the same type in the reference Eagle simulation. 
This result suggests that this galactic outflow model promotes the formation of
disc-dominated galaxies when a resolution at least higher than AqC6 simulations is considered (unless 
further recalibration of parameters). 

As for the FB3 galactic outflow model (Section \ref{sec:FB3}), we here
compare the evolution of SFRs and the circularity diagrams at $z=0$
(Figs \ref{AppB_sfrFB3} and \ref{AppB_jcircFB3}, respectively) for
three simulations: AqC5-FB3\_fk2.0, AqC5-FB3\_fk0.5, and
AqC5-FB3\_fk0.25.  For the sake of conciseness, we present results
only for the variation of the feedback efficiency $f_{\rm fb, kin}$
that spans the values 2.0, 0.5 and 0.25, respectively, while keeping
the mass loading factor $\eta=3.0$ fixed (see also Section \ref{FBtune}).

We only vary the $f_{\rm fb, kin}$ parameter for two reasons: first,
it is the parameter to which the model is most sensitive. Second,
\citet{SpringelHernquist2003}, whose model this galactic outflow scheme
is inspired, adopted $f_{\rm fb, kin}=0.25$ for most of their
simulations, even if they admitted that this feedback efficiency can
assume values of order unity or even larger.  Fig. \ref{AppB_sfrFB3}
shows that a feedback efficiency larger than $f_{\rm fb, kin}=0.25$ is
needed in order to avoid a too high SFR, both at $z>1$ -- so as to quench 
SF in the bulge -- and at low redshift. In fact, Fig. \ref{AppB_jcircFB3}
demonstrates that simulated galaxies obtained for values of the
feedback efficiency smaller than the reference value $f_{\rm fb, kin}=2$
are mostly dominated by a dispersion-supported component
($J_{\rm z}/J_{\rm circ}$=0) rather than by a rotationally supported one
($J_{\rm z}/J_{\rm circ}$=1).

%%%%%%%%%%%%%%%%%%%%%%%%%%%%%%%%%%%%%%%%%%%%%%%%%%
%%%%%%%%%%%%%%%%%%%%%%%%%%%%%%%%%%%%%%%%%%%%%%%%%%

% Don't change these lines
\bsp	% typesetting comment
\label{lastpage}
\end{document}